\documentclass[a4paper,twoside,11pt]{article}
\usepackage{authblk}
\usepackage[labelfont=bf]{caption}
\usepackage{subcaption}
\usepackage{pgf,pgfarrows,pgfnodes,pgfautomata,pgfheaps}
\usepackage{pstricks}
\usepackage{pstricks-add}
\usepackage{sidecap}
\usepackage{etex}
\usepackage{colortbl}
\usepackage{pst-node}
\usepackage{pst-grad}
\usepackage{pst-text}
\usepackage{pst-rel-points}
\usepackage{amssymb}
\usepackage{mathptm}
\usepackage{amsmath}
\usepackage{arydshln}
\usepackage{rotating}
\usepackage[normalem]{ulem}
\usepackage{fancyhdr}
\usepackage{fancybox}
\usepackage{fancyvrb}
\usepackage{lmodern}	
\usepackage{url}
\usepackage{datetime}
\makeatletter
\DeclareMathSizes{10.95}{10.95}{10}{9}
\makeatother
\usepackage{paralist}
\usepackage{breakcites}
\usepackage{cite}
\usepackage[breaklinks=true]{hyperref}
\hypersetup{colorlinks,linkcolor={black},citecolor={black},urlcolor={black}}
\usepackage{breakurl}
\usepackage{setspace}
\usepackage{sectsty}
\newdateformat{titledate}{\THEDAY\ \monthname[\THEMONTH], \THEYEAR}
\newdateformat{mydate}{\monthname[\THEMONTH] \THEYEAR}
\emergencystretch=\maxdimen
\newcommand{\figrule}{\arrayrulecolor{gray} 
	\nointerlineskip \hrulefill \arrayrulecolor{black} }
\newcommand{\vrulesep}{\unskip \vrule depth -16pt width 1.5pt}

\newcommand{\myrightarrow}{.}
\newcommand{\lptr}{{\bfseries\footnotesize lptr}}
\newcommand{\rptr}{{\bfseries\footnotesize rptr}}
\newcommand{\Inn}{\text{\ttfamily In}} 
\newcommand{\Outn}{{\text{\ttfamily Out}}} 
\newcommand{\ls}{{\text{\ttfamily list}}}
\newcommand{\tree}{{\text{\ttfamily tree}}}
\newcommand{\dlist}{{\text{\ttfamily dlist}}}
\newcommand{\new}{{\text{\ttfamily new}}}
\newcommand{\Null}{{\text{\ttfamily null}}}

\newcommand{\cons}{{\text{\ttfamily cons}}}
\newcommand{\cdr}{{\text{\ttfamily cdr}}}
\newcommand{\car}{{\text{\ttfamily car}}}
\newcommand{\class}{{\text{\ttfamily class}}}
\newcommand{\struct}{{\text{\ttfamily union}}}
\newcommand{\union}{{\text{\ttfamily struct}}}
\newcommand{\malloc}{{\text{\ttfamily malloc}}}
\newcommand{\deref}{\text {\ttfamily deref}}

\newcommand{\p}{\tt p}
\newcommand{\T}{\tt t}
\newcommand{\U}{\tt u}
\newcommand{\V}{\tt v}
\newcommand{\W}{\tt w}
\newcommand{\X}{{\tt x}}
\newcommand{\XX}{\tt X}
\newcommand{\Y}{{\tt y}}

\newcommand{\Z}{\tt z}
\newcommand{\A}{\tt A}
\newcommand{\B}{\tt B}
\newcommand{\C}{\tt C}
\newcommand{\D}{\tt D}
\newcommand{\f}{\tt f}
\newcommand{\g}{\tt g}

\newcommand{\OO}{\tt o}
\renewcommand{\l}{\darkgray l}
\newcommand{\R}{\tt r}
\newcommand{\E}{\tt e}
\newcommand{\LEFT}{\tt {left}}
\newcommand{\RIGHT}{\tt {right}}
\newcommand{\HEAD}{\tt {head}}
\newcommand{\TAIL}{\tt {tail}}

\begin{document}
\title{\textbf{Heap Abstractions for Static Analysis}}
\author{{\Large Vini Kanvar} and {\Large Uday P. Khedker}}
\affil{Department of Computer Science and Engineering \\
Indian Institute of Technology Bombay \\
Email: \tt\{{vini,uday\}@cse.iitb.ac.in}}
\titledate
\maketitle
\thispagestyle{empty}

\begin{abstract}
Heap data is potentially unbounded and seemingly arbitrary. As a consequence,
unlike stack and static memory, heap memory cannot be abstracted directly in
terms of a fixed set of source variable names appearing in the program being
analysed. This makes it an interesting topic of study and there is an abundance
of literature employing heap abstractions. Although most studies have addressed
similar concerns, their formulations and formalisms often seem dissimilar and
some times even unrelated. Thus, the insights gained in one description of heap
abstraction may not directly carry over to some other description. This survey
is a result of our quest for a unifying theme in the existing descriptions of
heap abstractions.  In particular, our interest lies in the abstractions and
not in the algorithms that construct them.

In our search of a unified theme, we view a heap abstraction as consisting of
two features: a {\em heap model} to represent the heap memory and a {\em
summarization} technique for bounding the heap representation. We classify the
models as storeless, store based, and hybrid. We describe various summarization
techniques based on $k$-limiting, allocation sites, patterns, variables, other
generic instrumentation predicates, and higher-order logics. This approach
allows us to compare the insights of a large number of seemingly dissimilar
heap abstractions and also paves way for creating new abstractions by
mix-and-match of models and summarization techniques.
\end{abstract}

\section{Heap Analysis: Motivation}
\label{sec:motivation}
Heap data is potentially unbounded and seemingly arbitrary. Although there is a
plethora of literature on heap, the formulations and formalisms often seem
dissimilar. This survey is a result of our quest for a unifying theme in the
existing descriptions of heap.

\subsection{Why Heap?}

Unlike stack or static memory, heap memory allows on-demand memory allocation
based on the statements in a program (and not just variable declarations). Thus
it facilitates creation of flexible data structures which can outlive the
procedures that create them and whose sizes can change during execution. 
With processors becoming faster and memories becoming larger as well as faster,
the ability of creating large and flexible data structures increases. Thus
the role of heap memory in user programs
as well as design and implementation of programming languages becomes more
significant.

\subsection{Why Heap Analysis?}

The increasing importance of the role of heap memory naturally leads to a myriad
requirements of its analysis. Although heap data has been subjected to static as
well as dynamic analyses, in this paper, we restrict ourselves to static
analysis.

Heap analysis, at a generic level, provides useful information about heap data, i.e.
heap pointers or references. Additionally, it helps in discovering control flow
through dynamic dispatch resolution. Specific applications that can benefit
from heap analysis include program understanding, program refactoring,
verification, debugging, enhancing security, improving performance, compile time
garbage collection, instruction scheduling, parallelization etc. Further, some
of the heap related questions asked during various applications include
whether a heap variable points to null, does a
program cause memory leaks, are two pointer expressions aliased, is a heap
location reachable from a variable, are two data structures disjoint, 
and many others. Section~\ref{app:heap-analysis-applications}
provides an overview of applications of heap analysis.

\subsection{Why Heap Abstraction?}

Answering heap related questions using compile time heap analysis is a
challenge because of the temporal and spatial structure of heap memory
characterized by the following aspects. 

\begin{itemize}
\item {\em Unpredictable lifetime.} 
The lifetime of a heap object may not be restricted to the scope in which it is
created.  Although the creation of a heap object is easy to discover in a static
analysis, the last use of a heap object, and hence the most appropriate point of
its deallocation, is not easy to discover.

\item {\em Unbounded number of allocations.} Heap locations are created
on-demand as a consequence of the execution of certain statements. Since these
statement may appear in loops or recursive procedures, the size of a heap
allocated data structure may be unbounded.  Further, since the execution
sequence is not known at compile time, heap seems to have an arbitrary
structure.

\item {\em Unnamed locations.} Heap locations cannot be named in programs, only
their pointers can be named.  A compile time analysis of a heap manipulating
program therefore, needs to create appropriate symbolic names for heap memory
locations. This is non-trivial because unlike stack and static data, the
association between symbolic names and memory locations cannot remain fixed.

In principle, a program that is restricted only to stack and static data, can
be rewritten without using pointers. However, the use of pointers is
unavoidable for heap data because the locations are unnamed.  Thus a heap
analysis inherits all challenges of a pointer analysis of stack and static
data\footnote{ Pointer analysis is undecidable~\cite{r94,c03}. It is inherently
difficult because a memory location can be accessed in more than one way i.e.
via pointer aliases.  Therefore, pointer analysis requires uncovering indirect
manipulations of data and control flow. Additionally, modern features such as
dynamic typing, field accesses, dynamic field additions and deletions, implicit
casting, pointer arithmetic, etc., make pointer analysis even harder.} and adds
to them because of unpredictable lifetimes and unbounded number of allocations.

\end{itemize}

Observe that none of these aspects are applicable to stack or static memory
because their temporal and spatial structures are far easier to discover.  Thus
an analysis of stack and static data does not require building sophisticated
abstractions of the memory. Analysis of heap requires us to create abstractions
to represent unbounded allocations of unnamed memory locations which have
statically unpredictable lifetimes.  As described in
Section~\ref{sec:model-n-appr}, two features common to all heap abstractions
are:
\begin{itemize}
\item models of heap which represent the structure of heap memory, and
\item summarization techniques to bound the representations.
\end{itemize}
We use this theme to survey the heap abstractions found in the 
static analysis literature.

\subsection{Organization of the paper}
Section~\ref{sec:basic_concepts} presents the basic concepts.
Section~\ref{sec:model-n-appr} defines heap abstractions in terms of models and
summarization techniques. We categorize heap models as storeless, store based,
or hybrid and describe various summarization techniques.  These are generic
ideas which are then used in Sections~\ref{sec:storeless},
\ref{sec:storebased}, and~\ref{sec:hybrid} to describe the related
investigations in the literature in terms of the interactions between the heap
models and summarization techniques.  Section~\ref{sec:perspective} compares
the models and summarization techniques to explore the design choices and
provides some guidelines.  Section~\ref{app:heap-analysis-applications}
describes major heap analyses and their applications.
Section~\ref{app:engg.approx} mentions some notable engineering approximations
used in heap analysis.  Section~\ref{sec:related} highlights some literature
survey papers and book chapters on heap analysis.
Section~\ref{sec:conclusions} concludes the paper by observing the overall
trend.  Appendix~\ref{app:cjava} compares the heap memory view of C/C++ and
Java. 

\section{Basic Concepts}
\label{sec:basic_concepts}

In this section, we build the basic concepts required to explain the heap
abstractions in later sections. We assume Java like programs, which use program
statements: $\X := \new$, $\X := \Null$, $\X := \Y$, $\X.\f := \Y$, and $\X :=
\Y.\f$. We also allow program statements $\X.\f := \new$ and $\X.\f := \Null$ as
syntactic sugar. The dot followed by a field represents field dereference by a
pointer variable. For ease of understanding, we draw our programs as control flow
graphs. $\Inn_n$ and $\Outn_n$ denote the program point before and after program
statement $n$ respectively.

\subsection{Examples of Heap Related Information}
\label{sec:prop}
Two most important examples of heap information are aliasing and points-to
relations because the rest of the questions are often answered using them.
\begin{itemize}
\item In {\em alias analysis}, two pointer expressions are said to be {\em
      aliased} to each other if they evaluate to the set of same memory locations.
      There are three possible cases of aliases between two pointer expressions: 
      \begin{itemize}
      \item The two pointer expressions {\em cannot\/} alias in any execution instance of the
            program.
      \item The two pointer expressions {\em must\/} alias in every execution instance of the
            program.
      \item The two pointer expressions {\em may\/} alias in some execution instances but not
            necessarily in all execution instances.  
      \end{itemize}
\item A {\em points-to analysis} attempts to determine the addresses that a
      pointer holds. A points-to information also has three possible cases: {\em
      must-points-to}, {\em may-points-to}, and {\em cannot-points-to}.  
\end{itemize}
An analysis is said to perform a {\em strong update} if in some situations it
can remove some alias/points-to information on processing an assignment
statement involving indirections on the left hand side (for example, {\ttfamily
*x}  or {\ttfamily x}$\verb+->+${\ttfamily n} in C, or {\ttfamily x.n} in
Java).  It is said to perform a {\em weak update} if no information can be
removed. Strong updates require the use of must-alias/must-points-to
information whereas weak updates can be performed using may-alias/may-points-to
information in a flow-sensitive analysis\footnote{A flow-sensitive heap
analysis computes, at each program point, an abstraction of the memory, which
is a safe approximation of the memory created along all control flow paths
reaching the program point}.

\subsection{Soundness and Precision of Heap Analysis}
\label{sec:analysis-prop}

A static analysis computes information representing the runtime behaviour of
the program being analysed. Two important considerations in a static analysis
of a program are {\em soundness\/} and {\em precision\/}.  Soundness guarantees
that the effects of all possible executions of the program have been included
in the information computed. Precision is a qualitative measure of the amount
of spurious information which is the information that cannot correspond to any
execution instance of the program; lesser the spurious information, more
precise is the information.

Applications involving program transformations require sound analyses because
the transformations must be valid for all execution instances.  Similarly
applications involving verification require a sound approximation of the
behaviour of all execution instances. On the other hand error detection or
validation applications can afford to compromise on soundness and may not cover
all possible execution paths.

When an analysis computes information that must hold for all execution instances
of a program, soundness is ensured by under-approximation of the information.
When it computes information that may hold in some execution instances,
soundness is ensured by over-approximation of the information. Precision is
governed by the extent of over- or under-approximation introduced in the
process.


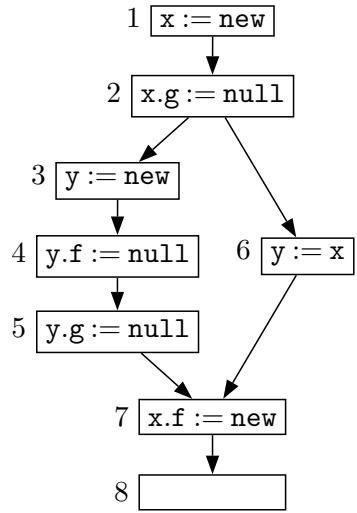
\begin{figure}[t]
\begin{center}
  {
    \psset{unit=.25mm}
    \psset{linewidth=.2mm}
	\psset{linewidth=.2mm,arrowsize=5pt,arrowinset=0}
    \begin{pspicture}(-60,-198)(114,75)

    \putnode{a}{origin}{40}{60}{1 \psframebox{$\X := \new$}\white\ 1}
    \putnode{b}{a}{0}{-40}{2 \psframebox{$\X.\g := \Null$}\white\ 2}
    \putnode{c}{b}{-50}{-45}{3 \psframebox{$\Y := \new$}\white\ 2}
    \putnode{d}{c}{0}{-40}{4 \psframebox{$\Y.\f := \Null$}\white\ 3}
    \putnode{e}{d}{0}{-40}{5 \psframebox{$\Y.\g := \Null$}\white\ 4}
    \putnode{f}{d}{100}{0}{6 \psframebox{$\Y := \X$}\white\ 6}
    \putnode{g}{e}{50}{-45}{7 \psframebox{$\X.\f := \new$}\white\ 7}
    \putnode{h}{g}{0}{-40}{8 \psframebox{\white $\X.\f := \new$}\white\ 7}
    
    \ncline{->}{a}{b}
    \ncline{->}{b}{c}
    \ncline{->}{c}{d}
    \ncline{->}{d}{e}
    \ncline{->}{b}{f}
    \ncline{->}{f}{g}
    \ncline{->}{e}{g}
    \ncline{->}{g}{h}

    \end{pspicture}
   }
\end{center}
\caption{Example to illustrate soundness and precision of information computed
by may and must analyses.}
\label{fig:safe-precise}
\figrule
\end{figure}


Consider the program in Figure~\ref{fig:safe-precise}. Let us consider a
may-null
(must-null) analysis whose result is a set of pointers that may (must) be null in
order to report possible (guaranteed) occurrences of null-dereference at
statement 8.  Assume that we restrict ourselves to the set $\{\X.\f, \X.\g,
\Y.\f, \Y.\g\}$. We know that both $\X.\g$ and $\Y.\g$ are guaranteed to be null
along all executions of the program. However, $\X.\f$ is guaranteed to be
non-null because of the assignment in statement 7 and $\Y.\f$ may or may not be
null depending on the execution of the program.  

\begin{enumerate}[(a)]
\item 
Consider the set $\{\X.\g, \Y.\g\}$ reported by an analysis at statement 8.
This set is:
	\begin{itemize}
	\item Sound for a must-null analysis because it includes all pointers
              that are guaranteed to be null at statement 8. Since it includes only those
              pointers that are guaranteed to be null, it is also precise.
              Any under-approximation of this set (i.e. a proper subset of this set)
              is sound but imprecise for a must-null analysis. An over-approximation of this
              set (i.e. a proper superset of this set) is unsound for must-null
              analysis because it would include a pointer which is not
              guaranteed to be null as explained in (b) below.
	\item Unsound for a may-null analysis because it excludes $\Y.\f$ which
	      may be null at statement~8. 
	\end{itemize}
\item 
On the other hand, the set $\{\X.\g, \Y.\g, \Y.\f\}$ reported  at statement 8
is:

	\begin{itemize}
	\item Sound for a may-null analysis because it includes all pointers
              that may be null at statement 8. Since it includes only those
              pointers that may be null, it is also precise.
              Any over-approximation of this set (i.e. a proper superset of this set)
              is sound but imprecise for a may-null analysis. Any under-approximation of this
              set (i.e. a proper subset of this set) is unsound for a may-null analysis
              because it would exclude a pointer which may be null as explained
in (a) above.
	\item Unsound for a must-null analysis because it includes $\Y.\f$ which
	      is not guaranteed be null at statement 8. 
	\end{itemize}
	\end{enumerate}

\section{Heap Abstractions}
\label{sec:model-n-appr}

In this section we define some generic ideas which are then used in the
subsequent sections to describe the work reported in the literature.

\subsection{Defining Heap Abstractions}

The goal of static analysis of heap memory is to abstract it at compile time to
derive useful information. We define a {\em heap abstraction} as the {\em heap
modeling} and {\em summarization} of the heap memory which  are introduced
below 

\begin{itemize}
\item
Let a snapshot of the runtime memory created by a program be called a concrete
memory. A {\em heap model} is a representation of one or more concrete
memories. It abstracts away less useful details and retains information that is
relevant to an application or analysis~\cite{mssf13}. For example, one may
retain only the reachable states in the abstract memory model.

We categorize the models as {\em storeless}, {\em store based}, and {\em
hybrid\/}.  They are defined in Section~\ref{sec:model}.

\item
Deriving precise runtime information of non-trivial programs, in general, is
not computable within finite time and memory (Rice theorem~\cite{r53}). For static
analysis of heap information, we need to {\em summarize} the modeled
information. Summarization should meet the following crucial requirements:
\begin{inparaenum}[(a)]
\item it should make the problem computable,
\item it should compute a sound approximation of the information corresponding
to any runtime instance, and
\item it should retain enough precision required by the application.
\end{inparaenum}

The summarizations are categorized based on
using {\em allocation sites}, {\em $k$-limiting}, {\em patterns}, {\em
variables}, other {\em generic instrumentation predicates}, or {\em
higher-order logics}.  
They are defined in Section~\ref{sec:appr}.
\end{itemize}


Some combinations of models and summarization techniques in common
heap abstractions are illustrated in Figure~\ref{fig:model-approximations}.


\newcommand{\Gheap}
{{
{\psset{unit=.25mm}
\psset{linewidth=.2mm,arrowsize=4pt,arrowinset=0}
		\begin{pspicture}(-30,-70)(380,25)
		\psrelpoint{origin}{q}{0}{-2}
		\rput(\x{q},\y{q}){\rnode{q}{
				\begin{tabular}{c}
				$\ \ \ \l_1$ \\
				$\X$
				\psframebox[fillstyle=solid,
				fillcolor=white,framesep=5]{$\l_3$} \end{tabular}}}
		\psrelpoint{origin}{j}{0}{-50}
		\rput(\x{j},\y{j}){\rnode{j}{
				\begin{tabular}{c}
				$\ \ \ \l_2$ \\
				$\Y$
				\psframebox[fillstyle=solid,
				fillcolor=white,framesep=5]{$\l_7$} \end{tabular}}}
		\psrelpoint{q}{s}{65}{-25}
		\rput(\x{s},\y{s}){\rnode{s}{
				\begin{tabular}{c}
				$\l_3$ \\
				\psframebox[fillstyle=solid,
				fillcolor=white,framesep=7]{
					$\f$ \psframebox[fillstyle=solid,
					fillcolor=white,framesep=5]{$\l_4$}
				} \end{tabular}}}
		\psrelpoint{s}{r}{70}{0}
		\rput(\x{r},\y{r}){\rnode{r}{
				\begin{tabular}{c}
				$\l_4$ \\
				\psframebox[fillstyle=solid,
				fillcolor=white,framesep=7]{
					$\f$ \psframebox[fillstyle=solid,
					fillcolor=white,framesep=5]{$\l_5$}
				} \end{tabular}}}
		\psrelpoint{r}{t}{70}{0}
		\rput(\x{t},\y{t}){\rnode{t}{
				\begin{tabular}{c}
				$\l_5$ \\
				\psframebox[fillstyle=solid,
				fillcolor=white,framesep=7]{
					$\f$ \psframebox[fillstyle=solid,
					fillcolor=white,framesep=5]{$\l_6$}
				} \end{tabular}}}
		\psrelpoint{t}{u}{70}{0}
		\rput(\x{u},\y{u}){\rnode{u}{
				\begin{tabular}{c}
				$\l_6$ \\
				\psframebox[fillstyle=solid,
				fillcolor=white,framesep=7]{
					$\f$ \psframebox[fillstyle=solid,
					fillcolor=white,framesep=5]{$\l_7$}
				} \end{tabular}}}
		\psrelpoint{u}{v}{70}{0}
		\rput(\x{v},\y{v}){\rnode{v}{
				\begin{tabular}{c}
				$\l_7$ \\
				\psframebox[fillstyle=solid,
				fillcolor=white,framesep=7]{
					$\f$ \fbox{\begin{tabular}{@{}c@{}} {\white 1} \\[-3ex] \gray ...\\[-2.5ex] {\white 1} \end{tabular}}
				} \end{tabular}}}
		
		\end{pspicture}
		}
}}

\newcommand{\Gheapone}
{\scalebox{.8}{
\large
{\psset{unit=.9mm}
\psset{linewidth=.3mm,arrowsize=5pt,arrowinset=0}
		\begin{pspicture}(-3,-25)(198,35)
		\psrelpoint{origin}{q}{10}{-2}
		\rput(\x{q},\y{q}){\rnode{q}{\pscirclebox[fillstyle=solid,
				fillcolor=white,framesep=0]{\begin{tabular}{c}$1$\\$n_0$\end{tabular}}}}
		\psrelpoint{origin}{j}{10}{22}
		\rput(\x{j},\y{j}){\rnode{j}{$\X$}}
		\ncline[linewidth=.4,arrowsize=7pt,arrowinset=0]{->}{j}{q}
		\psrelpoint{q}{s}{32}{0}
		\rput(\x{s},\y{s}){\rnode{s}{\pscirclebox[fillstyle=solid,
				fillcolor=white,framesep=0]{\begin{tabular}{c}$4$\\$n_1$\end{tabular}}}}
		\ncline{->}{q}{s}
		\naput{$\f$}
		\psrelpoint{s}{r}{32}{0}
		\rput(\x{r},\y{r}){\rnode{r}{\pscirclebox[fillstyle=solid,
				fillcolor=white,framesep=0]{\begin{tabular}{c}$6$\\$n_2$\end{tabular}}}}
		\ncline{->}{s}{r}
		\naput{$\f$}
		\psrelpoint{r}{k}{0}{22}
		\rput(\x{k},\y{k}){\rnode{k}{$\Y$}}
		\ncline[linewidth=.4,arrowsize=7pt,arrowinset=0]{->}{k}{r}

		\psrelpoint{r}{t}{32}{0}
		\rput(\x{t},\y{t}){\rnode{t}{\pscirclebox[fillstyle=solid,
				fillcolor=white,framesep=0]{\begin{tabular}{c}$4$\\$n_3$\end{tabular}}}}
		\ncline{->}{r}{t}
		\naput{$\f$}
		\psrelpoint{t}{u}{32}{0}
		\rput(\x{u},\y{u}){\rnode{u}{\pscirclebox[fillstyle=solid,
				fillcolor=white,framesep=0]{\begin{tabular}{c}$6$\\$n_4$\end{tabular}}}}
		\ncline{->}{t}{u}
		\naput{$\f$}
		\psrelpoint{u}{l}{0}{22}
		\rput(\x{l},\y{l}){\rnode{l}{$\Y$}}
		\ncline[linewidth=.4,arrowsize=7pt,arrowinset=0]{->}{l}{u}
		\psrelpoint{u}{m}{32}{0}
		\rput(\x{m},\y{m}){\rnode{m}{$\dots$}}
		\ncline[linewidth=.4,arrowsize=7pt,arrowinset=0]{->}{u}{m}

		\end{pspicture}
		}
}}

\newcommand{\Gheapthirteen}
{\scalebox{.8}{
\large
{\psset{unit=.9mm}
\psset{linewidth=.2mm,arrowsize=5pt,arrowinset=0}
		\begin{pspicture}(0,-10)(110,25)
		\psrelpoint{origin}{q}{10}{-2}
		\rput(\x{q},\y{q}){\rnode{q}{\pscirclebox[fillstyle=solid,
				fillcolor=white,framesep=5]{}}}
		\psrelpoint{origin}{j}{10}{11}
		\rput(\x{j},\y{j}){\rnode{j}{$\X$}}
		\ncline[linewidth=.4,arrowsize=7pt,arrowinset=0]{->}{j}{q}
		\psrelpoint{q}{s}{15}{0}
		\rput(\x{s},\y{s}){\rnode{s}{\pscirclebox[fillstyle=solid,
				fillcolor=white,framesep=5]{}}}
		\ncline{->}{q}{s}
		\naput{$\f$}
		\psrelpoint{s}{r}{15}{0}
		\rput(\x{r},\y{r}){\rnode{r}{\pscirclebox[fillstyle=solid,
				fillcolor=white,framesep=5]{}}}
		\ncline{->}{s}{r}
		\naput{$\f$}
		\psrelpoint{r}{k}{0}{13}
		\rput(\x{k},\y{k}){\rnode{k}{$\Y$}}
		\ncline[linewidth=.4,arrowsize=7pt,arrowinset=0]{->}{k}{r}
		\psrelpoint{r}{t}{15}{0}
		\rput(\x{t},\y{t}){\rnode{t}{\pscirclebox[fillstyle=solid,
				fillcolor=white,framesep=5]{}}}
		\ncline{->}{r}{t}
		\naput{$\f$}
		\psrelpoint{t}{u}{15}{0}
		\rput(\x{u},\y{u}){\rnode{u}{\pscirclebox[fillstyle=solid,
				fillcolor=white,framesep=5]{}}}
		\ncline{->}{t}{u}
		\naput{$\f$}
		\psrelpoint{u}{l}{0}{13}
		\rput(\x{l},\y{l}){\rnode{l}{$\Y$}}
		\ncline[linewidth=.4,arrowsize=7pt,arrowinset=0]{->}{l}{u}
		\psrelpoint{u}{m}{15}{0}
		\rput(\x{m},\y{m}){\rnode{m}{$\dots$}}
		\ncline[linewidth=.4,arrowsize=7pt,arrowinset=0]{->}{u}{m}

		\end{pspicture}
		}
}}
\newcommand{\Gheapfourteen}
{{
{\psset{unit=.25mm}
\psset{linewidth=.2mm,arrowsize=4pt,arrowinset=0}
		\begin{pspicture}(0,-15)(60,10)
		\psrelpoint{origin}{q}{10}{-2}
		\rput(\x{q},\y{q}){\rnode{q}{\psovalbox[fillstyle=solid,
				fillcolor=white,framesep=9]{}}}
		\psrelpoint{origin}{j}{10}{31}
		\rput(\x{j},\y{j}){\rnode{j}{$\X$}}
		\ncline{->}{j}{q}
		\psrelpoint{q}{s}{35}{0}
		\rput(\x{s},\y{s}){\rnode{s}{\psovalbox[fillstyle=solid,
				fillcolor=white,framesep=9]{}}}
		\ncline{->}{q}{s}
		\naput{$\f$}
		\psrelpoint{s}{r}{35}{0}
		\rput(\x{r},\y{r}){\rnode{r}{\psovalbox[fillstyle=solid,
				fillcolor=white,framesep=9]{}}}
		\ncline{->}{s}{r}
		\naput{$\f$}
		\nccurve[angleA=-45,angleB=-135,nodesep=-1.5,ncurv=2.5]{->}{r}{r}
		\naput{$\f$}
		\psrelpoint{r}{k}{0}{33}
		\rput(\x{k},\y{k}){\rnode{k}{$\Y$}}
		\ncline{->}{k}{r}

		\end{pspicture}
		}
}}

\newcommand{\Gheaptwo}
{{
{\psset{unit=.25mm}
\psset{linewidth=.2mm,arrowsize=4pt,arrowinset=0}
		\begin{pspicture}(0,-10)(110,25)
		\psrelpoint{origin}{q}{10}{-2}
		\rput(\x{q},\y{q}){\rnode{q}{\pscirclebox[fillstyle=solid,
				fillcolor=white,framesep=9]{}}}
		\psrelpoint{origin}{j}{10}{31}
		\rput(\x{j},\y{j}){\rnode{j}{$\X$}}
		\ncline{->}{j}{q}
		\psrelpoint{q}{s}{38}{0}
		\rput(\x{s},\y{s}){\rnode{s}{\pscirclebox[fillstyle=solid,
				fillcolor=white,framesep=9]{}}}
		\ncline{->}{q}{s}
		\naput{$\f$}
		\psrelpoint{s}{r}{38}{0}
		\rput(\x{r},\y{r}){\rnode{r}{\pscirclebox[fillstyle=solid,
				fillcolor=white,framesep=9]{}}}
		\ncline{->}{s}{r}
		\naput{$\f$}
		\psrelpoint{r}{k}{0}{33}
		\rput(\x{k},\y{k}){\rnode{k}{$\Y$}}
		\ncline{->}{k}{r}
		\psrelpoint{r}{t}{38}{0}
		\rput(\x{t},\y{t}){\rnode{t}{\pscirclebox[fillstyle=solid,
				fillcolor=white,framesep=9]{}}}
		\ncline{->}{r}{t}
		\naput{$\f$}
		\psrelpoint{t}{u}{38}{0}
		\rput(\x{u},\y{u}){\rnode{u}{\pscirclebox[fillstyle=solid,
				fillcolor=white,framesep=9]{}}}
		\ncline{->}{t}{u}
		\naput{$\f$}
		\psrelpoint{u}{l}{0}{33}
		\rput(\x{l},\y{l}){\rnode{l}{$\Y$}}
		\ncline{->}{l}{u}
		\psrelpoint{u}{m}{38}{0}
		\rput(\x{m},\y{m}){\rnode{m}{$\dots$}}
		\ncline{->}{u}{m}

		\end{pspicture}
		}
}}

\newcommand{\Gheapthree}
{{
{\psset{unit=.25mm}
\psset{linewidth=.2mm,arrowsize=4pt,arrowinset=0}
		\begin{pspicture}(0,-15)(60,10)
		\psrelpoint{origin}{q}{10}{-2}
		\rput(\x{q},\y{q}){\rnode{q}{\pscirclebox[fillstyle=solid,
				fillcolor=white,framesep=3]{$1$}}}
		\psrelpoint{origin}{j}{10}{31}
		\rput(\x{j},\y{j}){\rnode{j}{$\X$}}
		\ncline{->}{j}{q}
		\psrelpoint{q}{s}{35}{0}
		\rput(\x{s},\y{s}){\rnode{s}{\pscirclebox[fillstyle=solid,
				fillcolor=white,framesep=3]{$3$}}}
		\ncline{->}{q}{s}
		\naput{$\f$}
		\psrelpoint{s}{r}{35}{0}
		\rput(\x{r},\y{r}){\rnode{r}{\pscirclebox[fillstyle=solid,
				fillcolor=white,framesep=3]{$5$}}}
		\ncline{->}{s}{r}
		\naput{$\f$}
		\nccurve[angleA=-45,angleB=-135,nodesep=-1,ncurv=1]{->}{r}{s}
		\naput{$\f$}
		\psrelpoint{r}{k}{0}{33}
		\rput(\x{k},\y{k}){\rnode{k}{$\Y$}}
		\ncline{->}{k}{r}

		\end{pspicture}
		}
}}

\newcommand{\Gheapten}
{{
{\psset{unit=.25mm}
\psset{linewidth=.2mm,arrowsize=4pt,arrowinset=0}
		\begin{pspicture}(0,-10)(55,17)
		\psrelpoint{origin}{q}{10}{-2}
		\rput(\x{q},\y{q}){\rnode{q}{\pscirclebox[fillstyle=solid,
				fillcolor=white,framesep=9]{}}}
		\psrelpoint{origin}{j}{10}{32}
		\rput(\x{j},\y{j}){\rnode{j}{$\X$}}
		\ncline{->}{j}{q}
		\psrelpoint{q}{s}{43}{0}
		\rput(\x{s},\y{s}){\rnode{s}{\pscirclebox[fillstyle=solid,
				linewidth=.2mm,linestyle=dashed,dash=2.5pt 1.5pt,fillcolor=white,framesep=9]{}}}
		\ncline[linewidth=.2mm,linestyle=dashed,dash=2.5pt 1.5pt]{->}{q}{s}
		\naput{$\f$}
		\nccurve[linewidth=.2mm,linestyle=dashed,dash=2.5pt 1.5pt,angleA=-45,angleB=-135,nodesep=-1,ncurv=2]{->}{s}{s}
		\naput{$\f$}
		\psrelpoint{s}{r}{43}{0}
		\rput(\x{r},\y{r}){\rnode{r}{\pscirclebox[fillstyle=solid,
				fillcolor=white,framesep=9]{}}}
		\ncline[linewidth=.2mm,linestyle=dashed,dash=2.5pt 1.5pt]{->}{s}{r}
		\naput{$\f$}
		\psrelpoint{r}{k}{0}{34}
		\rput(\x{k},\y{k}){\rnode{k}{$\Y$}}
		\ncline{->}{k}{r}

		\end{pspicture}
		}
}}

\newcommand{\Gheapeleven}
{{
{\psset{unit=.25mm}
\psset{linewidth=.2mm,arrowsize=4pt,arrowinset=0}
		\begin{pspicture}(-3,-10)(198,10)
		\psrelpoint{origin}{q}{10}{-2}
		\rput(\x{q},\y{q}){\rnode{q}{\pscirclebox[fillstyle=solid,
				fillcolor=white,framesep=-4]{\begin{tabular}{c}$\langle{\X}\rangle$\end{tabular}}}}
		\psrelpoint{origin}{j}{10}{35}
		\rput(\x{j},\y{j}){\rnode{j}{$\X$}}
		\ncline{->}{j}{q}
		\psrelpoint{q}{s}{50}{0}
		\rput(\x{s},\y{s}){\rnode{s}
				{\pscirclebox[fillstyle=solid,
				fillcolor=white,framesep=-8]{\begin{tabular}{c}$\langle{\X}.\f\rangle$ \end{tabular}}}}
		\ncline{->}{q}{s}
		\naput{$\f$}
		\psrelpoint{s}{r}{60}{0}
		\rput(\x{r},\y{r}){\rnode{r}{\pscirclebox[fillstyle=solid,
				fillcolor=white,framesep=-8]
				{\begin{tabular}{c}$\langle{\X}.\f.\f,$ \\ $\Y\rangle$\end{tabular}}}}
		\ncline{->}{s}{r}
		\naput{$\f$}
		\psrelpoint{r}{n}{0}{47}
		\rput(\x{n},\y{n}){\rnode{n}{$\Y$}}
		\ncline{->}{n}{r}


		\psrelpoint{r}{t}{70}{0}
		\rput(\x{t},\y{t}){\rnode{t}{\pscirclebox[fillstyle=solid,
				fillcolor=white,framesep=-8]
				{\begin{tabular}{c}$\langle{\X}.\f.\f.\f$ \\ $\Y.\f\rangle$\end{tabular}}}}
		\ncline{->}{r}{t}
		\naput{$\f$}
		\psrelpoint{t}{u}{80}{0}
		\rput(\x{u},\y{u}){\rnode{u}{\pscirclebox[fillstyle=solid,
				fillcolor=white,framesep=-10]
				{\begin{tabular}{c}$\langle{\X}.\f.\f.\f.\f,$ \\ $\Y.\f.\f,\Y\rangle$\end{tabular}}}}
		\ncline{->}{t}{u}
		\naput{$\f$}
		\psrelpoint{u}{l}{0}{57}
		\rput(\x{l},\y{l}){\rnode{l}{$\Y$}}
		\ncline{->}{l}{u}
		\psrelpoint{u}{m}{62}{0}
		\rput(\x{m},\y{m}){\rnode{m}{$\dots$}}
		\ncline{->}{u}{m}
		\naput{$\f$}

		\end{pspicture}
		}
}}

\newcommand{\Gheaptwelve}
{{
{\psset{unit=.25mm}
\psset{linewidth=.2mm,arrowsize=4pt,arrowinset=0}
		\begin{pspicture}(-3,-10)(108,10)
		\psrelpoint{origin}{q}{10}{-2}
		\rput(\x{q},\y{q}){\rnode{q}{\pscirclebox[fillstyle=solid,
				fillcolor=white,framesep=-4]{\begin{tabular}{c}$\langle{\X}\rangle$\end{tabular}}}}
		\psrelpoint{origin}{j}{10}{35}
		\rput(\x{j},\y{j}){\rnode{j}{$\X$}}
		\ncline{->}{j}{q}
		\psrelpoint{q}{s}{60}{0}
		\rput(\x{s},\y{s}){\rnode{s}{\pscirclebox[fillstyle=solid,
				linewidth=.2mm,linestyle=dashed,dash=2.5pt 1.5pt,fillcolor=white,framesep=-9]
				{\begin{tabular}{c}$\langle{\X}(.\f)^+\rangle$\end{tabular}}}}
		\ncline[linewidth=.2mm,linestyle=dashed,dash=2.5pt 1.5pt]{->}{q}{s}
		\naput{$\f$}
		\nccurve[linewidth=.2mm,linestyle=dashed,dash=2.5pt 1.5pt,angleA=45,angleB=135,nodesep=-6,ncurv=2]{->}{s}{s}
		\nbput{$\f$}
		\psrelpoint{s}{u}{75}{0}
		\rput(\x{u},\y{u}){\rnode{u}{\pscirclebox[fillstyle=solid,
				fillcolor=white,framesep=-10]{\begin{tabular}{c}$\langle{\X}(.\f)^+.\f,$ \\ $\ \ \Y\rangle$\end{tabular}}}}
		\ncline[linewidth=.2mm,linestyle=dashed,dash=2.5pt 1.5pt]{->}{s}{u}
		\naput{$\f$}
		\psrelpoint{u}{l}{0}{55}
		\rput(\x{l},\y{l}){\rnode{l}{$\Y$}}
		\ncline{->}{l}{u}

		\end{pspicture}
		}
}}

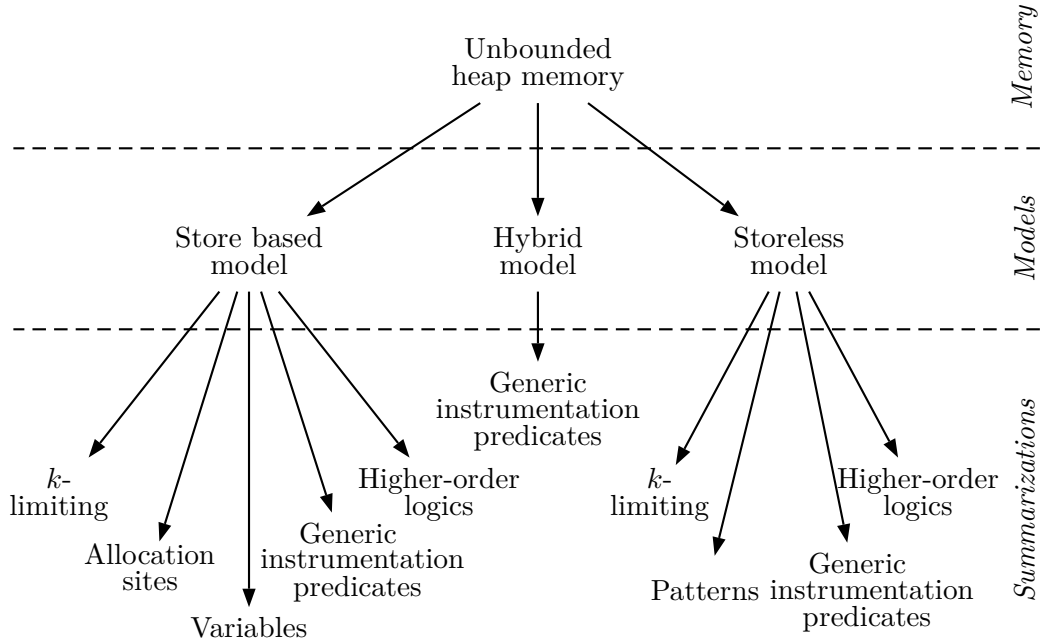
\begin{figure}[t]
\begin{center}
\psset{unit=1mm}
\begin{pspicture}(14,-12)(150,72)
\putnode{a}{origin}{84}{65}{\psframebox[linestyle=none]{%
			\renewcommand{\arraystretch}{.7}%
			\begin{tabular}{@{}c@{}}
			Unbounded\\
			heap memory\\
			\end{tabular}}}
\putnode{b1}{a}{-38}{-25}{\psframebox[linestyle=none]{%
			\renewcommand{\arraystretch}{.7}%
			\begin{tabular}{@{}c@{}}
			Store based\\
			model \\
			\end{tabular}}}
\putnode{c1}{a}{0}{-25}{\psframebox[linestyle=none]{%
			\renewcommand{\arraystretch}{.7}%
			\begin{tabular}{@{}c@{}}
			Hybrid\\
			model \\
			\end{tabular}}}
\putnode{d1}{a}{33}{-25}{\psframebox[linestyle=none]{%
			\renewcommand{\arraystretch}{.7}%
			\begin{tabular}{@{}c@{}}
			Storeless \\
			model \\
			\end{tabular}}}
\putnode{p1}{b1}{-25}{-32}{\psframebox[linestyle=none]{%
			\renewcommand{\arraystretch}{.7}%
			\begin{tabular}{@{}c@{}}
			$k$- \\
			limiting 
			\end{tabular}}}
\putnode{p2}{b1}{-13}{-42}{\psframebox[linestyle=none]{%
			\renewcommand{\arraystretch}{.7}%
			\begin{tabular}{@{}c@{}}
			Allocation \\
			sites 
			\end{tabular}}}
\putnode{p3}{b1}{0}{-50}{\psframebox[linestyle=none]{%
			\renewcommand{\arraystretch}{.7}%
			\begin{tabular}{@{}c@{}}
			Variables \\
			\end{tabular}}}
\putnode{p4}{b1}{13}{-41}{\psframebox[linestyle=none]{%
			\renewcommand{\arraystretch}{.7}%
			\begin{tabular}{@{}c@{}}
			Generic \\
			\ \ \ instrumentation \\
			\ \ predicates
			\end{tabular}}}
\putnode{p5}{b1}{25}{-32}{\psframebox[linestyle=none]{%
			\renewcommand{\arraystretch}{.7}%
			\begin{tabular}{@{}c@{}}
			Higher-order \\
			logics \\
			\end{tabular}}}
\putnode{q1}{c1}{0}{-21}{\psframebox[linestyle=none]{%
			\renewcommand{\arraystretch}{.7}%
			\begin{tabular}{@{}c@{}}
			Generic \\
			instrumentation \\
			predicates
			\end{tabular}}}
\putnode{r1}{d1}{-17}{-32}{\psframebox[linestyle=none]{%
			\renewcommand{\arraystretch}{.7}%
			\begin{tabular}{@{}c@{}}
			$k$- \\
			limiting
			\end{tabular}}}
\putnode{r2}{d1}{-11}{-45}{\psframebox[linestyle=none]{%
			\renewcommand{\arraystretch}{.7}%
			\begin{tabular}{@{}c@{}}
			\\ [-.4em]
			Patterns \\
			\\ [-.4em]
			\end{tabular}}}
\putnode{r3}{d1}{9}{-45}{\psframebox[linestyle=none]{%
			\renewcommand{\arraystretch}{.7}%
			\begin{tabular}{@{}c@{}}
			Generic \\
			\ \ \ instrumentation \\
			\ \ predicates
			\end{tabular}}}
\putnode{r4}{d1}{17}{-32}{\psframebox[linestyle=none]{%
			\renewcommand{\arraystretch}{.7}%
			\begin{tabular}{@{}c@{}}
			Higher-order \\
			logics \\
			\end{tabular}}}
{
\psset{arrowsize=1.75mm,arrowinset=0}
\ncline{->}{a}{b1}
\ncline{->}{a}{c1}
\ncline{->}{a}{d1}
\ncline{->}{b1}{p1}
\ncline[nodesepB=-1]{->}{b1}{p2}
\ncline{->}{b1}{p3}
\ncline{->}{b1}{p4}
\ncline{->}{b1}{p5}
\ncline{->}{c1}{q1}
\ncline[nodesepB=-1]{->}{d1}{r1}
\ncline{->}{d1}{r2}
\ncline{->}{d1}{r3}
\ncline{->}{d1}{r4}
}
\putnode{e1}{a}{64}{1}{\rotatebox{90}{\em Memory}}
\putnode{e2}{e1}{0}{-24}{\rotatebox{90}{\em Models}}
\putnode{e3}{e2}{0}{-34}{\rotatebox{90}{\em Summarizations }}
\putnode{p1}{e1}{2}{-12}{}
\putnode{p2}{p1}{-135}{0}{}
\ncline[linestyle=dashed]{p1}{p2}
\putnode{p1}{e2}{2}{-12}{}
\putnode{p2}{p1}{-135}{0}{}
\ncline[linestyle=dashed]{p1}{p2}
\end{pspicture}

\end{center}
\caption{Heap memory can be modeled as storeless, store based, or hybrid.  These
models are summarized using allocation sites, $k$-limiting, patterns, variables,
other generic instrumentation predicates, or higher-order logics.}
\label{fig:model-approximations}
\figrule
\end{figure}


\begin{figure}[t]
\begin{subfigure}[c]{.2\textwidth}
\begin{center}
  {
    \psset{unit=.25mm}
    \psset{linewidth=.2mm}
	\psset{linewidth=.2mm,arrowsize=4pt,arrowinset=0}
    \begin{pspicture}(-25,-155)(85,80)

    \putnode{a}{origin}{40}{60}{1 \psframebox{$\X := \new$}\white\ 1}
    \putnode{b}{a}{0}{-35}{2 \psframebox{$\Y := \X$}\white\ 2}
    \putnode{c}{b}{0}{-40}{3 \psframebox{$\Y \myrightarrow \f := \new$}\white\ 3}
    \putnode{d}{c}{0}{-35}{4 \psframebox{$\Y := \Y \myrightarrow \f$}\white\ 4}
    \putnode{e}{d}{0}{-35}{5 \psframebox{$\Y \myrightarrow \f := \new$}\white\ 5}
    \putnode{f}{e}{0}{-35}{6 \psframebox{$\Y := \Y \myrightarrow \f$}\white\ 6}
    
    \ncline{->}{a}{b}
    \ncline{->}{b}{c}
    \ncline{->}{c}{d}
    \ncline{->}{d}{e}
    \ncline{->}{e}{f}

    \ncloop[angleA=270,angleB=90,loopsize=60,linearc=5,offsetB=-15,arm=14]{->}{f}{c}

    \end{pspicture}
   }
\end{center}
\caption{Example}
\label{eg-model-appr}
\end{subfigure}
~
\begin{subfigure}[c]{.8\textwidth}
\begin{center}
\Gheap
\end{center}
\caption{\parbox[t]{.9\textwidth}{Execution snapshot showing an unbounded heap
graph at $\Outn_6$ of the program in Figure~\ref{eg-model-appr}. Here we have
shown the heap graph after iterating twice over the loop. Stack locations $\X$
and $\Y$ point to heap locations $l_3$ and $l_7$, respectively. Heap locations
$l_3$, $l_4$, $l_5$, and so on point to heap locations $l_4$, $l_5$, $l_6$, and
so on, respectively.}}
\label{fig:unbounded-heap}
\end{subfigure}
\caption{Running example to illustrate heap models and summarizations, which
have been shown in Figures~\ref{fig:storebased-appr}, \ref{fig:storeless-appr},
and \ref{fig:hybrid-appr}. In the program we have purposely duplicated
the program statements in order to create a heap graph where variable $\Y$ is at
even number of indirections from variable $\X$ after each iteration of the loop.
Not all summarization techniques are able to capture this information.}
\label{fig:demo}
\figrule
\end{figure}
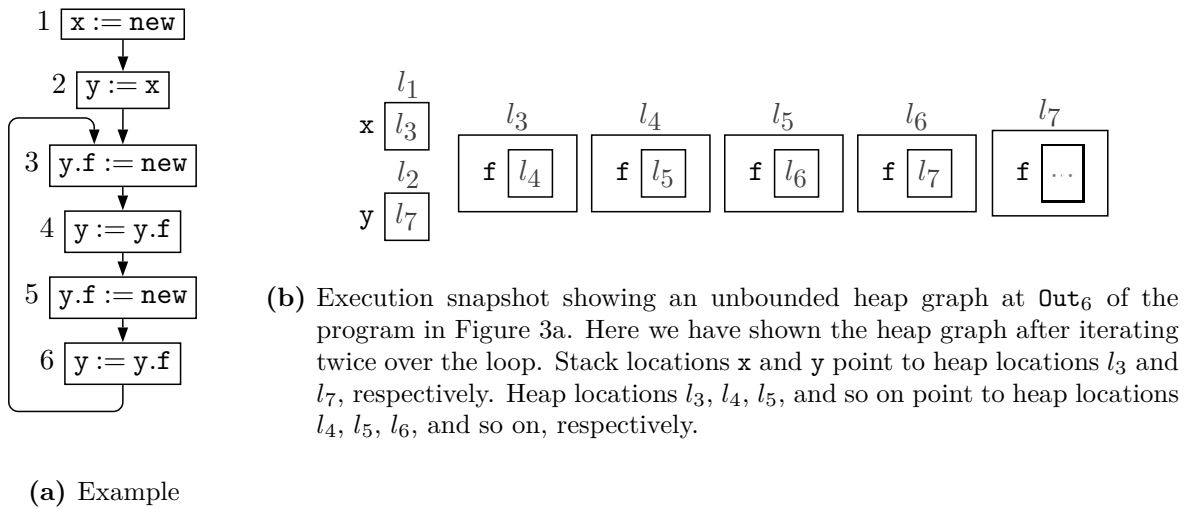


\begin{figure}[t]
\begin{center}
\psset{unit=1mm}
\begin{pspicture}(14,-20)(150,72)
\putnode{a}{origin}{84}{65}{\psframebox[linestyle=none]{%
			\renewcommand{\arraystretch}{.7}%
			\begin{tabular}{@{}c@{}}
			Unbounded\\
			heap memory\\
			\\ [-.7em]
			{\small (Figure~\ref{fig:unbounded-heap})} \\
			\end{tabular}}}
\putnode{b1}{a}{-38}{-25}{\psframebox[linestyle=none]{%
			\renewcommand{\arraystretch}{.7}%
			\begin{tabular}{@{}c@{}}
			Store based\\
			model \\
			\\ [-.7em]
			{\small (Figure~\ref{fig:storebased-appr})} \\
			\end{tabular}}}
\putnode{c1}{a}{0}{-25}{\psframebox[linestyle=none]{%
			\renewcommand{\arraystretch}{.7}%
			\begin{tabular}{@{}c@{}}
			Hybrid\\
			model \\
			\\ [-.7em]
			{\small (Figure~\ref{fig:hybrid-appr})} \\
			\end{tabular}}}
\putnode{d1}{a}{33}{-25}{\psframebox[linestyle=none]{%
			\renewcommand{\arraystretch}{.7}%
			\begin{tabular}{@{}c@{}}
			Storeless \\
			model \\
			\\ [-.7em]
			{\small (Figure~\ref{fig:storeless-appr})} \\
			\end{tabular}}}
\putnode{p1}{b1}{-25}{-34}{\psframebox[linestyle=none]{%
			\renewcommand{\arraystretch}{.7}%
			\begin{tabular}{@{}c@{}}
			$k$- \\
			limiting \\
			\\ [-.7em]
			{\small (Figure~\ref{fig:k-limiting-storebased})} \\
			\end{tabular}}}
\putnode{p2}{b1}{-13}{-47}{\psframebox[linestyle=none]{%
			\renewcommand{\arraystretch}{.7}%
			\begin{tabular}{@{}c@{}}
			Allocation \\
			sites \\
			\\ [-.7em]
			{\small (Figure~\ref{fig:alloc-storebased})} \\
			\end{tabular}}}
\putnode{p3}{b1}{0}{-58}{\psframebox[linestyle=none]{%
			\renewcommand{\arraystretch}{.7}%
			\begin{tabular}{@{}c@{}}
			Variables \\
			\\ [-.7em]
			{\small (Figure~\ref{fig:variables-storebased})} \\
			\end{tabular}}}
\putnode{p4}{b1}{13}{-48}{\psframebox[linestyle=none]{%
			\renewcommand{\arraystretch}{.7}%
			\begin{tabular}{@{}c@{}}
			Generic \\
			\ \ \ instrumentation \\
			\ \ predicates \\
			\\ [-.7em]
			\\
			\end{tabular}}}
\putnode{p5}{b1}{25}{-37}{\psframebox[linestyle=none]{%
			\renewcommand{\arraystretch}{.7}%
			\begin{tabular}{@{}c@{}}
			Higher-order \\
			logics \\
			\\
			\\ [-.7em]
			\\
			\end{tabular}}}
\putnode{q1}{c1}{0}{-22}{\psframebox[linestyle=none]{%
			\renewcommand{\arraystretch}{.7}%
			\begin{tabular}{@{}c@{}}
			Generic \\
			instrumentation \\
			predicates \\
			\\ [-.7em]
			{\small (Figure~\ref{fig:hybrid-2})} \\
			\end{tabular}}}
\putnode{r1}{d1}{-17}{-32}{\psframebox[linestyle=none]{%
			\renewcommand{\arraystretch}{.7}%
			\begin{tabular}{@{}c@{}}
			$k$- \\
			limiting \\
			\\ [-.7em]
			{\small (Figure~\ref{fig:k-limiting-storeless})} \\
			\end{tabular}}}
\putnode{r2}{d1}{-11}{-50}{\psframebox[linestyle=none]{%
			\renewcommand{\arraystretch}{.7}%
			\begin{tabular}{@{}c@{}}
			Patterns \\ 
			\\ [-.7em]
			{\small (Figure~\ref{fig:pattern-storeless})} \\
			\\
			\end{tabular}}}
\putnode{r3}{d1}{10}{-50}{\psframebox[linestyle=none]{%
			\renewcommand{\arraystretch}{.7}%
			\begin{tabular}{@{}c@{}}
			Generic \\
			\ \ instrumentation \\
			\ predicates \\
			\\ [-.7em]
			\\
			\end{tabular}}}
\putnode{r4}{d1}{17}{-34}{\psframebox[linestyle=none]{%
			\renewcommand{\arraystretch}{.7}%
			\begin{tabular}{@{}c@{}}
			Higher-order \\
			logics \\
			\\
			\\ [-.7em]
			\\
			\end{tabular}}}
{
\psset{arrowsize=1.75mm,arrowinset=0}
\ncline{->}{a}{b1}
\ncline{->}{a}{c1}
\ncline{->}{a}{d1}
\ncline{->}{b1}{p1}
\ncline[nodesepB=-1]{->}{b1}{p2}
\ncline{->}{b1}{p3}
\ncline{->}{b1}{p4}
\ncline{->}{b1}{p5}
\ncline{->}{c1}{q1}
\ncline[nodesepB=-1]{->}{d1}{r1}
\ncline{->}{d1}{r2}
\ncline{->}{d1}{r3}
\ncline{->}{d1}{r4}
}
\putnode{e1}{a}{64}{1}{\rotatebox{90}{\em Memory}}
\putnode{e2}{e1}{0}{-24}{\rotatebox{90}{\em Models}}
\putnode{e3}{e2}{0}{-34}{\rotatebox{90}{\em Summarizations }}
\putnode{p1}{e1}{2}{-12}{}
\putnode{p2}{p1}{-135}{0}{}
\ncline[linestyle=dashed]{p1}{p2}
\putnode{p1}{e2}{2}{-12}{}
\putnode{p2}{p1}{-135}{0}{}
\ncline[linestyle=dashed]{p1}{p2}
\end{pspicture}

\end{center}
\caption{Figures illustrating
various heap models and their summarizations for the program in
Figure~\ref{fig:demo}.}
\label{fig:model-approximations-with-illustrating-figures}
\figrule
\end{figure}
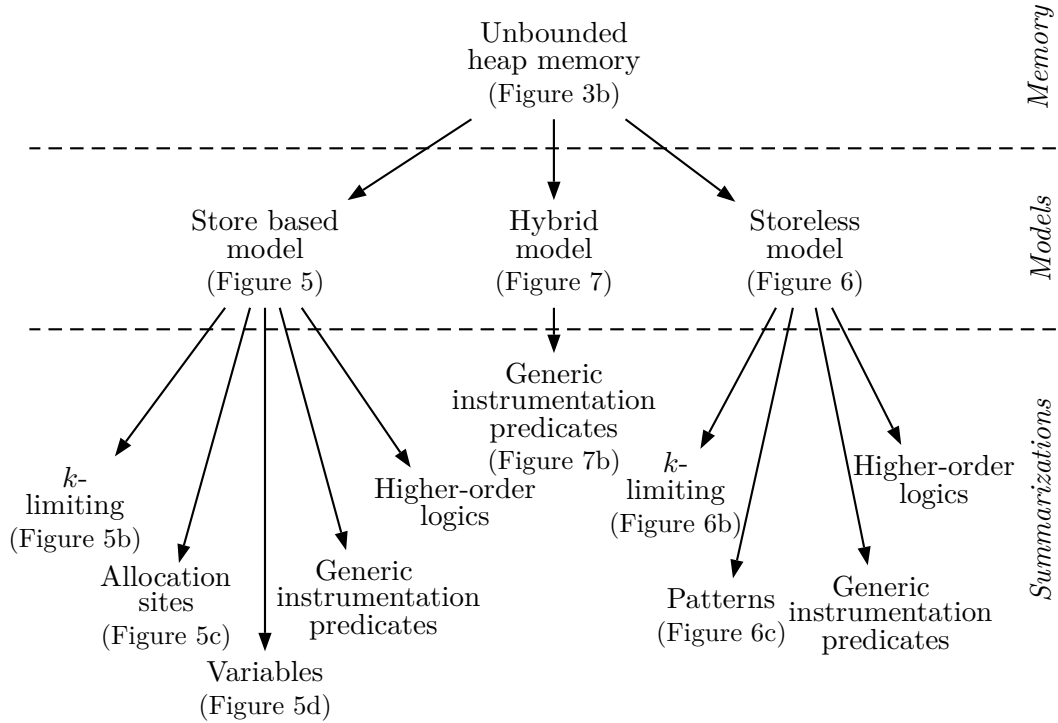


\subsection{Heap Models} 
\label{sec:model}
Heap objects are dynamically allocated, are unbounded in number, and do not
have fixed names. Hence, various schemes are used to name them at compile time.
The choice of naming them gives rise to different views of heap.  We define the
resulting models and explain them using a running example in
Figure~\ref{fig:demo}.
Figure~\ref{fig:model-approximations-with-illustrating-figures} associates the
models with the figures that illustrate them for our example program.

\begin{figure}[t]
\begin{center}
\begin{subfigure}[b]{1\textwidth}
\captionsetup{justification=centering}
\begin{center}
  {
    \psset{unit=.25mm}
    \begin{pspicture}(-60,-5)(150,50)

    \putnode{b}{origin}{0}{22}{\Gheaptwo}
    \end{pspicture}
   }
\end{center}
\caption{Unbounded store based model.}
\label{fig:unbounded-storebased}
\end{subfigure}

\begin{subfigure}[b]{.30\textwidth}
\captionsetup{justification=centering}
\begin{center}
  {
    \psset{unit=.25mm}
    \begin{pspicture}(-40,-15)(60,75)

    \putnode{b}{origin}{0}{22}{\Gheapfourteen}
    \end{pspicture}
   }
\end{center}
\caption{$k$-limiting $(k=2)$ summarization.}
\label{fig:k-limiting-storebased}
\end{subfigure}
~
\vrulesep
\begin{subfigure}[b]{.30\textwidth}
\captionsetup{justification=centering}
\begin{center}
  {
    \psset{unit=.25mm}
    \begin{pspicture}(-35,-18)(58,58)

    \putnode{b}{origin}{0}{22}{\Gheapthree}
    \end{pspicture}
   }
\end{center}
\caption{Allocation site based summarization.}

\label{fig:alloc-storebased}
\end{subfigure}
\vrulesep
~
\begin{subfigure}[b]{.30\textwidth}
\captionsetup{justification=centering}
\begin{center}
  {
    \psset{unit=.25mm}
    \begin{pspicture}(-30,-15)(80,65)

    \putnode{b}{origin}{0}{24}{\Gheapten}
    \end{pspicture}
   }
\end{center}
\caption{Variable based summarization.}
\label{fig:variables-storebased}
\end{subfigure}
\end{center}
\caption{Store based heap graphs at $\Outn_6$ for the program in
Figure~\ref{eg-model-appr}. Figures~\ref{fig:k-limiting-storebased},
\ref{fig:alloc-storebased}, and \ref{fig:variables-storebased} are bounded
representations of heap information in Figure~\ref{fig:unbounded-storebased}.
The numbers inside the graph nodes indicate the object's allocation sites in
the program in Figure~\ref{eg-model-appr}.}
\label{fig:storebased-appr}
\figrule
\end{figure}
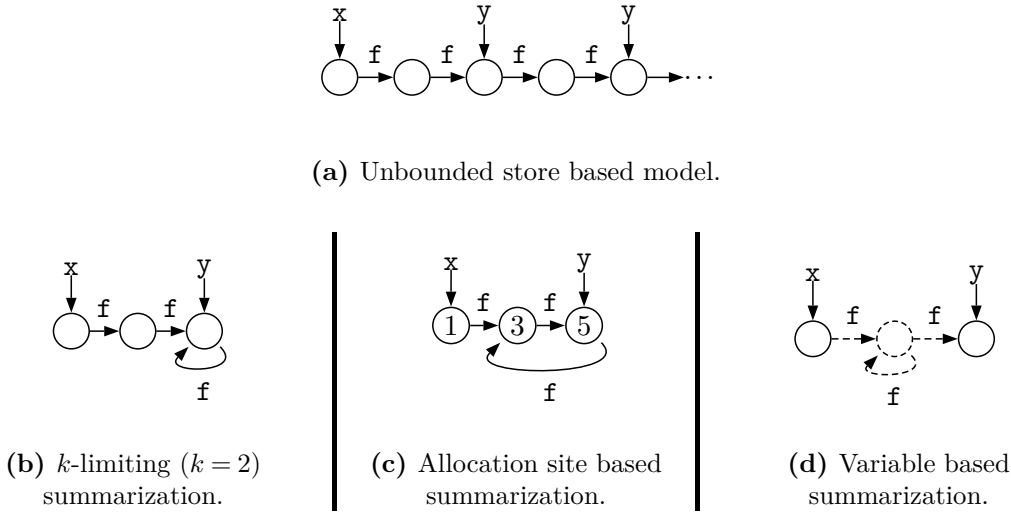

\begin{itemize}
\item {\em Store based model.} A store based model explicates heap locations in
terms of their addresses and generally represents the heap memory as a directed
graph~\cite{isy88,cwz90,r97,srw99,mrr02,shrvlgg00,bbhimv06,dhy06,gbc06,sbl11,cdoy11}.
The nodes of the graph represent locations or objects in the memory. An edge
\text{$x\rightarrow o_1$} in the graph denotes the fact that the pointer
variable $x$ may hold the address of object $o_1$. Since objects may have
fields that hold the addresses, we can also have a labelled edge
\text{$x\stackrel{f}{\rightarrow} o_1$} denoting the fact that the field $f$ of
object $x$ may hold the address of object $o_1$.  Let $V$ be the set of root
variables, $F$ be the set of fields names, and $O$ be the set of heap objects.
Then a concrete heap memory graph can be viewed as a collection of two
mappings: \text{${V \mapsto O}$} and \text{${O \times F \mapsto O}$}.  Observe
that this formalization assumes that $O$ is not fixed and is unbounded.  It is
this feature that warrants summarization techniques.

An abstract heap memory graph\footnote{In the rest of the paper, we refer to an
abstract heap memory graph simply by a memory graph.} is an approximation of
concrete heap memory graph which collects together all addresses that a
variable or a field may hold 
\begin{itemize}
\item at all execution instances of the same program point, or
\item across all execution instances of all program points.
\end{itemize}
Hence the ranges in the  mappings have to be extended to $2^O$ for an abstract
memory graph.  Thus a memory graph can be viewed as a collection of
mappings\footnote{In principle a graph may be represented in many ways. We
choose a collection of mappings for convenience.}
\text{${V \mapsto 2^O}$} and \text{${O \times F \mapsto 2^O}$}.

Figure~\ref{fig:demo} shows our running example and an execution snapshot of the
heap memory created and accessed by it. The execution snapshot shows stack
locations $\X$ and $\Y$ and heap locations with the addresses $l_3$, $l_4$,
$l_5$, $l_6$, and $l_7$. The address inside each box denotes the location that
the box points to. This structure is represented using a store based model in
Figure~\ref{fig:storebased-appr}.  Here the root variable $\Y$ points to a heap
location that is at even number of indirections via $\f$ from $\X$ after each
iteration of the loop in the program in Figure~\ref{eg-model-appr}.

\item {\em Storeless model.} The storeless model (originally proposed by
Jonkers \cite{j81}) views the heap as a collection of {\em access
paths}~\cite{j81,lr92,d94,gh96,as01,bil04,ksk07,ma12}. An {\em access path}
consists of a pointer variable which is followed by a sequence of
fields of a structure. The desired properties of both a concrete and an
abstract heap memory are stored as relations on access paths. The storeless
model does not explicate the memory locations or objects corresponding to these
access paths.  Given $V$ as the set of root variables and $F$ as the set of
field variable names, the set of access paths is defined as ${V \times F^*}$.
For example, access path $\X.\f.\f.\f.\f$ represents a memory location
reachable from $\X$ via four indirections of field $\f$. Observe that the
number of access paths is potentially infinite and the length of each access
path is unbounded. It is this feature that warrants summarization techniques. 

The heap memory at $\Outn_6$ of our running example (Figure~\ref{fig:demo}) is
represented using storeless model in Figure~\ref{fig:storeless-appr}.  The alias
information is stored as a set of equivalence classes containing access paths
that are aliased. Access paths $\X \myrightarrow \f \myrightarrow \f
\myrightarrow \f \myrightarrow \f$ and $\Y$ are put in the same equivalence
class at $\Outn_6$ because they are aliased at some point in the execution time
of the program.

\item {\em Hybrid model.} Chakraborty \cite{s12} describes a hybrid heap model which
represents heap structures using a combination of store based and storeless
models \cite{lh88,gh98,rbrsr05,dd12}. Heap memory of
Figure~\ref{fig:unbounded-heap} is represented using the hybrid model in
Figure~\ref{fig:hybrid-appr}. The model stores both objects (as in a store
based model) and access paths (as in a storeless model).

\end{itemize}


\subsection{Heap Summarization Techniques}
\label{sec:appr}

In the presence of loops and recursion, the size of graphs in a store based
model and the lengths of the access paths (and hence their number) in a
storeless model is potentially unbounded.
For fixpoint computation of heap information in a static
analysis, we need to approximate the potentially unbounded heap memory in terms of
summarized heap locations called {\em summarized objects}.  A summarized object
is a compile time representation of one or more runtime (aka concrete) heap
objects. 

\subsubsection{Summarization}

Summarized heap information is formally represented as Kleene closure or
wild card in regular expressions, summary node in heap graphs, or recursive
predicates.
\begin{itemize}
\item Summarized access paths are stored as regular expressions~\cite{d94} of
the form $\R \myrightarrow \E$, where $\R$ is a root variable and $\E$ is a
regular expression over field names defined in terms of concatenation
($\myrightarrow$), Kleene closure ($*$ and $+$ used as superscripts), and wild
card ($*$ used inline) operators.  For example, access path $\X \myrightarrow \f
\myrightarrow *$ represents an access path $\X \myrightarrow \f$ followed by
zero or more dereferences of any field.  Access path $\X (\myrightarrow \f)^*$
represents an access path $\X$ followed by any number of dereferences of field
$\f$.
\item Summarized heap graphs are stored by associating each graph node with a
boolean predicate indicating whether it is a summary node representing more than
one concrete heap location~\cite{cwz90}.  Observe that summary nodes may result
in spurious cycles in the graph if two objects represented by a summary node are
connected by an edge. 
\item Summarized collection of paths in the heap can also be stored in the
form of recursive predicates~\cite{as01,gbc06}.
\end{itemize}

\subsubsection{Materialization}

A collection of concrete nodes with the same property are summarized as a
summary node. However, after creation of a summary node, a program statement
could make a root variable point to one of the heap locations represented by
the summary node. Traditional summarization techniques~\cite{lh88,cwz90} do not
``un-summarize" this heap location from the summary node.  Thus in
traditional summarization techniques, a property discovered for a summarized
node may be satisfied by some of the represented heap locations and not
necessarily by all.  For example, when determining which pointer expressions
refer to the same heap location, all pointer expressions pointing to the same
summarized object will be recognized as possible candidates, even though some
of them may have been changed by new assignments. Therefore, a heap analysis
using this traditional summarization technique has a serious disadvantage: it
can answer only may-pointer questions. As a result traditional summarization
techniques cannot allow strong updates. In order to compute precise
must-pointer information, Sagiv et al. \cite{srw96} materialize
(``un-summarize") summarized objects (explained in Section~\ref{sec:var}).
Since this allows the locations that violate the common property to be removed
from the summary node and be represented by a newly created node, this opens up
the possibility that a summary node could represent a must property satisfied
by all locations represented by the summary node.  Performing strong updates is
an example of increased precision that can be facilitated by materialization.
Literature contains many approaches for must-pointer analysis, ranging from
relatively simple abstractions such as recency abstraction~\cite{br06} to
sophisticated shape analysis~\cite{srw96}. An analysis involving
materialization is expensive because of the additional examination required and
the possible increase in the size of the graph.

\begin{figure}[t]
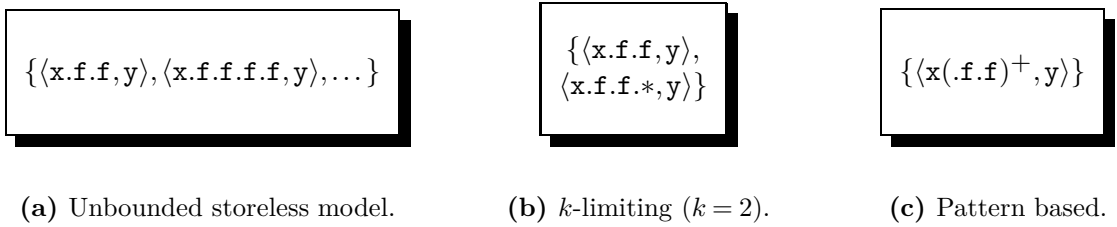

\begin{subfigure}[b]{.40\textwidth}
\begin{center}
  {
	\shadowbox{
	\begin{tabular}{@{}c@{}}
	 \\
	 $\{\langle \X.\f.\f, \Y \rangle,\langle \X.\f.\f.\f.\f, \Y \rangle, \dots\}$  \\
	\\ 
	\end{tabular}
	}
   }
\end{center}
\caption{Unbounded storeless model.}
\label{fig:unbounded-storeless}
\end{subfigure}
~
\begin{subfigure}[b]{.28\textwidth}
\begin{center}
  {
	\shadowbox{
	\begin{tabular}{@{}c@{}}
	\\ [-.5em]
	$\{\langle \X.\f.\f, \Y \rangle$, \\
	$\langle \X.\f.\f.*, \Y \rangle\}$  \\
	\\ [-.5em]
	\end{tabular}
	}
  }
\end{center}
\caption{$k$-limiting ($k=2$).}
\label{fig:k-limiting-storeless}
\end{subfigure}
~
\begin{subfigure}[b]{.28\textwidth}
\begin{center}
  {
	\shadowbox{
 	\begin{tabular}{@{}c@{}}
	 \\
	 $\{\langle\X(.\f.\f)^+, \Y \rangle\}$   \\
	\\ 
	\end{tabular}
	}
   }
\end{center}
\caption{Pattern based.}
\label{fig:pattern-storeless}
\end{subfigure}

\caption{Storeless model of heap graph at $\Outn_6$ of the program in
Figure~\ref{eg-model-appr}. Figures~\ref{fig:k-limiting-storeless} and
\ref{fig:pattern-storeless} are the bounded representations of heap information
in Figure~\ref{fig:unbounded-storeless}. Equivalence class of aliased access
paths is denoted by $\langle$ and $\rangle$.
}
\label{fig:storeless-appr}
\figrule
\end{figure}


\begin{figure}[t]
\begin{subfigure}[b]{.62\textwidth}
\begin{center}
  {
    \psset{unit=.25mm}
    \begin{pspicture}(-105,20)(290,127)

    \putnode{a}{origin}{0}{60}{\Gheapeleven}
    \end{pspicture}
   }
\end{center}
\caption{Unbounded hybrid model.}
\label{fig:hybrid-1}
\end{subfigure}
~
\vrulesep
\begin{subfigure}[b]{.38\textwidth}
\begin{center}
  {
    \psset{unit=.25mm}
    \begin{pspicture}(-60,20)(120,117)

    \putnode{a}{origin}{0}{60}{\Gheaptwelve}
    \end{pspicture}
   }
\end{center}
\caption{Variable based summarization.}
\label{fig:hybrid-2}
\end{subfigure}
\caption{Hybrid model of heap graph at $\Outn_6$ of the program in
Figure~\ref{eg-model-appr}. Figure~\ref{fig:hybrid-2} is the bounded
representation of the heap information in Figure~\ref{fig:hybrid-1}. Although
the access paths in the nodes can be inferred from the graph itself, they have
been denoted for simplicity.}
\label{fig:hybrid-appr}
\figrule
\end{figure}
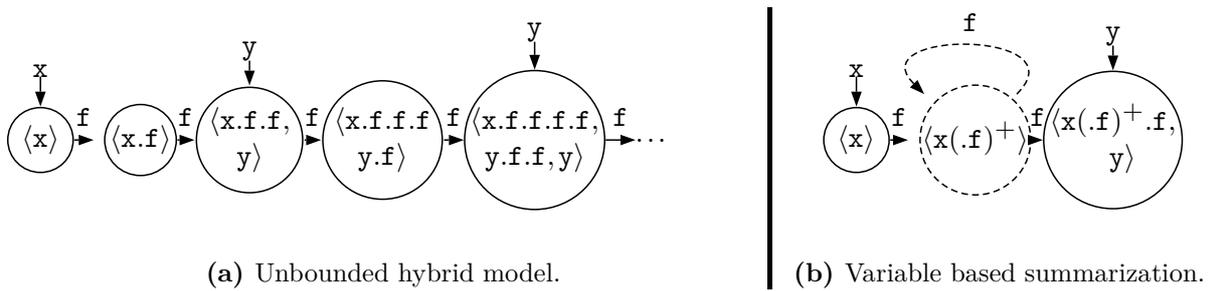


\subsubsection{Summarization Techniques}

We introduce below the six commonly found summarization techniques
using our running program of
Figure~\ref{eg-model-appr}. The figures illustrating these techniques have been
listed in Figure~\ref{fig:model-approximations-with-illustrating-figures}.
Note that our categorization is somewhat arbitrary in that some
techniques can be seen as special cases of some other techniques but we have
chosen to list them separately because of their prevalence.

The main distinction between various summarization techniques lies in how they map a
heap of potentially unbounded size to a bounded size. An implicit guiding
principle is to find a balance between precision and efficiency without
compromising on soundness.

\begin{enumerate}

\item {\em $k$-limiting summarization} distinguishes between the heap nodes
reachable by a sequence of up to $k$ indirections from a variable (i.e. it records
paths of length $k$ in the memory graph) and over-approximates the paths longer
than $k$.

$k$-limiting summarization has been performed on store based model~\cite{lh88}.
Figure~\ref{fig:k-limiting-storebased} represents a $k$-bounded representation
of the hybrid model in Figure~\ref{fig:unbounded-storebased}. For $k=2$, heap nodes
beyond two indirections are not stored. A self loop is created on the second
indirection (node corresponding to $\X \myrightarrow \f \myrightarrow \f$) to
over-approximate this information.  This stores spurious aliases for access
paths with more than $k=2$ indirections (for example, $\X \myrightarrow \f
\myrightarrow \f \myrightarrow \f$ and $\Y$ are spuriously marked as aliases at
$\Outn_6$).

$k$-limiting summarization has also been performed on storeless
model~\cite{jm79,lr92}.  This was proposed by Jones and Muchnick \cite{jm79}.
Figure~\ref{fig:k-limiting-storeless} represents a $k$-bounded representation
of the storeless model in Figure~\ref{fig:unbounded-storeless}. This also introduces
the same spurious alias pairs as in Figure~\ref{fig:k-limiting-storebased}.

\item {\em Summarization using allocation sites} merges heap objects that have
been allocated at the same program site. This technique is used for
approximating store based heap model~\cite{mrr02,br06} and hybrid
model~\cite{lh88}. It gives the same name to all objects allocated in a given
program statement.  The summarization is based on the premise that nodes
allocated at different allocation sites are manipulated differently, while the
ones allocated at the same allocation site are manipulated similarly.
Figure~\ref{fig:alloc-storebased} represents allocation site based
summarization heap graph of the store based model in
Figure~\ref{fig:unbounded-storebased}. Here all objects allocated at program
statements 3 and 5 are respectively clustered together. This summarization on
the given example does not introduce any spurious alias pairs.  We will show
spuriousness introduced due to this summarization in Section~\ref{sec:alloc-k}. 

\item {\em Summarization using patterns} merges access paths based on some
chosen patterns of occurrences of field names in the access paths. Pattern based
summarization has been used to bound the heap access
paths~\cite{d94,ksk07,ma12}.  Figure~\ref{fig:pattern-storeless} represents
pattern based summarization of the storeless model in
Figure~\ref{fig:unbounded-storeless}. For this example, it marks
every second dereference of field $\f$ (along the chain rooted by $\X$) as
aliased with $\Y$ which is precise.

\item {\em Summarization using variables} merges those heap objects that are
pointed to by the same set of root variables. For this, Sagiv et al. \cite{srw07} use the
predicate {\em pointed-to-by-x} on nodes for all variables $x$ to
denote whether a node is pointed to by variable $x$. Thus, all nodes with the
same pointed-to-by-$x$ predicate values are merged and represented by a summary
node.  Variable based summarization has been performed on store based heap
model~\cite{cwz90,srw96,srw98,bbhimv06}.  Figure~\ref{fig:variables-storebased}
represents variable based summarization of the store based model in
Figure~\ref{fig:unbounded-storebased}. After the first iteration of the loop of
the program in Figure~\ref{eg-model-appr}, there are three nodes---the first
pointed to by $\X$ and the third pointed to by $\Y$.  In the second iteration
of the loop, nodes reachable by access paths $\X \myrightarrow \f$, $\X
\myrightarrow \f \myrightarrow \f$, and $\X \myrightarrow \f \myrightarrow \f
\myrightarrow \f$ are not pointed to by any variable (as shown in
Figure~\ref{fig:unbounded-heap}). Therefore, they are merged together as a
summary node represented by dashed lines in
Figure~\ref{fig:variables-storebased} which shows the graphs after the first
and the second iterations of the loop. The dashed edges to and from summary
nodes denote indefinite connections between nodes. This graph also records $\X
\myrightarrow \f \myrightarrow \f \myrightarrow \f$ and $\Y$ as aliases at
$\Outn_6$ which is spurious.

Figure~\ref{fig:hybrid-2} is a variable based summarized representation of the
unbounded hybrid model in Figure~\ref{fig:hybrid-1}. A summary node (shown with
a dashed boundary in the figure) is created from nodes that are not pointed to
by any variable. Summarized access paths are appropriately marked on the nodes
in the hybrid model.

\item {\em Summarization using other generic instrumentation predicates} merge
those heap objects that satisfy a given
predicate~\cite{isy88,gh96,r97,srw99,shrvlgg00,wsr00,rbrsr05,br06,srw07}. 

Note that the
summarization techniques introduced above are all based on some predicate, as
listed below:
\begin{itemize}
\item $k$-limiting predicate: Is the heap location at most $k$ indirections
from a root variable?
\item Allocation site based predicate: Is the heap location allocated at a
particular program site? 
\item Pattern based predicate: Does the pointer expression to a heap location
have a particular pattern?
\item Variable based predicate: Is a heap location pointed to by a root
variable?
\end{itemize}
Since the above four are very commonly used predicates, we have separated them
out in our classification. 

Apart from these common predicates, summarization may
be based on other predicates too depending on the requirements of a client
analysis.  Some examples of these predicates are: is a heap location part of a
cycle, is a heap location pointed to by more than one object, is a heap location
allocated most recently at a particular allocation site, does the data in a heap 
node belong to a given type.  We group such possible predicates under {\em generic
instrumentation predicates}. A shape analysis framework~\cite{srw99,wsr00,srw07}
accepts any set of predicates as parameter to control the degree of efficiency
and precision in the summarization technique.

\item {\em Summarization using higher-order logics} includes those logics that
have more expressive power than first order (predicate) logic. Classical logics,
like Hoare logic~\cite{h69}, fail when they are used to reason about programs
that manipulate the heap. This is because classical logics assume that each
storage location has a distinct variable name~\cite{s07}, i.e. there are no
aliases in the memory. However, heap memory contains aliases of variables and
becomes difficult to analyse.  Therefore, heap specialized logics, such as
those listed below, have been used for heap abstraction. 
\begin{itemize}
\item Separation Logic~\cite{dhy06,gbc06,cdoy11}, 
\item Pointer Assertion Logic (PAL)~\cite{as01}, 
\item Weak Alias Logic (wAL)~\cite{bil04}, 
\item Flag Abstraction Language~\cite{lkr05,klzr06}, 
\end{itemize}
These are extensions of classical logics. For example, separation logic adds
separating connectives to classical logic to 
allow separate reasoning for independent parts of heap~\cite{bil03,s07}. 
Summarizations using higher-order logics differ from
summarizations using generic instrumentation predicates in the following sense:
the former use formal reasoning in logics specialized for heap memory.  Unlike
the latter, these techniques may be highly inefficient and even include
undecidable logics; therefore, in order to ensure their termination, they
generally need support with program annotations in the form of assertions and
invariants~\cite{jjsk97}.  In the following sections, we illustrate separation
logic, which is generally based on store based heap models, and  PAL, wAL, and
Flag Abstraction Language, which have been used on storeless heap modes.

\end{enumerate}

\newcommand{\Galiasone}
{{
{\psset{unit=.25mm}
\psset{linewidth=.2mm,arrowsize=4pt,arrowinset=0}
		\begin{pspicture}(-3,-3.5)(38,0)
		\psrelpoint{origin}{q}{8}{-2}
		\rput(\x{q},\y{q}){\rnode{q}{\pscirclebox[fillstyle=solid,
				fillcolor=white,framesep=6]{$\X$}}}
		\psrelpoint{origin}{j}{-20}{-2}
		\rput(\x{j},\y{j}){\rnode{j}{}}
		\ncline[linewidth=1,doubleline=true,arrowsize=4pt,arrowinset=0]{->}{j}{q}
		\psrelpoint{q}{s}{40}{0}
		\rput(\x{s},\y{s}){\rnode{s}{\pscirclebox[fillstyle=solid,
				fillcolor=white,framesep=2.6]{$\f_{5}$}}}
		\ncline{->}{q}{s}
		\psrelpoint{s}{r}{40}{0}
		\rput(\x{r},\y{r}){\rnode{r}{\pscirclebox[fillstyle=solid,
				fillcolor=white,framesep=2.6]{$\f_{6}$}}}
		\ncline{->}{s}{r}
		\nccurve[angleA=45,angleB=135,nodesep=-3,ncurv=1]{->}{r}{s}
		\psrelpoint{r}{t}{40}{0}
		\rput(\x{t},\y{t}){\rnode{t}{\pscirclebox[fillstyle=solid,
				fillcolor=white,framesep=2.6]{$\g_{3}$}}}
		\ncline{->}{r}{t}
		\nccurve[angleA=-45,angleB=-135,nodesep=-3,ncurv=.7]{->}{q}{t}

		\end{pspicture}
		}
}}

\newcommand{\Galiasfour}
{{
{\psset{unit=.25mm}
\psset{linewidth=.2mm,arrowsize=4pt,arrowinset=0}
		\begin{pspicture}(-3,-3.5)(38,0)
		\psrelpoint{origin}{q}{12}{-2}
		\rput(\x{q},\y{q}){\rnode{q}{\psovalbox[fillstyle=solid,
				fillcolor=white,framesep=9]{}}}
		\psrelpoint{origin}{j}{-23}{-2}
		\rput(\x{j},\y{j}){\rnode{j}{$\X$}}
		\ncline{->}{j}{q}
		\psrelpoint{q}{s}{45}{0}
		\rput(\x{s},\y{s}){\rnode{s}{\psovalbox[fillstyle=solid,
				fillcolor=white,framesep=9]{}}}
		\ncline{->}{q}{s}
		\naput{$\f$}
		\psrelpoint{s}{r}{45}{0}
		\rput(\x{r},\y{r}){\rnode{r}{\psovalbox[fillstyle=solid,
				fillcolor=white,framesep=9]{}}}
		\ncline{->}{s}{r}
		\naput{$\f$}
		\psrelpoint{r}{m}{0}{45}
		\rput(\x{m},\y{m}){\rnode{m}{$\Y$}}
		\ncline[linestyle=dotted,dotsep=1.5pt,linewidth=.4mm]{->}{m}{r}
		\psrelpoint{r}{t}{45}{0}
		\rput(\x{t},\y{t}){\rnode{t}{\psovalbox[fillstyle=solid,
				fillcolor=white,framesep=9]{}}}
		\ncline{->}{r}{t}
		\naput{$\f$}
		\psrelpoint{t}{u}{45}{0}
		\rput(\x{u},\y{u}){\rnode{u}{\psovalbox[fillstyle=solid,
				fillcolor=white,framesep=9]{}}}
		\ncline{->}{t}{u}
		\naput{$\f$}
		\psrelpoint{u}{n}{0}{45}
		\rput(\x{n},\y{n}){\rnode{n}{$\Y$}}
		\ncline[linestyle=dotted,dotsep=1.5pt,linewidth=.4mm]{->}{n}{u}
		\psrelpoint{u}{d}{45}{0}
		\rput(\x{d},\y{d}){\rnode{d}{{$\dots$}}}
		\ncline{->}{u}{d}
		\naput{$\f$}
		\psrelpoint{q}{v}{0}{-40}
		\rput(\x{v},\y{v}){\rnode{v}{\psovalbox[fillstyle=solid,
				fillcolor=white,framesep=9]{}}}
		\ncline{->}{q}{v}
		\naput{$\g$}
		\psrelpoint{r}{w}{0}{-40}
		\rput(\x{w},\y{w}){\rnode{w}{\psovalbox[fillstyle=solid,
				fillcolor=white,framesep=9]{}}}
		\ncline{->}{r}{w}
		\naput{$\g$}
		\psrelpoint{u}{x}{0}{-40}
		\rput(\x{x},\y{x}){\rnode{x}{\psovalbox[fillstyle=solid,
				fillcolor=white,framesep=9]{}}}
		\ncline{->}{u}{x}
		\naput{$\g$}

		\psrelpoint{v}{y}{0}{-40}
		\rput(\x{y},\y{y}){\rnode{y}{\psovalbox[fillstyle=solid,
				fillcolor=white,framesep=9]{}}}
		\ncline{->}{y}{v}
		\naput{$\g$}
		\psrelpoint{w}{z}{0}{-40}
		\rput(\x{z},\y{z}){\rnode{z}{\psovalbox[fillstyle=solid,
				fillcolor=white,framesep=9]{}}}
		\ncline{->}{z}{w}
		\naput{$\g$}
		\ncline{->}{y}{z}
		\naput{$\f$}
		\psrelpoint{z}{p}{0}{-45}
		\rput(\x{p},\y{p}){\rnode{p}{$\Z$}}
		\ncline[linestyle=dotted,dotsep=1.5pt,linewidth=.4mm]{->}{p}{z}
		\psrelpoint{x}{o}{0}{-40}
		\rput(\x{o},\y{o}){\rnode{o}{\psovalbox[fillstyle=solid,
				fillcolor=white,framesep=9]{}}}
		\ncline{->}{o}{x}
		\naput{$\g$}
		\ncline{->}{z}{o}
		\naput{$\f$}
		\psrelpoint{o}{l}{0}{-45}
		\rput(\x{l},\y{l}){\rnode{l}{$\Z$}}
		\ncline[linestyle=dotted,dotsep=1.5pt,linewidth=.4mm]{->}{l}{o}
		\psrelpoint{o}{p}{45}{0}
		\rput(\x{p},\y{p}){\rnode{p}{{$\dots$}}}
		\ncline{->}{o}{p}
		\naput{$\f$}

		\psrelpoint{y}{p}{-35}{0}
		\rput(\x{p},\y{p}){\rnode{p}{{$\W$}}}
		\ncline{->}{p}{y}

		\end{pspicture}
		}
}}

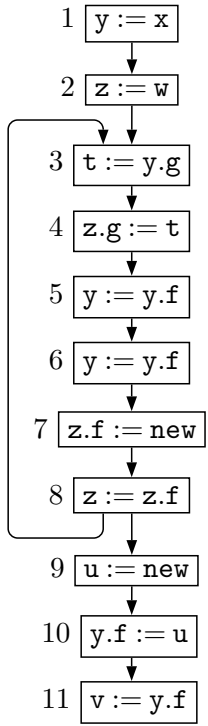
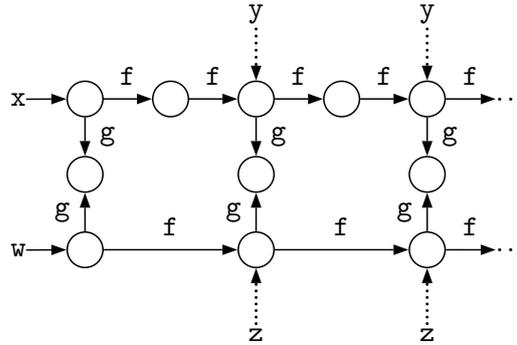
\begin{figure}[t]
\begin{subfigure}[c]{0.3\textwidth}
\begin{center}
  {
    \psset{unit=.25mm}
    \psset{linewidth=.2mm}
    \psset{linewidth=.2mm,arrowsize=4pt,arrowinset=0}
    \begin{pspicture}(-30,-310)(90,75)

    \putnode{a}{origin}{40}{60}{1 \psframebox{$\Y := \X$}\white\ 1}
    \putnode{b}{a}{0}{-35}{2 \psframebox{$\Z := \W$}\white\ 2}
    \putnode{c}{b}{0}{-40}{3 \psframebox{$\T := \Y \myrightarrow \g$}\white\ 3}
    \putnode{d}{c}{0}{-35}{4 \psframebox{$\Z \myrightarrow \g := \T$}\white\ 4}
    \putnode{e}{d}{0}{-35}{5 \psframebox{$\Y := \Y \myrightarrow \f$}\white\ 5}
    \putnode{f}{e}{0}{-35}{6 \psframebox{$\Y := \Y \myrightarrow \f$}\white\ 6}
    \putnode{g}{f}{0}{-35}{7 \psframebox{$\Z \myrightarrow \f := \new$}\white\ 7}
    \putnode{h}{g}{0}{-35}{8 \psframebox{$\Z := \Z \myrightarrow \f$}\white\ 7}
    \putnode{i}{h}{0}{-40}{\ 9 \psframebox{$\U := \new$}\white\ 7}
    \putnode{j}{i}{0}{-35}{10 \psframebox{$\Y \myrightarrow \f := \U$}\white\ 7}
    \putnode{k}{j}{0}{-35}{11 \psframebox{$\V := \Y \myrightarrow \f$}\white\ 7}
    
    \ncline{->}{a}{b}
    \ncline{->}{b}{c}
    \ncline{->}{c}{d}
    \ncline{->}{d}{e}
    \ncline{->}{e}{f}
    \ncline{->}{f}{g}
    \ncline{->}{g}{h}
    \ncline{->}{h}{i}
    \ncline{->}{i}{j}
    \ncline{->}{j}{k}

    \ncloop[angleA=270,angleB=90,loopsize=50,linearc=5,offset=-15,arm=13]{->}{h}{c}

    \end{pspicture}
   }
\end{center}
\caption{Example}
\label{fig:eg1-prog}
\end{subfigure}
~
\begin{subfigure}[c]{.7\textwidth}
\begin{center}
  {
    \psset{unit=.25mm}
    \psset{linewidth=.3mm}
	\psset{linewidth=.3mm,arrowsize=5pt,arrowinset=0}
    \begin{pspicture}(-15,-75)(265,115)

    \putnode{a}{origin}{40}{60}{\Galiasfour}
    \end{pspicture}
   }
\end{center}
\caption{\parbox[t]{.9\textwidth}{Execution snapshot showing an unbounded heap
graph at {\Outn}$_8$ of the program in Figure~\ref{fig:eg1-prog}. $\Y$ points to
$\X \myrightarrow \f \myrightarrow \f$ and $\Z$ points to $\W \myrightarrow \f$
in the first iteration of the loop. In the second iteration, $\Y$ points to $\X
\myrightarrow \f \myrightarrow \f \myrightarrow \f \myrightarrow \f$ and $\Z$
points to $\W \myrightarrow \f \myrightarrow \f$.}}
\label{heap}
\end{subfigure}
\caption{Running example to illustrate various heap summarization techniques.
We assume that all variables are predefined in the program.  Summarized
representations of the heap memory in Figure~\ref{heap} are shown on a storeless
model in Figure~\ref{eg1}, and are shown on a hybrid model in Figures~\ref{lh88}
and \ref{dd12}.}
\label{eg-1}
\figrule
\end{figure}



These heap summarization techniques can be combined judiciously.
Most investigations indeed involve multiple summarization techniques and their
variants by using additional ideas. Section~\ref{sec:perspective} outlines
the common factors influencing the possible choices and some broad guidelines.

\section{Summarization in Storeless Heap Model}
\label{sec:storeless}

As described in Section~\ref{sec:model}, a storeless heap model views the heap
memory as a collection of access paths. By contrast, the store based model
views the memory as a graph in which nodes are heap objects and edges are
fields containing addresses.  The view of storeless model may seem to be a
secondary view of memory that builds on a primary view of memory created by the
store based model.  In this section we present the summarization techniques for
a storeless model. Sections~\ref{sec:storebased} and~\ref{sec:hybrid} present
techniques for a store based model and a hybrid model, respectively.

\subsection{$k$-Limiting Summarization}
\label{sec:less-k}

\begin{figure}[t]
\begin{subfigure}[b]{.34\textwidth}
\begin{center}
  {
	\shadowbox{
	\begin{tabular}{@{}l@{}}
	$\langle \X \myrightarrow \g, \W \myrightarrow \g \rangle$ \\
	$\langle \X \myrightarrow \f \myrightarrow \f \myrightarrow \g, 
		\W \myrightarrow \f \myrightarrow \g \rangle$ \\
	$\langle \X \myrightarrow \f \myrightarrow \f \myrightarrow \f \myrightarrow *, 
		\W \myrightarrow \f \myrightarrow \f \myrightarrow * \rangle$
	\end{tabular}
	}
   }
\end{center}
\caption{\parbox[t]{.8\textwidth}{Alias pairs for variables $\X$ and $\W$ at
{\Outn}$_8$ for $k=3$~\cite{lr92}. }}
\label{lr92}
\end{subfigure}
~
\vrulesep
\begin{subfigure}[b]{.33\textwidth}
\begin{center}
  {
	\shadowbox{
	\begin{tabular}{@{}l@{}}
	$\langle \Y, \X ( \myrightarrow \f \myrightarrow \f)^+ \rangle$ \\
	$\langle \Z, \W ( \myrightarrow \f)^+ \rangle$ \\
	$\langle \X(.\f.\f)^+.\g,\W(.\f)^+.\g \rangle$
	\end{tabular}
	}
  }
\end{center}
\caption{\parbox[t]{.8\textwidth}{Aliases at {\Outn}$_8$~\cite{ma12}.\\}}
\label{ma12}
\end{subfigure}
\vrulesep
~
\begin{subfigure}[b]{.32\textwidth}
\begin{center}
  {
	\shadowbox{
	\begin{tabular}{@{}l@{}}
	$\langle \X \myrightarrow \f^{2i} \myrightarrow \g,\W \myrightarrow \f^i \myrightarrow \g \rangle$
	\end{tabular}
	}
  }
\end{center}
\caption{\parbox[t]{.8\textwidth}{Parameterised alias pairs for variables $\X$
and $\W$ at {\Outn}$_8$~\cite{d94}.}}
\label{d94}
\end{subfigure}

\begin{tabular}{c}
\end{tabular}

\begin{subfigure}[b]{.34\textwidth}
\begin{center}
  {
    \psset{unit=.25mm}
    \begin{pspicture}(0,20)(170,95)

    \putnode{a}{origin}{40}{60}{\Galiasone}
    \end{pspicture}
   }
\end{center}
\caption{\parbox[t]{.8\textwidth}{Live access graph at {\Inn}$_1$ when variable
$\T$ is live at $\Outn_{11}$~\cite{ksk07}.}}
\label{ksk07}
\end{subfigure}
~
\vrulesep
\begin{subfigure}[b]{.33\textwidth}
\begin{center}
  {
	\begin{tabular}{@{}ll@{}}
 	\begin{tabular}{|l|l|l|}
	\cline{2-3}
	\cline{2-3}
	\multicolumn{1}{c|}{}& \cellcolor{gray}$\X$ & \cellcolor{gray}$\Y$ \\
	\hline
	\cellcolor{gray}$\X$ & 1 & 1 \\ \hline
	\cellcolor{gray}$\Y$ & 0 & 1 \\ 
	\hline
	\end{tabular}
	&
	\begin{tabular}{|l|l|l|}
	\cline{2-3}
	\multicolumn{1}{c|}{}& \cellcolor{gray}$\X$ & \cellcolor{gray}$\Y$ \\
	\hline
	\cellcolor{gray}$\X$ & 1 & 1 \\ \hline
	\cellcolor{gray}$\Y$ & 1 & 1 \\ 
	\hline
	\end{tabular} \\
	Direction & Interference \\
	Matrix & Matrix
	\end{tabular}
   }
\end{center}
\caption{\parbox[t]{.8\textwidth}{Direction and interference matrices for
variables $\X$ and $\Y$ at $\Outn_8$~\cite{gh96}.}}
\label{matrix}
\end{subfigure}
\vrulesep
~
\begin{subfigure}[b]{.32\textwidth}
\begin{center}
  {
	\begin{tabular}{@{}l@{}}
 	\begin{tabular}{|c|l|c|}
	\cline{2-3}
	\cline{2-3}
	\multicolumn{1}{c|}{}& \cellcolor{gray}$\X$ & \cellcolor{gray}$\Y$ \\
	\hline
	\cellcolor{gray}$\X$ & $S$\hspace*{.05cm} & $\f^{\tiny{+}}$\hspace*{-.2cm} \\ \hline
	\cellcolor{gray}$\Y$ &  & $S$ \\ 
	\hline
	\end{tabular}\\
	Path Matrix
	\\
	\\
	\end{tabular}
   }
\end{center}
\caption{\parbox[t]{.85\textwidth}{Path matrix for
variables $\X$ and $\Y$ at $\Outn_8$~\cite{hn89}.\\}}
\label{path-matrix}
\end{subfigure}
\caption{Summarization techniques on a storeless model for the program in
Figure~\ref{fig:eg1-prog}: $k$-limiting (Figure~\ref{lr92}), pattern based
(Figure~\ref{ma12}, \ref{d94}, \ref{ksk07}), and other generic instrumentation
predicates based (Figure~\ref{matrix}, \ref{path-matrix}) summarization
techniques are shown.  (Equivalence class of aliased access paths is denoted by
$\langle$ and $\rangle$ in Figures~\ref{lr92}, \ref{ma12}, and \ref{d94}.)}
\label{eg1}
\figrule
\end{figure}
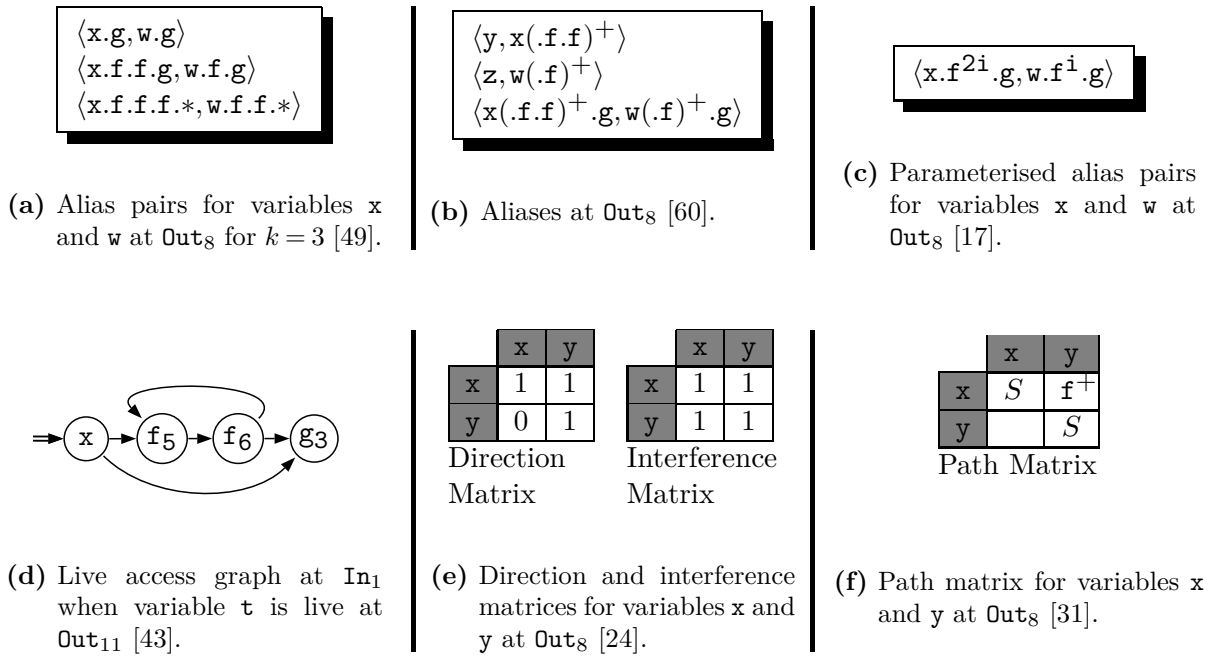

May-aliases have been represented as equivalence classes of $k$-limited access
paths~\cite{lr92}. For the program in Figure~\ref{fig:eg1-prog}, with its
unbounded memory graph in Figure~\ref{heap}, information bounded using
$k$-limiting summarization of access paths is shown in Figure~\ref{lr92} (alias
pairs of variables $\Y$ and $\Z$ are not shown for simplicity). The method
records alias pairs precisely up to $k$ indirections and approximates beyond
that. For $k=3$, fields up to three indirections from the root variables in the
access paths are recorded; those beyond three indirections are summarized with
a wild card (symbol $*$). Observe that this summarization induces the spurious
alias relationship $\langle \X \myrightarrow \f \myrightarrow \f \myrightarrow
\f, \W \myrightarrow \f \myrightarrow \f \rangle$. 
 

\subsection{Summarization Using Patterns}
\label{sec:less-patt}
A common theme in the literature has been to construct expressions consisting of
access paths approximated and stored either as a regular expression or a context
free language.

\begin{itemize}

\item
Consider the possibility of representing access paths in terms of regular
expressions~\cite{ma12}.  For example, let $\p$ be the initial access path
outside a program loop. After each iteration of the loop, if the value of $\p$
is advanced into the heap relative to its previous value via the field ${\LEFT}$ or
${\RIGHT}$, then the access path can be represented as $\p(.\LEFT \mid
.\RIGHT)^*$. The bounded alias information for the unbounded memory graph of
Figure~\ref{heap} is shown in Figure~\ref{ma12}. The example illustrates that
the method is able to identify $(\myrightarrow \f \myrightarrow \f)$ as the
repeating sequence of dereferences in the access path rooted at $\X$ and
$(\myrightarrow \f)$ as the repeating sequence of dereferences in the access
path rooted at $\W$.  The alias $\langle \X(.\f.\f)^*.\g,\W(.\f)^*.\g \rangle$
at $\Outn_8$, indicates that $\X.\f.\f.\g$ is aliased to $\W.\g$, which is
spurious.

The general problem of detecting possible iterative accesses can be undecidable
in the worst case~\cite{ma12}.  This is because a repeated advance into the heap
may arise from an arbitrarily long cycle of pointer relations. Therefore, the
focus in the work by  Matosevic and Abdelrahman \cite{ma12} remains on detecting only consecutive
repetitions of the same type of field accesses. For efficiency, finite state
automata are used to compactly represent sets of access paths that share common
prefixes.

\item
On similar lines, repetition of field dereferences in program loops can be
identified more efficiently and precisely by using the statement numbers where
the field dereference has occurred~\cite{ksk07}. This has been used to perform liveness
based garbage collection by computing live access graphs of the program. A live
access graph is a summarized representation of the live access paths\footnote{An
access path is live at a program point if it is possibly used after the program
point.} in the form of a graph; here a node denotes both a field name and the
statement number where the field dereference has occurred; the edges are used to
identify field names in an access path. 

A live access graph is illustrated in Figure~\ref{ksk07} for the program in
Figure~\ref{fig:eg1-prog}. Let us assume that variable $\T$ is live at
$\Outn_{11}$ in the program i.e. it is being used after statement 11. This
implies that access path $\Y.\g$ (or $\X(.\f.\f)^*.\g$) is live at $\Inn_3$ since
it is being accessed via variable $\T$ in the program loop. Therefore, access
paths $\X(.\f.\f)^*.\g$ are live at $\Inn_1$. These access paths are represented
as a summarized live access graph in Figure~\ref{ksk07}. The cycle over nodes
$\f_5$ and $\f_6$ denotes the Kleene closure in the access paths
$\X(.\f.\f)^*.\g$. This illustrates that the method is able to identify
$(.\f.\f)$ as a repeating sequence in the live access paths at $\Inn_1$. 

Basically, this is achieved by assigning the same name to the objects that are
dereferenced by a field at the same statement number. For example, the last
field in each of the access paths, $\X.\f$, $\X.\f.\f.\f$., and so on, is
dereferenced in statement 5; therefore, all these fields $\f$ (dereferenced in
statement 5) are represented by the same node $\f_5$ in Figure~\ref{ksk07}.
Similarly, the last fields $\f$ in each of the access paths, $\X(.\f.\f)^*.\g$,
are represented by the same node $\f_6$ because each of them is dereferenced in
statement 6. With the use of statement numbers, unlike the method by
Matosevic and Abdelrahman \cite{ma12}, this method can identify even non-consecutive repetitions of
fields efficiently.

In somewhat similar lines, liveness based garbage collection for functional
programs has been performed using a store based model by Inoue et al. \cite{isy88} and
Asati et al. \cite{askm14} (see Section~\ref{sec:other}).

\item
More precise expressions of access paths compared to those in the above methods
are constructed by parameterising the expressions with a counter to denote the
number of unbounded repetitions of the expression~\cite{d94}. Right-regular
equivalence relation on access paths helps in performing an exact summary of the
may-aliases of the program. The precisely bounded information for the unbounded
memory graph of Figure~\ref{heap} is illustrated in Figure~\ref{d94}.  The key
idea of the summarization is to represent the position of an element in a
recursive structure by counters denoting the number of times each recursive
component of the structure has to be unfolded to give access to this element.
This records the fact that the object reached after dereferencing $2i$ number of
$\f$ fields on access path $\X$ is aliased with the object reached after
dereferencing $i$ number of $\f$ fields on the access path $\W$. Due to the
parameterisation with $2i$ and $i$ on field $\f$ of both aliased access paths
which are rooted at variables $\X$ and $\W$ respectively, the method excludes
the spurious alias pairs derived from the alias information in
Figure~\ref{ma12}. 

\end{itemize}

\subsection{Summarization Using Generic Instrumentation Predicates}
\label{sec:less-pred}

Since the identification of patterns can be undecidable in the worst
case~\cite{ma12}, the power of summarization using patterns is limited by the
set of patterns that the algorithm chooses to identify. Instead of using a
fixed set of patterns, summarization using generic instrumentation predicates
enables a richer set of possibilities. We review this approach in this section.

\subsubsection*{A digression on shape analysis}
The use of heap analysis to determine shapes of the heap memory dates back
to the work by 
Jones and Muchnick \cite{jm79}.  Some of the notable works which also determine
shapes are enlisted below.
\begin{itemize}[--]
\item Analysis to determine shape using a storeless model has been presented by
Jones and Muchnick \cite{jm79}, Hendren and Nicolau \cite{hn89}, Ghiya and
Hendren \cite{gh96}, and others (presented in this section).
\item Analysis to determine shape using a store based model has been presented
by Chase et al. \cite{cwz90}, Sagiv et al. \cite{srw96,srw99,srw07}, Distefano
et al. \cite{dhy06}, Gotsman et al. \cite{gbc06}, Calcagno et al.
\cite{cdoy11}, and others  (see Section~\ref{sec:storebased}).
\item Analysis to determine shape using a hybrid model has been presented by
Rinetzky et al. \cite{rbrsr05} and others  (see Section~\ref{sec:hybrid}).
\end{itemize}

This study of the structure and shape of the heap has been called {\em shape
analysis}. Below we discuss shape analysis techniques used on a storeless
model.

\begin{itemize}
\item 
Hendren and Nicolau \cite{hn89} and Ghiya and Hendren \cite{gh96} classify the
shapes of the heap into tree, DAG, and cyclic graph, and choose to use the
following predicates on a storeless model.
\begin{enumerate}[(a)]
\item Direction relationship, which is true from pointer $\X$ to pointer $\Y$,
if $\X$ can reach $\Y$ via field indirections.
\item Interference relationship, which is true for pointers $\X$ and $\Y$, if a
common heap object can be accessed starting from $\X$ and $\Y$. This is a
symmetric relationship.
\end{enumerate}
Direction and interference relationships are stored in terms of
matrices as shown in Figure~\ref{matrix} for the
program in Figure~\ref{fig:eg1-prog}. Here, the heap has been encoded as access
paths in path matrices (direction and interference) at each program statement.
Direction relationship between pointers $\X$ and $\Y$ is true (represented by 1
in the direction matrix), since $\X$ reaches $\Y$ via indirections of field $\f$
at $\Outn_8$ of the program in Figure~\ref{fig:eg1-prog}. Since $\Y$ cannot
reach a node pointed to by $\X$ at $\Outn_8$, 0 is marked in the corresponding
entry of the direction matrix. Here, from the direction relationship, we can
derive that objects pointed by $\X$ and $\Y$ are not part of a cycle, since $\X$
has a path to $\Y$, but not vice versa.  Interference relationship between
pointers $\X$ and $\Y$ is true, since a common heap object can be accessed
starting from $\X$ and $\Y$. 

\item 
Storeless heap abstraction using reachability matrices can also be summarized
using regular expressions of path relationships between pointer
variables~\cite{hn89}. This is used to identify tree and DAG shaped heap data
structures by discovering definite and possible path relationships in the form
of path matrices at each program point. For variables $\X$ and $\Y$, an entry
in the path matrix, denoted as $p[\X, \Y]$, describes the path relationship
from $\X$ to $\Y$. In other words, each entry in the path matrix is a set of
path expressions of field dereferences made for pointer $\X$ to reach pointer
$\Y$.  Figure~\ref{path-matrix} shows the summarized path matrix for pointers
$\X$ and $\Y$ at $\Outn_8$ of the program in Figure~\ref{fig:eg1-prog}. Entry
$p[\X,\X] = \{S\}$ denotes that source and destination pointer variables are
the same. Entry $p[\X,\Y]=\{\f^+\}$ denotes that there exists a path from $\X$
to $\Y$ via one or more indirections of field $\f$. An empty entry $p[\Y,\X]$
denotes that there is no path from pointer $\Y$ to pointer $\X$.

This analysis calculates the part of the data structure that is between two
variables at each program point. The analysis can differentiate between a tree
and a DAG by the number of paths to a variable calculated in the path matrix.
The information is used for interference detection and parallelism extraction.
This approach is, however, restricted to acyclic data structures. Some follow-up
methods~\cite{h90,hn90,hdegjs93} also use path matrices for alias analysis of
heap allocated data structures.

\end{itemize}

\subsection{Summarization Using Higher-Order Logics}

To describe heap specific properties, various formalisms like Pointer Assertion
Logic, Weak Alias Logic, and Flag Abstraction Language have been proposed in the
literature.

\begin{itemize}
\item 
PALE (Pointer Assertion Logic Engine)~\cite{as01} is a tool that provides a
technique to check the partial correctness of programs annotated manually by
the programmer using PAL (Pointer Assertion Logic). The programmer encodes
everything in PAL including the program code, heap data structures, pre- and
post-conditions of various modules of the program, and loop invariants.  PAL is
an assertion language which is a monadic second-order logic, or more precisely,
WS2S (weak monadic second-order logic with two successors). Unlike first-order
logic, ordinary second-order logic allows quantification
(existential/universal) over predicates. ``Monadic" means that quantification
of only monadic predicates, i.e. sets, is allowed.  ``Weak" means that
quantification over finite sets is allowed. ``With two successors" means that
the formulae are interpreted over a tree domain (which is infinite). Although
it is technically ``two" successors, it is trivial to encode any fan-out.

\newcommand{\TLEFT}{{\bfseries left}}
\newcommand{\TRIGHT}{{\bfseries right}}
\newcommand{\TREE}{{\bfseries Tree}}
\newcommand{\ROOT}{{\bfseries root}}
Here is an example~\cite{as01} of a specification of type binary tree using PAL. 
\\[-1.4em]
\begin{center}
\begin{tabular}{l}
\hline
\\[-.8em]
{\ttfamily type \TREE \ = \{ } \\
{\ttfamily \ \ \quad \quad data \TLEFT,\TRIGHT:\TREE;} \\
{\ttfamily \ \ \quad \quad pointer \ROOT:\TREE[\ROOT$\langle$(\TLEFT+\TRIGHT)$^*\rangle$this}\\
{\ttfamily \ \ \quad \quad \quad \quad  \quad \quad \quad \quad \&
empty(\ROOT\^{}\TREE.\TLEFT \ union \ROOT\^{}\TREE.\TRIGHT)];} \\
\}
\\[-.8em]
\\ \hline
\\ [-1em]
\end{tabular}
\end{center}
{\white .}\\[-1em]
\renewcommand{\TLEFT}{{\text {\ttfamily left}}}
\renewcommand{\TRIGHT}{{\text {\ttfamily right}}}
\renewcommand{\TREE}{{\text {\ttfamily Tree}}}
\renewcommand{\ROOT}{{\text {\ttfamily root}}}
A memory graph consists of a ``backbone" which represents a spanning tree of
the underlying heap data structure. The memory links of the backbone are
encoded using {\ttfamily data} fields in PAL. Other memory links of the data
structure are encoded in PAL using {\ttfamily pointer} fields that are defined
on top of the backbone\footnote{Anders M{\o}ller, 04 May 2015, personal
communication.}.  The above example defines a heap location of type $\TREE$
which consists of $\TLEFT$, $\TRIGHT$, and $\ROOT$ links. The $\ROOT$ link is
an extra pointer which points to the root of the tree. It is defined with a
formula specified between the square brackets which is explained below. 
\begin{itemize}
\item The formula {\ttfamily $\ROOT\langle(\TLEFT$+$\TRIGHT)^*\rangle$this}
specifies that the $\ROOT$ location reaches {\ttfamily this} location via a
sequence of $\TLEFT$ or $\TRIGHT$ fields. The Kleene closure in this regular
expression helps in summarizing unbounded information. 
\item In PAL, formula $x$\^{}$T.p$ can be read as $x$\^{}$(T.p)$, where
\^{}$(T.p)$ represents a step upwards in the backbone i.e. backwards along field
$p$ from a location of type $T$ in order to reach a location pointed to by $x$.
In the above example, formulae $\ROOT$\^{}$\TREE.\TLEFT$ and
$\ROOT$\^{}$\TREE.\TRIGHT$ denote that location $\ROOT$ can be reached by
moving a step upwards in the backbone along $\TLEFT$ and $\TRIGHT$ fields from
a location of type $\TREE$.  The {\ttfamily empty()} formula above specifies
that locations having $\TLEFT$ or $\TRIGHT$ pointers to the $\ROOT$ location
must be empty.
\end{itemize}

Once the data structures, loop invariants, pre- and post-conditions are
specified by the programmer in PAL, PALE passes these PAL annotations to the
MONA tool~\cite{mona} for automatic verification of the program.  MONA reports
null-pointer dereferences, memory leaks, violations of assertions, graph type
errors, and verifies shape properties of data structures. 

Let us take an example of the predicates used in MONA logic.  Consider statement
5, $\Z:=\Y.\f$ which is executed on the linked list of the program in
Figure~\ref{eg2-prog}. The linked list can be specified in PAL as:
{\mbox{\ttfamily type Node = \{data f: Node;\}}}. For program points $i=\Inn_5$
and $j=\Outn_5$, MONA code~\cite{as01} generated for this statement is
\\[-1.4em]
\begin{center}
\begin{tabular}{l}
\hline
\\[-.8em]
$
\begin{aligned}
\text{\ttfamily memfailed\_$j$ () =} & 
	\text{ \ttfamily memfailed\_$i$() | null\_y\_$i$()} \\
\text{\ttfamily ptr\_z\_$j$($v$) =} & 
	\text{ \ttfamily ex2 $w$: ptr\_y\_$i$($w$) \& succ\_Node\_f\_$i$($w$,$v$)} \\
 \text{\ttfamily null\_z\_$j$() =} & 
	\text{ \ttfamily ex2 $w$: ptr\_y\_$i$($w$) \& null\_Node\_f\_$i$($w$)}
\end{aligned}
$
\\[-.8em]
\\ \hline
\\ [-1em]
\end{tabular}
\end{center}
{\white .}\\[-1em]
MONA code uses the following predicates in the above formula: 
\begin{itemize}
\item ${\text{\ttfamily memfailed}}()$ is true if a null dereference has occurred.
\item ${\text{\ttfamily null}}\_p()$ is true if pointer variable $p$ is null.
\item ${\text{\ttfamily ptr}}\_p(v)$ is true if the destination of pointer variable
$p$ is object $v$.
\item ${\text{\ttfamily succ}}\_t\_f(v,w)$ is true if object $v$ of type $t$ reaches
location $w$ via pointer field $f$.
\item ${\text{\ttfamily null}}\_t\_f(v)$ is true if object $v$ of type $t$ reaches a
null location via pointer field $f$.
\end{itemize}
The predicates in the above MONA code for statement 5 have been indexed with the
program points. For example, for program point $i$, the value of predicate
${\text{\ttfamily memfailed}}()$ is ${\text{\ttfamily memfailed}}\_i()$. Also,
${\text{\ttfamily ex}}2$ is an existential quantifier used for {\ttfamily Node}
object $w$ in the above MONA code.

In the above MONA code, the first line specifies that program point $j$ is in a
state of memory failure if either there was a memory failure at program point
$i$ or variable $\Y$ was null at $i$. The second line specifies that if object
$w$ is the destination of variable $\Y$, and $w$ reaches object $v$ via
pointer field $\f$, then $v$ is the destination of variable $\Z$. The third
line specifies that if object $w$ is the destination of variable $\Y$, and $w$
reaches a null location via pointer field $\f$, then variable $\Z$ is null.

Since MONA's logic is decidable, PALE will definitely reach a fixpoint. Due
to the overhead of manually adding annotations to the program, the technique is
suited for small input programs only.

\item 
Unlike PAL, which describes a restricted class of graphs, Weak Alias Logic (wAL)
deal with unrestricted graphs~\cite{bil04}. The user annotates the program with
pre- and post-conditions and loop invariants using wAL. The annotations are then
automatically verified for correctness. wAL is an undecidable monadic second
order logic that can describe the shapes of most recursive data structures like
lists, trees, and dags. Let $X$ and $Y$ be heap structures represented as a set
of regular expressions or access paths, and let $\rho$ be a regular expression
or a set of regular expressions. In wAL, $\langle X \rangle \rho$ specifies
that $X$ is bound to the heap, which is described by $\rho$. Also the formula
$X^{-1}Y$ in wAL denotes all the paths from $X$ to $Y$. Given below are
some predicates in wAL~\cite{bil04}.
\newcommand{\REACH}{{\text {\ttfamily reach}}}
\newcommand{\SHARE}{{\text {\ttfamily share}}}
\renewcommand{\TREE}{{\text {\ttfamily tree}}}
\renewcommand{\ROOT}{{\text {\ttfamily root}}}
\\[-1.4em]
\begin{center}
\begin{tabular}{c}
\\ [-.5em]
\hline
\\ [-.8em]
$\begin{aligned}
\REACH(X,Y) =& \ \langle Y \rangle X \Sigma^+\\
\SHARE(X,Y) =& \ \exists Z. \REACH(X,Z) \wedge \REACH(Y,Z)\\
\TREE(\ROOT)\  =& \ \forall X. \langle X \rangle \ROOT \Rightarrow \forall Y,Z.
(\REACH(X,Y) \wedge \REACH(X,Z) \Rightarrow \neg \SHARE(Y,Z))
\end{aligned}$
\\ \\[-.8em] \hline
\\ [-.5em]
\\ [-1em]
\end{tabular}
\end{center}
{\white .}\\[-1em]
These predicates are explained below.
\begin{itemize}
\item Predicate $\REACH(X,Y)$ states that location $Y$ is reachable from
location $X$ via a non-empty path $\Sigma^+$. The Kleene closure over the set of
pointer fields $\Sigma$ helps in summarizing unbounded information. 
\item Predicate $\SHARE(X,Y)$ states that locations $X$ and $Y$ reach a common
location via non-empty paths, respectively.
\item Predicate $\TREE(\ROOT)$ describes the shape of a tree structure pointed
to by a variable $\ROOT$. It states that sharing is absent in a tree structure.
\end{itemize}

\newcommand{\NEXT}{{\text {\ttfamily next}}}
Let us derive the pre-condition for statement 3, $\Y.\f:=\X$ of the program in
Figure~\ref{eg2-prog}, when its post-condition is given.
\newcommand{\ACLIST}{{\text{\ttfamily aclist}}}
\\[-1.4em]
\begin{center}
\begin{tabular}{c}
\\ [-.5em]
\hline
\\ [-.8em]
$\begin{aligned}
&{\text{pre-condition:}} & 
\{\ACLIST(\X) \wedge \forall X,Y{\text{\ttfamily [}}\langle X \rangle \X \wedge \langle Y \rangle \Y 
\Rightarrow X^{-1} Y = \emptyset{\text{\ttfamily ]}}\} \\
&{\text{assignment statement:}} &
\Y.\f := \X \\
&{\text{post-condition:}} & 
\{\ACLIST(\Y)\}
\end{aligned}$
\\ \\[-.8em] \hline
\\ [-.5em]
\\ [-1em]
\end{tabular}
\end{center}
{\white .}\\[-1em]
The post-condition for the assignment statement $\Y.\f:=\X$ specifies that
variable $\Y$ points to an acyclic linked list (denoted by predicate
$\ACLIST(\Y)$). The pre-condition for the assignment statement is that variable
$\X$ should be an acyclic linked list and that there should be no path from
$\X$ to $\Y$ (otherwise the assignment statement would create a cycle,
invalidating the post-condition $\ACLIST(\Y)$). 

Bozga et al. \cite{bil04} have also designed pAL (Propositional Alias Logic), which is a
decidable subset of wAL.  However, pAL can describe only finite graphs and does
not have the ability to describe properties like list-ness, circularity, and
reachability.


\item 
Hob~\cite{lkr05-hob} is a program analysis framework, which allows a developer
to use multiple analysis plugins for the same program. Each procedure can be
verified by a different analysis plugin; therefore, an efficient analysis plugin
can be chosen for each procedure depending on the properties of the procedure
that the developer wishes to verify. The Hob project has been plugged with the
following three analysis plugins~\cite{klzr06}:
\begin{enumerate}
\item Flag Abstraction Language plugin~\cite{lkr05} uses first order boolean
algebra extended with cardinality constraints. It is used to infer loop
invariants.
\item PALE plugin~\cite{as01} uses monadic second order logic to verify
properties of tree like data structures.
\item Theorem proving plugin uses higher-order logic to handle all data
structures.
\end{enumerate}

\end{itemize}


\newcommand{\Galiasoutsix}
{{
{\psset{unit=.25mm}
\psset{linewidth=.2mm,arrowsize=4pt,arrowinset=0}
		\begin{pspicture}(-3,-3.5)(38,0)
		\psrelpoint{origin}{q}{10}{-2}
		\rput(\x{q},\y{q}){\rnode{q}{{$\X$}}}
		\psrelpoint{origin}{j}{-3}{-2}
		\rput(\x{j},\y{j}){\rnode{j}{}}
		\psrelpoint{q}{s}{0}{38}
		\rput(\x{s},\y{s}){\rnode{s}{\pscirclebox[fillstyle=solid,
				fillcolor=white,framesep=9]{}}}
		\ncline{->}{q}{s}
		\psrelpoint{s}{v}{40}{0}
		\rput(\x{v},\y{v}){\rnode{v}{\pscirclebox[fillstyle=solid,
				fillcolor=white,framesep=9]{}}}
		\ncline{->}{s}{v}
		\naput{$\f$}
		\psrelpoint{v}{d}{43}{0}
		\rput(\x{d},\y{d}){\rnode{d}{{$\dots$}}}
		\ncline[nodesepB=3]{->}{v}{d}
		\naput{$\f$}
		\psrelpoint{d}{t}{43}{0}
		\rput(\x{t},\y{t}){\rnode{t}{\pscirclebox[fillstyle=solid,
				fillcolor=white,framesep=9]{}}}
		\ncline[nodesepA=3]{->}{d}{t}
		\naput{$\f$}
		\psrelpoint{t}{y}{40}{0}
		\rput(\x{y},\y{y}){\rnode{y}{\pscirclebox[fillstyle=solid,
				fillcolor=white,framesep=9]{}}}
		\ncline{->}{t}{y}
		\naput{$\f$}
		\psrelpoint{t}{u}{0}{38}
		\rput(\x{u},\y{u}){\rnode{u}{{$\Y$}}}
		\ncline{->}{u}{t}
		\psrelpoint{t}{x}{0}{-38}
		\rput(\x{x},\y{x}){\rnode{x}{{$\Z$}}}
		\ncline{->}{x}{t}
		\psrelpoint{y}{w}{40}{0}
		\rput(\x{w},\y{w}){\rnode{w}{{$\dots$}}}
		\ncline[nodesepB=3]{->}{y}{w}
		\naput{$\f$}

		\end{pspicture}
		}
}}

\newcommand{\Galiasoutfour}
{{
{\psset{unit=.25mm}
\psset{linewidth=.2mm,arrowsize=4pt,arrowinset=0}
		\begin{pspicture}(-3,-3.5)(38,0)
		\psrelpoint{origin}{q}{10}{-2}
		\rput(\x{q},\y{q}){\rnode{q}{{$\X$}}}
		\psrelpoint{origin}{j}{-3}{-2}
		\rput(\x{j},\y{j}){\rnode{j}{}}
		\psrelpoint{q}{s}{0}{38}
		\rput(\x{s},\y{s}){\rnode{s}{\pscirclebox[fillstyle=solid,
				fillcolor=white,framesep=9]{}}}
		\ncline{->}{q}{s}
		\psrelpoint{s}{t}{40}{0}
		\rput(\x{t},\y{t}){\rnode{t}{\pscirclebox[fillstyle=solid,
				fillcolor=white,framesep=9]{}}}
		\ncline{->}{s}{t}
		\naput{$\f$}
		\psrelpoint{t}{d}{43}{0}
		\rput(\x{d},\y{d}){\rnode{d}{{$\dots$}}}
		\ncline[nodesepB=3]{->}{t}{d}
		\naput{$\f$}
		\psrelpoint{s}{u}{0}{38}
		\rput(\x{u},\y{u}){\rnode{u}{{$\Y$}}}
		\ncline{->}{u}{s}

		\end{pspicture}
		}
}}

\renewcommand{\Galiasone}
{{
{\psset{unit=.25mm}
\psset{linewidth=.2mm,arrowsize=4pt,arrowinset=0}
		\begin{pspicture}(-3,-3.5)(38,0)
		\psrelpoint{origin}{q}{10}{-2}
		\rput(\x{q},\y{q}){\rnode{q}{{$\X$}}}
		\psrelpoint{origin}{j}{-3}{-2}
		\rput(\x{j},\y{j}){\rnode{j}{}}
		\psrelpoint{q}{s}{0}{35}
		\rput(\x{s},\y{s}){\rnode{s}{\pscirclebox[fillstyle=solid,
				fillcolor=white,framesep=9]{}}}
		\ncline{->}{q}{s}
		\psrelpoint{s}{r}{55}{0}
		\rput(\x{r},\y{r}){\rnode{r}{\pscirclebox[fillstyle=solid,
				fillcolor=white,framesep=9,linewidth=.2mm,linestyle=dashed,dash=2.5pt 1.5pt]{}}}
		\ncline[linestyle=dashed,dash=2.5pt 1.5pt,linewidth=.2mm]{->}{s}{r}
		\nbput{$\f$}
		\nccurve[linewidth=.2mm,linestyle=dashed,dash=2.5pt 1.5pt,angleB=-45,angleA=-135,nodesep=-2,ncurv=2.5]{->}{r}{r}
		\nbput{$\f$}
		\psrelpoint{r}{t}{55}{0}
		\rput(\x{t},\y{t}){\rnode{t}{\pscirclebox[fillstyle=solid,
				fillcolor=white,framesep=9]{}}}
		\ncline[linestyle=dashed,dash=2.5pt 1.5pt,linewidth=.2mm]{->}{r}{t}
		\nbput{$\f$}
		\psrelpoint{t}{u}{-15}{-35}
		\rput(\x{u},\y{u}){\rnode{u}{{$\Y$}}}
		\ncline{->}{u}{t}
		\psrelpoint{t}{x}{15}{-35}
		\rput(\x{x},\y{x}){\rnode{x}{{$\Z$}}}
		\ncline{->}{x}{t}
		\nccurve[linewidth=.2mm,linestyle=dashed,dash=2.5pt 1.5pt,angleB=45,angleA=135,nodesep=-2,ncurv=1]{->}{t}{r}
		\nbput{$\f$}
		\nccurve[angleB=105,angleA=75,nodesep=-1,ncurv=1]{->}{s}{t}
		\naput{$\f$}

		\end{pspicture}
		}
}}

\newcommand{\Galiasoneb}
{{
{\psset{unit=.25mm}
\psset{linewidth=.2mm,arrowsize=4pt,arrowinset=0}
		\begin{pspicture}(-3,-3.5)(38,0)
		\psrelpoint{origin}{q}{10}{-2}
		\rput(\x{q},\y{q}){\rnode{q}{{$\X$}}}
		\psrelpoint{origin}{j}{-3}{-2}
		\rput(\x{j},\y{j}){\rnode{j}{}}
		\psrelpoint{q}{s}{0}{35}
		\rput(\x{s},\y{s}){\rnode{s}{\pscirclebox[fillstyle=solid,
				fillcolor=white,framesep=9]{}}}
		\ncline{->}{q}{s}
		\psrelpoint{s}{r}{55}{0}
		\rput(\x{r},\y{r}){\rnode{r}{\pscirclebox[fillstyle=solid,
				fillcolor=white,framesep=9,linewidth=.2mm,linestyle=dashed,dash=2.5pt 1.5pt]{}}}
		\ncline[linestyle=dashed,dash=2.5pt 1.5pt,linewidth=.2mm]{->}{s}{r}
		\nbput{$\f$}
		\nccurve[linewidth=.2mm,linestyle=dashed,dash=2.5pt 1.5pt,angleB=-45,angleA=-135,nodesep=-3,ncurv=2.5]{->}{r}{r}
		\nbput{$\f$}
		\psrelpoint{r}{t}{55}{0}
		\rput(\x{t},\y{t}){\rnode{t}{\pscirclebox[fillstyle=solid,
				fillcolor=white,framesep=9]{}}}
		\ncline[linestyle=dashed,dash=2.5pt 1.5pt,linewidth=.2mm]{->}{r}{t}
		\nbput{$\f$}
		\psrelpoint{t}{u}{-15}{-35}
		\rput(\x{u},\y{u}){\rnode{u}{{$\Y$}}}
		\ncline{->}{u}{t}
		\psrelpoint{t}{x}{15}{-35}
		\rput(\x{x},\y{x}){\rnode{x}{{$\Z$}}}
		\ncline{->}{x}{t}
		\nccurve[angleB=135,angleA=45,nodesep=-3,ncurv=1.5]{->}{s}{t}
		\naput{$\f$}
		\psrelpoint{t}{v}{55}{0}
		\rput(\x{v},\y{v}){\rnode{v}{\pscirclebox[fillstyle=solid,
				fillcolor=white,framesep=9,linewidth=.2mm,linestyle=dashed,dash=2.5pt 1.5pt]{}}}
		\nccurve[linewidth=.2mm,linestyle=dashed,dash=2.5pt 1.5pt,angleB=-45,angleA=-135,nodesep=-3,ncurv=2.5]{->}{v}{v}
		\nbput{$\f$}
		\ncline[linestyle=dashed,dash=2.5pt 1.5pt,linewidth=.2mm]{->}{t}{v}
		\nbput{$\f$}

		\psrelpoint{s}{r1}{0}{30}
		\rput(\x{r1},\y{r1}){\rnode{r1}{{${\R}_{\X}$}}}
		\ncline{->}{r1}{s}
		\psrelpoint{r}{m}{0}{30}
		\rput(\x{m},\y{m}){\rnode{m}{{${\R}_{\X}$}}}
		\ncline{->}{m}{r}
		\psrelpoint{t}{r3}{0}{40}
		\rput(\x{r3},\y{r3}){\rnode{r3}{{\begin{tabular}{@{}c@{}c@{}}${\R}_{\X},$\\[-.9ex]${\R}_{\Y},{\R}_{\Z}$\end{tabular}}}}
		\ncline{->}{r3}{t}
		\psrelpoint{v}{r4}{0}{40}
		\rput(\x{r4},\y{r4}){\rnode{r4}{{\begin{tabular}{@{}c@{}c@{}}${\R}_{\X},$\\[-.9ex]${\R}_{\Y},{\R}_{\Z}$\end{tabular}}}}
		\ncline{->}{r4}{v}

		\end{pspicture}
		}
}}

\newcommand{\Galiastraditional}
{{
{\psset{unit=.25mm}
\psset{linewidth=.2mm,arrowsize=4pt,arrowinset=0}
		\begin{pspicture}(-3,-3.5)(38,0)
		\psrelpoint{origin}{q}{10}{-2}
		\rput(\x{q},\y{q}){\rnode{q}{{$\X$}}}
		\psrelpoint{q}{j}{23}{0}
		\rput(\x{j},\y{j}){\rnode{j}{$\Y$}}
		\psrelpoint{q}{s}{10}{38}
		\rput(\x{s},\y{s}){\rnode{s}{\psovalbox[fillstyle=solid,
				fillcolor=white,framesep=4]{{\em Site }2}}}
		\ncline{->}{q}{s}
		\ncline{->}{j}{s}
		\nccurve[linewidth=.2mm,angleA=45,angleB=135,nodesep=-1,ncurv=2.5]{->}{s}{s}
		\nbput{$\f$}

		\end{pspicture}
		}
}}

\newcommand{\Galiastwo}
{{
{\psset{unit=.25mm}
\psset{linewidth=.2mm,arrowsize=4pt,arrowinset=0}
		\begin{pspicture}(-3,-3.5)(38,0)
		\psrelpoint{origin}{q}{10}{-2}
		\rput(\x{q},\y{q}){\rnode{q}{{$\X$}}}
		\psrelpoint{q}{j}{23}{0}
		\rput(\x{j},\y{j}){\rnode{j}{$\Y$}}
		\psrelpoint{q}{s}{10}{38}
		\rput(\x{s},\y{s}){\rnode{s}{\psovalbox[fillstyle=solid,
				fillcolor=white,framesep=4]{{\em Site }2}}}
		\ncline{->}{q}{s}
		\ncline{->}{j}{s}
		\psrelpoint{s}{r}{92}{0}
		\rput(\x{r},\y{r}){\rnode{r}{\psovalbox[fillstyle=solid,
				fillcolor=white,framesep=4,linewidth=.2mm,linestyle=dashed,dash=2.5pt 1.5pt]{{\em Site }2}}}
		\ncline[linewidth=.2mm,linestyle=dashed,dash=2.5pt 1.5pt]{->}{s}{r}
		\naput{$\f$}
		\nccurve[linewidth=.2mm,linestyle=dashed,dash=2.5pt 1.5pt,angleA=45,angleB=135,nodesep=-1,ncurv=2.5]{->}{r}{r}
		\nbput{$\f$}

		\end{pspicture}
		}
}}

\newcommand{\Galiasthree}
{{
{\psset{unit=.25mm}
\psset{linewidth=.2mm,arrowsize=4pt,arrowinset=0}
		\begin{pspicture}(-3,-3.5)(38,0)
		\psrelpoint{origin}{q}{10}{-2}
		\rput(\x{q},\y{q}){\rnode{q}{{$\X$}}}
		\psrelpoint{origin}{j}{-3}{-2}
		\rput(\x{j},\y{j}){\rnode{j}{}}
		\psrelpoint{q}{s}{0}{38}
		\rput(\x{s},\y{s}){\rnode{s}{\psovalbox[fillstyle=solid,
				fillcolor=white,framesep=4]{{\em Site }2}}}
		\ncline{->}{q}{s}
		\psrelpoint{s}{r}{92}{0}
		\rput(\x{r},\y{r}){\rnode{r}{\psovalbox[fillstyle=solid,
				fillcolor=white,framesep=4,linewidth=.2mm,linestyle=dashed,dash=2.5pt 1.5pt]{{\em Site }2}}}
		\ncline[linewidth=.2mm,linestyle=dashed,dash=2.5pt 1.5pt]{->}{s}{r}
		\naput{$\f$}
		\nccurve[linewidth=.2mm,linestyle=dashed,dash=2.5pt 1.5pt,angleA=45,angleB=135,nodesep=-1,ncurv=2.5]{->}{r}{r}
		\nbput{$\f$}
		\psrelpoint{r}{t}{15}{-38}
		\rput(\x{t},\y{t}){\rnode{t}{{$\Z$}}}
		\ncline[linewidth=.2mm,linestyle=dashed,dash=2.5pt 1.5pt]{->}{t}{r}
		\psrelpoint{r}{u}{-15}{-38}
		\rput(\x{u},\y{u}){\rnode{u}{{$\Y$}}}
		\ncline[linewidth=.2mm,linestyle=dashed,dash=2.5pt 1.5pt]{->}{u}{r}

		\end{pspicture}
		}
}}

\newcommand{\Gnodea}
{{
{\psset{unit=.25mm}
\psset{linewidth=.2mm,arrowsize=4pt,arrowinset=0}
		\begin{pspicture}(-6,50)(27,-11)
		\psframe[linewidth=.4mm,linestyle=dotted,dotsep=1.5pt](-6,50)(27,-11)
		\psrelpoint{origin}{q}{10}{-3}
		\rput(\x{q},\y{q}){\rnode{q}{{$\X,\Y$}}}
		\psrelpoint{origin}{j}{-3}{-2}
		\rput(\x{j},\y{j}){\rnode{j}{}}
		\psrelpoint{q}{s}{0}{35}
		\rput(\x{s},\y{s}){\rnode{s}{\pscirclebox[fillstyle=solid,
				fillcolor=white,framesep=4]{$i$}}}
		\ncline[nodesepA=1]{->}{q}{s}
		\end{pspicture}
		}
}}

\newcommand{\Gnodeb}
{{
{\psset{unit=.25mm}
\psset{linewidth=.2mm,arrowsize=4pt,arrowinset=0}
		\begin{pspicture}(-6,56)(65,-11)
		\psframe[linewidth=.4mm,linestyle=dotted,dotsep=1.5pt](-6,56)(65,-11)
		\psrelpoint{origin}{q}{10}{-3}
		\rput(\x{q},\y{q}){\rnode{q}{{$\X,\Y$}}}
		\psrelpoint{origin}{j}{-3}{-2}
		\rput(\x{j},\y{j}){\rnode{j}{}}
		\psrelpoint{q}{s}{0}{35}
		\rput(\x{s},\y{s}){\rnode{s}{\pscirclebox[fillstyle=solid,
				fillcolor=white,framesep=2]{$j$}}}
		\ncline[nodesepA=1]{->}{q}{s}
		\psrelpoint{s}{t}{39}{0}
		\rput(\x{t},\y{t}){\rnode{t}{\pscirclebox[fillstyle=solid,
				fillcolor=white,framesep=4]{$i$}}}
		\psrelpoint{t}{u}{0}{-35}
		\rput(\x{u},\y{u}){\rnode{u}{{$\Z$}}}
		\ncline[nodesepA=1]{->}{u}{t}

		\ncline{->}{s}{t}
		\naput{$\f$}

		\end{pspicture}
		}
}}

\newcommand{\Gnodec}
{{
{\psset{unit=.25mm}
\psset{linewidth=.2mm,arrowsize=4pt,arrowinset=0}
		\begin{pspicture}(-6,55)(65,-11)
		\psframe[linewidth=.4mm,linestyle=dotted,dotsep=1.5pt](-6,55)(65,-11)
		\psrelpoint{origin}{q}{10}{-3}
		\rput(\x{q},\y{q}){\rnode{q}{{$\X$}}}
		\psrelpoint{origin}{j}{-3}{-3}
		\rput(\x{j},\y{j}){\rnode{j}{}}
		\psrelpoint{q}{s}{0}{35}
		\rput(\x{s},\y{s}){\rnode{s}{\pscirclebox[fillstyle=solid,
				fillcolor=white,framesep=2]{$j$}}}
		\ncline[nodesepA=1]{->}{q}{s}
		\psrelpoint{s}{t}{39}{0}
		\rput(\x{t},\y{t}){\rnode{t}{\pscirclebox[fillstyle=solid,
				fillcolor=white,framesep=4]{$i$}}}
		\psrelpoint{t}{u}{0}{-35}
		\rput(\x{u},\y{u}){\rnode{u}{{$\Y,\Z$}}}
		\ncline[nodesepA=1]{->}{u}{t}

		\ncline{->}{s}{t}
		\naput{$\f$}

		\end{pspicture}
		}
}}

\newcommand{\Gnoded}
{{
{\psset{unit=.25mm}
\psset{linewidth=.2mm,arrowsize=4pt,arrowinset=0}
		\begin{pspicture}(-6,55)(105,-11)
		\psframe[linewidth=.4mm,linestyle=dotted,dotsep=1.5pt](-6,55)(105,-11)
		\psrelpoint{origin}{q}{10}{-3}
		\rput(\x{q},\y{q}){\rnode{q}{{$\X$}}}
		\psrelpoint{origin}{j}{-3}{-3}
		\rput(\x{j},\y{j}){\rnode{j}{}}
		\psrelpoint{q}{s}{0}{35}
		\rput(\x{s},\y{s}){\rnode{s}{\pscirclebox[fillstyle=solid,
				fillcolor=white,framesep=3]{$j$}}}
		\ncline[nodesepA=1]{->}{q}{s}
		\psrelpoint{s}{t}{39}{0}
		\rput(\x{t},\y{t}){\rnode{t}{\pscirclebox[fillstyle=solid,
				fillcolor=white,framesep=4]{$k$}}}
		\psrelpoint{t}{u}{0}{-35}
		\rput(\x{u},\y{u}){\rnode{u}{{$\Y$}}}
		\ncline[nodesepA=1]{->}{u}{t}

		\ncline{->}{s}{t}
		\naput{$\f$}
		\psrelpoint{t}{v}{39}{0}
		\rput(\x{v},\y{v}){\rnode{v}{\pscirclebox[fillstyle=solid,
				fillcolor=white,framesep=4]{$i$}}}
		\psrelpoint{v}{w}{0}{-35}
		\rput(\x{w},\y{w}){\rnode{w}{{$\Z$}}}
		\ncline[nodesepA=1]{->}{w}{v}

		\ncline{->}{t}{v}
		\naput{$\f$}

		\end{pspicture}
		}
}}

\renewcommand{\Galiasfour}
{{
{\psset{unit=.25mm}
\psset{linewidth=.2mm,arrowsize=4pt,arrowinset=0}
		\begin{pspicture}(-1,20)(48,-7)
		\psrelpoint{origin}{k}{40}{-2}
		\rput(\x{k},\y{k}){\rnode{k}{{\Gnodea}}}
		\psrelpoint{k}{l}{0}{-150}
		\rput(\x{l},\y{l}){\rnode{l}{{\Gnodeb}}}
		\ncline{->}{k}{l}
		\naput{\begin{tabular}{l}$[i>1]$\\$i:=i-1$\\$j:=1$\end{tabular}}
		\psrelpoint{l}{m}{0}{-110}
		\rput(\x{m},\y{m}){\rnode{m}{{\Gnodec}}}
		\ncline{->}{l}{m}
		\psrelpoint{l}{n}{150}{0}
		\rput(\x{n},\y{n}){\rnode{n}{{\Gnoded}}}
		\ncline[nodesepB=-6]{->}{n}{m}
		\naput{$j:=j+k$}
    		\ncloop[angleA=270,angleB=90,loopsize=-226,linearc=2,offsetB=3,arm=19]{->}{m}{n}
		\nbput{\begin{tabular}{l}$[i>1]$\\$i:=i-1$\\$k:=1$\end{tabular}}
		\psrelpoint{k}{d}{-40}{0}
		\rput(\x{d},\y{d}){\rnode{d}{{$\Inn_5$}}}
		\psrelpoint{l}{o}{-65}{0}
		\rput(\x{o},\y{o}){\rnode{o}{{$\Outn_5$}}}
		\psrelpoint{l}{c}{65}{0}
		\rput(\x{c},\y{c}){\rnode{c}{{$\Outn_5$}}}
		\psrelpoint{m}{p}{-65}{0}
		\rput(\x{p},\y{p}){\rnode{p}{{$\Outn_6$}}}

		\end{pspicture}
		}
}}

\newcommand{\Gcwzoutfour}
{{
{\psset{unit=.25mm}
\psset{linewidth=.2mm,arrowsize=4pt,arrowinset=0}
		\begin{pspicture}(-3,-3.5)(38,0)
		\psrelpoint{origin}{q}{10}{-2}
		\rput(\x{q},\y{q}){\rnode{q}{{$\X$}}}
		\psrelpoint{origin}{j}{-3}{-2}
		\rput(\x{j},\y{j}){\rnode{j}{}}
		\psrelpoint{q}{s}{0}{38}
		\rput(\x{s},\y{s}){\rnode{s}{\pscirclebox[fillstyle=solid,
				fillcolor=white,framesep=3]{$2$}}}
		\ncline{->}{q}{s}
		\psrelpoint{s}{t}{55}{0}
		\rput(\x{t},\y{t}){\rnode{t}{\pscirclebox[fillstyle=solid,
				fillcolor=white,framesep=3]{$2$}}}
		\ncline{->}{s}{t}
		\naput{$\f$}
		\nccurve[linewidth=.2mm,angleB=45,angleA=135,nodesep=-3,ncurv=3]{->}{t}{t}
		\naput{$\f$}
		\psrelpoint{s}{u}{0}{38}
		\rput(\x{u},\y{u}){\rnode{u}{{$\Y$}}}
		\ncline{->}{u}{s}

		\end{pspicture}
		}
}}

\newcommand{\Gcwzoutsix}
{{
{\psset{unit=.25mm}
\psset{linewidth=.2mm,arrowsize=4pt,arrowinset=0}
		\begin{pspicture}(-3,-3.5)(38,0)
		\psrelpoint{origin}{q}{10}{-2}
		\rput(\x{q},\y{q}){\rnode{q}{{$\X$}}}
		\psrelpoint{origin}{j}{-3}{-2}
		\rput(\x{j},\y{j}){\rnode{j}{}}
		\psrelpoint{q}{s}{0}{38}
		\rput(\x{s},\y{s}){\rnode{s}{\pscirclebox[fillstyle=solid,
				fillcolor=white,framesep=3]{$2$}}}
		\ncline{->}{q}{s}
		\psrelpoint{s}{t}{55}{0}
		\rput(\x{t},\y{t}){\rnode{t}{\pscirclebox[fillstyle=solid,
				fillcolor=white,framesep=3]{$2$}}}
		\ncline{->}{s}{t}
		\naput{$\f$}
		\nccurve[linewidth=.2mm,angleB=45,angleA=-45,nodesep=-3,ncurv=3]{->}{t}{t}
		\nbput{$\f$}
		\psrelpoint{t}{u}{0}{38}
		\rput(\x{u},\y{u}){\rnode{u}{{$\Y$}}}
		\ncline{->}{u}{t}
		\psrelpoint{t}{v}{0}{-38}
		\rput(\x{v},\y{v}){\rnode{v}{{$\Z$}}}
		\ncline{->}{v}{t}

		\end{pspicture}
		}
}}

\newcommand{\Gsrwa}
{{
{\psset{unit=.25mm}
\psset{linewidth=.2mm,arrowsize=4pt,arrowinset=0}
		\begin{pspicture}(-3,-3.5)(38,0)
		\psrelpoint{origin}{q}{10}{-2}
		\rput(\x{q},\y{q}){\rnode{q}{{$\X$}}}
		\psrelpoint{origin}{j}{-3}{-2}
		\rput(\x{j},\y{j}){\rnode{j}{}}
		\psrelpoint{q}{s}{0}{38}
		\rput(\x{s},\y{s}){\rnode{s}{\pscirclebox[fillstyle=solid,
				fillcolor=white,framesep=9]{}}}
		\ncline{->}{q}{s}
		\psrelpoint{s}{t}{45}{0}
		\rput(\x{t},\y{t}){\rnode{t}{\pscirclebox[fillstyle=solid,
				fillcolor=white,framesep=9,linewidth=.2mm,linestyle=dashed,dash=2.5pt 1.5pt]{}}}
		\ncline[linestyle=dashed,dash=2.5pt 1.5pt,linewidth=.2mm]{->}{s}{t}
		\naput{$\f$}
		\nccurve[linewidth=.2mm,linestyle=dashed,dash=2.5pt 1.5pt,angleB=-45,angleA=-135,nodesep=-3,ncurv=3]{->}{t}{t}
		\nbput{$\f$}
		\psrelpoint{s}{u}{0}{38}
		\rput(\x{u},\y{u}){\rnode{u}{{$\Y$}}}
		\ncline{->}{u}{s}

		\end{pspicture}
		}
}}

\newcommand{\Gsrwb}
{{
{\psset{unit=.25mm}
\psset{linewidth=.2mm,arrowsize=4pt,arrowinset=0}
		\begin{pspicture}(-3,-3.5)(38,0)
		\psrelpoint{origin}{q}{10}{-2}
		\rput(\x{q},\y{q}){\rnode{q}{{$\X$}}}
		\psrelpoint{origin}{j}{-3}{-2}
		\rput(\x{j},\y{j}){\rnode{j}{}}
		\psrelpoint{q}{s}{0}{35}
		\rput(\x{s},\y{s}){\rnode{s}{\pscirclebox[fillstyle=solid,
				fillcolor=white,framesep=9]{}}}
		\ncline{->}{q}{s}
		\psrelpoint{s}{t}{45}{0}
		\rput(\x{t},\y{t}){\rnode{t}{\pscirclebox[fillstyle=solid,
				fillcolor=white,framesep=9]{}}}
		\ncline{->}{s}{t}
		\nbput{$\f$}
		\psrelpoint{t}{u}{-15}{-35}
		\rput(\x{u},\y{u}){\rnode{u}{{$\Y$}}}
		\ncline{->}{u}{t}
		\psrelpoint{t}{x}{15}{-35}
		\rput(\x{x},\y{x}){\rnode{x}{{$\Z$}}}
		\ncline{->}{x}{t}
		\psrelpoint{t}{v}{45}{0}
		\rput(\x{v},\y{v}){\rnode{v}{\pscirclebox[fillstyle=solid,
				fillcolor=white,framesep=9,linewidth=.2mm,linestyle=dashed,dash=2.5pt 1.5pt]{}}}
		\nccurve[linewidth=.2mm,linestyle=dashed,dash=2.5pt 1.5pt,angleB=-45,angleA=-135,nodesep=-3,ncurv=2.5]{->}{v}{v}
		\nbput{$\f$}
		\ncline[linestyle=dashed,dash=2.5pt 1.5pt,linewidth=.2mm]{->}{t}{v}
		\nbput{$\f$}

		\end{pspicture}
		}
}}

\newcommand{\Gsrwc}
{{
{\psset{unit=.25mm}
\psset{linewidth=.2mm,arrowsize=4pt,arrowinset=0}
		\begin{pspicture}(-3,-3.5)(38,0)
		\psrelpoint{origin}{q}{10}{-2}
		\rput(\x{q},\y{q}){\rnode{q}{{$\X$}}}
		\psrelpoint{origin}{j}{-3}{-2}
		\rput(\x{j},\y{j}){\rnode{j}{}}
		\psrelpoint{q}{s}{0}{35}
		\rput(\x{s},\y{s}){\rnode{s}{\pscirclebox[fillstyle=solid,
				fillcolor=white,framesep=9]{}}}
		\ncline{->}{q}{s}
		\psrelpoint{s}{r}{55}{0}
		\rput(\x{r},\y{r}){\rnode{r}{\pscirclebox[fillstyle=solid,
				fillcolor=white,framesep=9]{}}}
		\ncline{->}{s}{r}
		\nbput{$\f$}
		\psrelpoint{r}{t}{55}{0}
		\rput(\x{t},\y{t}){\rnode{t}{\pscirclebox[fillstyle=solid,
				fillcolor=white,framesep=9]{}}}
		\ncline{->}{r}{t}
		\nbput{$\f$}
		\psrelpoint{t}{u}{-15}{-35}
		\rput(\x{u},\y{u}){\rnode{u}{{$\Y$}}}
		\ncline{->}{u}{t}
		\psrelpoint{t}{x}{15}{-35}
		\rput(\x{x},\y{x}){\rnode{x}{{$\Z$}}}
		\ncline{->}{x}{t}
		\psrelpoint{t}{v}{55}{0}
		\rput(\x{v},\y{v}){\rnode{v}{\pscirclebox[fillstyle=solid,
				fillcolor=white,framesep=9,linewidth=.2mm,linestyle=dashed,dash=2.5pt 1.5pt]{}}}
		\nccurve[linewidth=.2mm,linestyle=dashed,dash=2.5pt 1.5pt,angleB=-45,angleA=-135,nodesep=-3,ncurv=2.5]{->}{v}{v}
		\nbput{$\f$}
		\ncline[linestyle=dashed,dash=2.5pt 1.5pt,linewidth=.2mm]{->}{t}{v}
		\nbput{$\f$}
		\nccurve[angleB=135,angleA=45,nodesep=-3,ncurv=1.5]{->}{s}{t}
		\naput{$\f$}

		\psrelpoint{s}{r1}{0}{30}
		\rput(\x{r1},\y{r1}){\rnode{r1}{{${\R}_{\X}$}}}
		\ncline{->}{r1}{s}
		\psrelpoint{r}{m}{0}{30}
		\rput(\x{m},\y{m}){\rnode{m}{{${\R}_{\X}$}}}
		\ncline{->}{m}{r}
		\psrelpoint{t}{r3}{0}{40}
		\rput(\x{r3},\y{r3}){\rnode{r3}{{\begin{tabular}{@{}c@{}c@{}}${\R}_{\X},$\\[-.9ex]${\R}_{\Y},{\R}_{\Z}$\end{tabular}}}}
		\ncline{->}{r3}{t}
		\psrelpoint{v}{r4}{0}{40}
		\rput(\x{r4},\y{r4}){\rnode{r4}{{\begin{tabular}{@{}c@{}c@{}}${\R}_{\X},$\\[-.9ex]${\R}_{\Y},{\R}_{\Z}$\end{tabular}}}}
		\ncline{->}{r4}{v}

		\end{pspicture}
		}
}}

\begin{figure}[t]
\begin{center}
\begin{subfigure}[b]{0.28\textwidth}
\begin{center}
  {
    \psset{unit=.25mm}
    \psset{linewidth=.2mm}
    \psset{linewidth=.2mm,arrowsize=4pt,arrowinset=0}
    \begin{pspicture}(-15,-165)(85,70)

    \putnode{a}{origin}{40}{60}{1 \psframebox{$\X := \text{\ttfamily null}$}\white\ 1}
    \putnode{b}{a}{0}{-40}{2 \psframebox{$\Y := \new$}\white\ 2}
    \putnode{c}{b}{0}{-35}{3 \psframebox{$\Y \myrightarrow \f := \X$}\white\ 3}
    \putnode{d}{c}{0}{-35}{4 \psframebox{$\X := \Y$}\white\ 4}
    \putnode{e}{d}{0}{-56}{5 \psframebox{$\Z := \Y \myrightarrow \f$}\white\ 6}
    \putnode{f}{e}{0}{-35}{6 \psframebox{$\Y:= \Z$}\white\ 7}
    
    \ncline{->}{a}{b}
    \ncline{->}{b}{c}
    \ncline{->}{c}{d}
    \ncline{->}{d}{e}
    \ncline{->}{e}{f}

    \ncloop[angleA=270,angleB=90,loopsize=47,linearc=3,offset=-8,arm=13]{->}{d}{b}
    \ncloop[angleA=270,angleB=90,loopsize=55,linearc=3,offsetB=-8,arm=13]{->}{f}{e}

    \end{pspicture}
   }
\end{center}
\caption{Example}
\label{eg2-prog}
\end{subfigure}
~
\begin{minipage}[b]{0.68\textwidth}
\begin{subfigure}[b]{1\textwidth}
\begin{center}
  {
    \psset{unit=.25mm}
    \begin{pspicture}(20,55)(125,145)

    \putnode{a}{origin}{40}{60}{\Galiasoutfour}
    \end{pspicture}
   }
\end{center}
\caption{\parbox[t]{.9\textwidth}{Execution snapshot showing an unbounded heap
graph at $\Outn_4$ of program in Figure~\ref{eg2-prog}.}}
\label{eg2-unbounded-1}
\end{subfigure}

\begin{subfigure}[b]{1\textwidth}
\begin{center}
  {
    \psset{unit=.25mm}
    \begin{pspicture}(20,55)(248,145)

    \putnode{a}{origin}{40}{60}{\Galiasoutsix}
    \end{pspicture}
   }
\end{center}
\caption{\parbox[t]{.9\textwidth}{Execution snapshot showing an unbounded heap
graph at $\Outn_6$ of program in Figure~\ref{eg2-prog}.}}
\label{eg2-unbounded-2}
\end{subfigure}
\end{minipage}
\end{center}

\caption{Running example to illustrate various heap summarization techniques.
Summarized representations of the heap memories in Figures~\ref{eg2-unbounded-1}
and \ref{eg2-unbounded-2} are shown on a store based model in
Figures~\ref{cwz90}, \ref{srw}, \ref{bbhimv06}, \ref{br06}, and \ref{sep}.}
\label{eg-2}
\figrule
\end{figure}
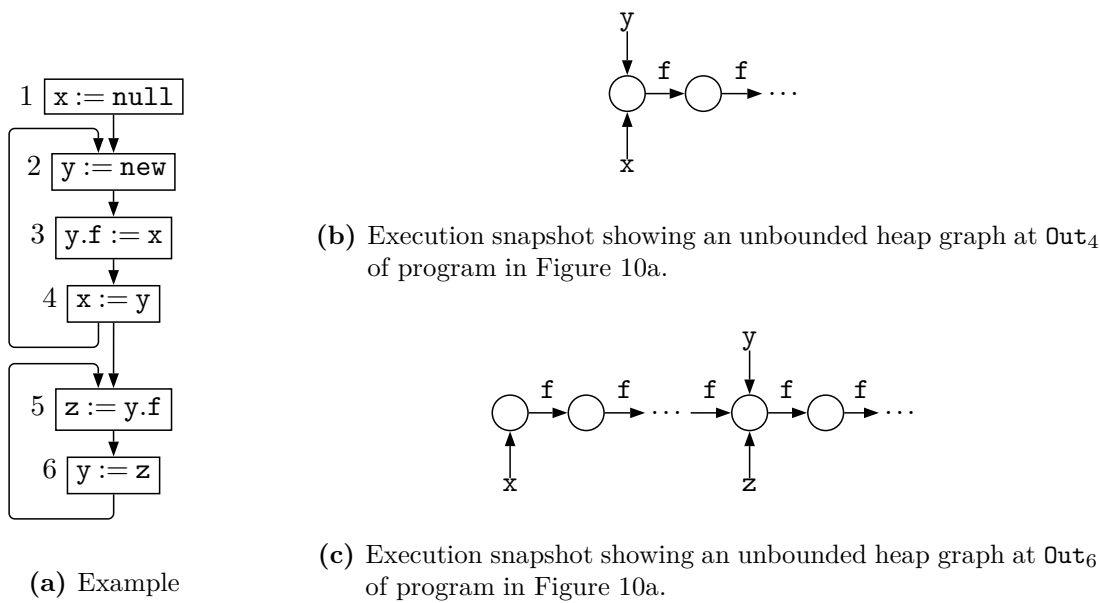


\section{Summarization in Store Based Heap Model}
\label{sec:storebased}
It is easier to visualize a memory graph as heap objects connected through
fields. This is the view of a store based heap model as introduced in
Section~\ref{sec:model}. The following sections summarize this unbounded view
using techniques involving a combination of allocation sites, variables, some
other generic instrumentation predicates, and higher-order logics.

\subsection{Summarization Using Allocation Sites and Variables}
\label{sec:alloc-var}

Chase et al. \cite{cwz90} were the first to summarize heap nodes using techniques involving
allocation sites and variables.  In their method, heap nodes with the following
properties are summarized:
\begin{enumerate}
\item heap nodes created at the same program point (i.e. allocation site) such
      that
\item they have the same pointed-to-by-$x$ predicate values for each
pointer variable $x$. 
\end{enumerate}

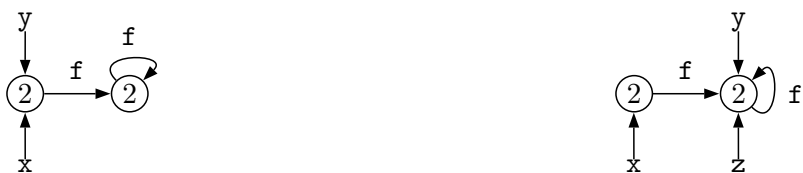
\begin{figure}[t]
\begin{subfigure}[b]{.5\textwidth}
\begin{center}
  {
    \psset{unit=.25mm}
    \begin{pspicture}(15,55)(105,145)

    \putnode{a}{origin}{40}{60}{\Gcwzoutfour}
    \end{pspicture}
   }
\end{center}
\caption{Summarized shape graph at $\Outn_4$.}
\label{cwz90-1}
\end{subfigure}
~
\begin{subfigure}[b]{.5\textwidth}
\begin{center}
  {
    \psset{unit=.25mm}
    \begin{pspicture}(15,55)(115,145)

    \putnode{a}{origin}{40}{60}{\Gcwzoutsix}
    \end{pspicture}
   }
\end{center}
\caption{Summarized shape graph at $\Outn_6$.}
\label{cwz90-2}
\end{subfigure}

\caption{Summarization using allocation sites and variables \protect
\cite{cwz90} for the program in Figure~\ref{eg2-prog}.}
\label{cwz90}
\figrule
\end{figure}

We illustrate this for the program in Figure~\ref{eg2-prog}. The unbounded
memory graphs at $\Outn_4$ and $\Outn_6$ are shown in Figures~\ref{eg2-unbounded-1} and
\ref{eg2-unbounded-2}, respectively. The corresponding summarized graphs created
using this method~\cite{cwz90} at $\Outn_4$ and $\Outn_6$ are shown in
Figures~\ref{cwz90-1} and~\ref{cwz90-2}, respectively.  In
Figure~\ref{cwz90-1}, we see that nodes have been named by their allocation
site, i.e.  statement 2. Also, since this method keeps nodes apart on the basis
of pointer variables, we get two abstract nodes---one node pointed to by pointer
variables $\X$ and $\Y$, and the other node not pointed to by any variable. The
self loop on the second node denotes the presence of unbounded number of nodes
that are not pointed to by any pointer variable.

This method analyses Lisp like programs and constructs shape graphs for heap
variables. It can determine the shape of the allocated heap as tree, simple
cycle, and doubly linked list. In case of lists and trees, if all the nodes are
allocated at the same site then the shape graph would contain a single summary
node with a self loop, making all the nodes aliased to each other. For example,
from the graph in Figure~\ref{cwz90-1}, it cannot be inferred whether the
structure is a linear list or it contains a cycle in the concrete heap memory.
To avoid this, each node is augmented with a reference count i.e. the number of
references to the corresponding heap location from other heap locations (and not
from stack variables). For example, the reference count of the summary node not
pointed to by any variable in Figure~\ref{cwz90-1} is one. A reference count of
less than or equal to one for each node indicates that the data structure is a
tree or a list; whereas, a reference count of more than one indicates that the
data structure is a graph with sharing or cycles. Therefore, this method can
identify at $\Outn_4$ that the program creates a linear list. 

However, the method cannot perform materialization of summary nodes. For
example, after analysing statements 5 and 6 of the program in
Figure~\ref{eg2-prog}, the abstract graph obtained at $\Outn_6$ is shown in
Figure~\ref{cwz90-2}.  It can be seen that the summary node (not pointed to by
any variable) in the graph at $\Outn_4$ in Figure~\ref{cwz90-1} has not been
materialized when $\Y$ and $\Z$ point to the heap locations corresponding to
this summary node. The graph in Figure~\ref{cwz90-2}, therefore, indicates that
$\Y$ and $\Z$ may possibly point to two different heap locations on a list
which is never true at $\Outn_6$ of the program. Additionally, due to the lack
of materialization, this method is not able to determine list reversal and list
insertion programs.  Finally, Sagiv et al. \cite{srw96} highlight, ``this method does not
perform strong updates for a statement of the form $\X.\f := \text{\ttfamily
null}$, except under very limited circumstances."


\subsection{Summarization Using Variables}
\label{sec:var}

Variable based summarization technique has been used in shape analysis. Shape
analysis encompasses all algorithms that compute the structure of heap
allocated storage with varying degrees of power and complexity~\cite{srw07}.
Heap nodes not pointed to by any root variable are summarized as a single
summary node. When a program statement creates a pointer from a new root
variable to one of the heap locations represented by the summary node, the
algorithm materializes the summary node. It creates two nodes---one
representing a single materialized heap node pointed to by the new root
variable and the other representing the remaining summary nodes not pointed to
by any root variable.
We describe below some shape analysis techniques that summarize using variables.

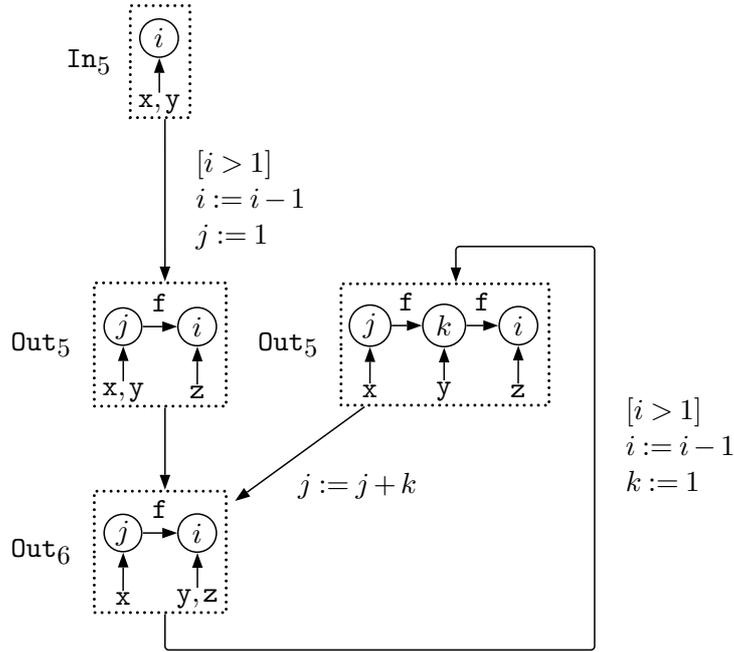
\begin{figure}[t]
\begin{center}
  {
    \psset{unit=.25mm}
    \begin{pspicture}(-40,-280)(375,95)
    \putnode{a}{origin}{40}{60}{\Galiasfour}
    \end{pspicture}
   }
\end{center}
\caption{Summarization using variables: Counter automaton \protect
\cite{bbhimv06} for the program statements 5 to 6 in Figure~\ref{eg2-prog} is
shown. States of the automaton denote the abstract heaps at the program points
shown. Edges of the automaton denote the condition of transition in the
automaton. Counter variables ($i,j,$ and $k$) corresponding to each abstract
node in the heap are depicted inside the node itself.}

\label{bbhimv06}
\figrule
\end{figure}

\begin{itemize}
\item
Sagiv et al.  \cite{srw96,srw98} distinguish between heap locations by
their pointed-to-by-$x$ predicate values for all variables $x$ in the
program\footnote{A generalized approach of shape analysis~\cite{srw96} is
TVLA~\cite{srw99}, which uses summarization using generic instrumentation
predicates (see Section~\ref{sec:other} and Figures~\ref{shape3}
and~\ref{srw99}).}.  We use the running program in Figure~\ref{eg2-prog} to
illustrate various forms of the shape analysis techniques. Unbounded memory
graphs of the program are shown in Figure~\ref{eg2-unbounded-1} and
Figure~\ref{eg2-unbounded-2}. Fixpoint computation of the bounded shape
graph~\cite{srw96} at $\Outn_6$ is shown in Figure~\ref{srw96}. Intermediate
steps are shown in Figures~\ref{shape1} and \ref{shape2}. Let us see how these
are obtained. Figure~\ref{shape1} shows a shape graph at $\Outn_4$ which
contains a node pointed to by both $\X$ and $\Y$.  This node in turn points to
a summary node through link $\f$ representing an unbounded number of
dereferences of field $\f$. At $\Outn_5$, $\Z$ points to a node $\Y
\myrightarrow \f$ of Figure~\ref{shape1}. For this, a node (pointed to by $\Z$)
is created by materializing the summary node $\Y \myrightarrow \f$. At
$\Outn_6$, $\Y$  points to this materialized node (pointed to by $\Z$) (shown
in Figure~\ref{shape2}). In the subsequent iteration of the loop, $\Y$ and $\Z$
point to a subsequent node (shown in Figure~\ref{srw96}). The remaining nodes
(not pointed to by any of $\X$, $\Y$, and $\Z$---those between $\X$ and $\Y$
and those beyond $\Y$) get summarized (represented using dashed lines) as shown
in Figure~\ref{srw96}. Here we see that node pointed to by $\X$ either directly
points to the node pointed to by $\Y$ (or $\Z$) via field $\f$ or points to an
unbounded number of nodes before pointing to the node pointed to by $\Y$ (or
$\Z$) via field $\f$.

Let us compare the shape graphs produced by Sagiv et al. \cite{srw96} (Figures~\ref{shape1}
and~\ref{srw96}) with those of Chase et al. \cite{cwz90} (Figures~\ref{cwz90-1} and
\ref{cwz90-2}). The graphs at $\Outn_4$ shown in Figure~\ref{cwz90-1} and
Figure~\ref{shape1} store identical information. However, the graph at $\Outn_6$
shown in Figure~\ref{srw96} is more precise than the graph at $\Outn_6$ in
Figure~\ref{cwz90-2}---unlike the latter, the former is
able to indicate that $\Y$ and $\Z$ always point to the same location on the
list due to materialization.


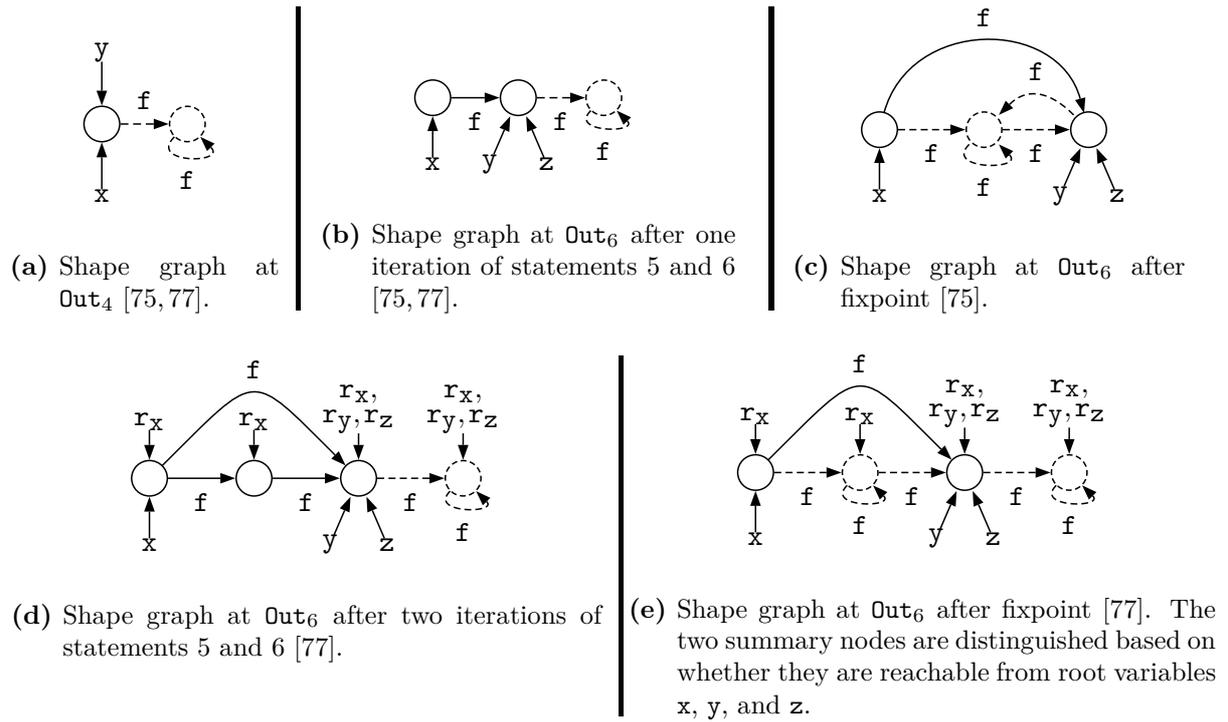
\begin{figure}[t]
\begin{subfigure}[b]{.23\textwidth}
\begin{center}
  {
    \psset{unit=.25mm}
    \begin{pspicture}(15,55)(90,145)

    \putnode{a}{origin}{40}{60}{\Gsrwa}
    \end{pspicture}
   }
\end{center}
\caption{\parbox[t]{.8\textwidth}{Shape graph at $\Outn_4$
\cite{srw96,srw99}.}}
\label{shape1}
\end{subfigure}
~
\vrulesep
\begin{subfigure}[b]{.38\textwidth}
\begin{center}
  {
    \psset{unit=.25mm}
    \begin{pspicture}(15,55)(145,115)

    \putnode{a}{origin}{40}{60}{\Gsrwb}
    \end{pspicture}
   }
\end{center}
\caption{\parbox[t]{.8\textwidth}{Shape graph at $\Outn_6$ after one iteration
of statements 5 and 6 \cite{srw96,srw99}.}}
\label{shape2}
\end{subfigure}
~
\vrulesep
\begin{subfigure}[b]{.36\textwidth}
\begin{center}
  {
    \psset{unit=.25mm}
    \begin{pspicture}(15,55)(160,160)

    \putnode{a}{origin}{40}{60}{\Galiasone}
    \end{pspicture}
   }
\end{center}
\caption{\parbox[t]{.8\textwidth}{Shape graph at $\Outn_6$ after
fixpoint~\cite{srw96}.}}
\label{srw96}
\end{subfigure}

\begin{subfigure}[b]{.5\textwidth}
\begin{center}
  {
    \psset{unit=.25mm}
    \begin{pspicture}(15,55)(210,160)

    \putnode{a}{origin}{40}{60}{\Gsrwc}
    \end{pspicture}
   }
\end{center}
\caption{\parbox[t]{.9\textwidth}{Shape graph at $\Outn_6$ after two iterations
of statements 5 and 6~\cite{srw99}.\\ \\}}
\label{shape3}
\end{subfigure}
~
\vrulesep
\begin{subfigure}[b]{.5\textwidth}
\begin{center}
  {
    \psset{unit=.25mm}
    \begin{pspicture}(15,55)(220,155)

    \putnode{a}{origin}{40}{60}{\Galiasoneb}
    \end{pspicture}
   }
\end{center}
\caption{\parbox[t]{.9\textwidth}{Shape graph at $\Outn_6$ after fixpoint
\cite{srw99}. The two summary nodes are distinguished based on whether they are
reachable from root variables $\X$, $\Y$, and $\Z$.}}
\label{srw99}
\end{subfigure}
\caption{Summarization using variables \protect \cite{srw96} is shown in
Figures~\ref{shape1}, \ref{shape2}, and \ref{srw96}. Summarization using generic
instrumentation predicates \protect \cite{srw99} is shown in
Figures~\ref{shape1}, \ref{shape2}, \ref{shape3}, and \ref{srw99} for the
program in Figure~\ref{eg2-prog}. Pointer ${\R}_{x}$ denotes whether any 
variable $x$ can transitively reach the node.  It can be seen that variable $\Z$
materializes the summary node pointed to by ${\Y \myrightarrow \f}$ in
Figures~\ref{shape1} and \ref{shape2}.}
\label{srw}
\figrule
\end{figure}


\item
An imprecision in shape analysis is that its summary nodes do not remember the
exact count of the number of concrete nodes represented by a summary node in
an abstract heap graph.  These counts are useful in checking termination of the
programs that needs to consider the size of the list being accessed.  An
interesting solution to this problem is the use of a counter with every such
summary node in the heap graph in order to denote the number of concrete heap
locations represented by the summary node~\cite{bbhimv06}.  This is used to
define a {\em counter automaton} abstraction of the state transition behaviour
of heap manipulating programs.  This is illustrated in Figure~\ref{bbhimv06} for
the program in Figure~\ref{eg2-prog}.  With the use of variables $i$, $j$, and
$k$ for counters, the algorithm ensures that the analysis is not unbounded. The
automaton starts with a heap graph containing one summary node (with counter
$i$), pointed to by $\X$ and $\Y$ at $\Inn_5$. It proceeds to $\Outn_5$ if
counter $i > 1$, and materializes the node into a unique node (with a new
counter $j=1$) pointed to by $\X$ and $\Y$, and the remaining summary node (with
counter $i$) pointed to by $\Z$. Here counter $i$ used at $\Inn_5$ is
decremented at $\Outn_5$. The graph at $\Outn_5$ is then transformed to
$\Outn_6$ under the influence of program statement 6. To further transform this
graph from $\Outn_6$ to $\Outn_5$ in the loop, if counter $i > 1$, it
materializes the summary node pointed to by $\Y$ at $\Outn_6$ into a new node
(with a new counter $k=1$) pointed to $\Y$, and the remaining summary node (with
counter $i$) pointed to by $\Z$. Here counter $i$ used at $\Outn_6$ is
decremented by one at $\Outn_5$. In the transformation from $\Outn_5$ to
$\Outn_6$, since $\Y$ will start to point to $\Z$, the node with counter $k$
will not be pointed to by any variable. Therefore, nodes with counters $k$ and
$j$ are merged, and their counter values updated (added up) at $\Outn_6$.
Bouajjani et al. \cite{bbhimv06} have used these counters for verifying safety
and termination of some sorting programs.

\end{itemize}


\subsection{Summarization Using Generic Instrumentation Predicates}
\label{sec:other}
We describe below some other generic instrumentation predicates based
summarization techniques, including TVLA, type propagation analyses, acyclic
call paths, and context free grammars that have been used for a  store based
heap model.

\begin{itemize}
\item
As an improvement over the summarization technique using only
variables~\cite{srw96} (see Section~\ref{sec:var}), the following predicates
are used in order to summarize heap nodes more
precisely~\cite{srw99,wsr00,srw07}.
\begin{itemize}
\item {\em pointed-to-by-$x$} property denotes whether a heap node is pointed
directly by variable $x$.
\item {\em reachable-from-$x$-via-$f$} property denotes whether variable $x$ can
transitively reach a heap node via field $f$.
\end{itemize}
We use the running program in Figure~\ref{eg2-prog} to illustrate the
summarization. Unbounded memory graphs at $\Outn_4$ and $\Outn_6$ of the program
are shown in Figures~\ref{eg2-unbounded-1} and~\ref{eg2-unbounded-2}. Fixpoint
computation of a bounded shape graph using predicates {\em pointed-to-by-$x$}
and {\em reachable-from-$x$-via-$f$} for summarization at $\Outn_6$ is shown in
Figure~\ref{srw99}. Intermediate steps are shown in
Figures~\ref{shape1},~\ref{shape2}, and~\ref{shape3}. We have already explained
the bounded shape graph obtained using only {\em pointed-to-by-$x$}
predicate~\cite{srw96} for summarization at $\Outn_6$ in Figure~\ref{srw96}
(see Section~\ref{sec:var}). Compare Figures~\ref{srw96} and~\ref{srw99} to
observe that the bounded shape graphs obtained are the same with respect to the
nodes pointed to by a root pointer variable; however, they differ with respect
to the summary nodes not pointed to by any root pointer variable. This is
because of the use of the additional predicate {\em
reachable-from-$x$-via-$f$}; this predicate is denoted as ${\R}_{\X}$,
${\R}_{\Y}$, and ${\R}_{\Z}$ in Figures~\ref{shape3} and~\ref{srw99}. To see
how Figure~\ref{srw99} is obtained, further observe the following in the
intermediate step shown in Figure~\ref{shape3}: the node pointed to by
${\R}_{\X}$ is kept separate from the summary node pointed to by ${\R}_{\X}$,
${\R}_{\Y}$, and ${\R}_{\Z}$.  Therefore, the shape graph in Figure~\ref{srw99}
represents unbounded dereferences of field $\f$ following root node $\X$ and
another sequence of unbounded dereferences of field $\f$ following root node
$\Y$ (or $\Z$). 

This paper builds a parametric framework, which allows the designer of shape
analysis algorithm to identify any desired heap property. The designer can
specify different predicates in order to obtain more useful and finer results,
depending on the kind of data structure used in a program. For example, the use
of predicate ``is shared" gives more precise sharing information, and the use of
predicate ``lies on cycle" gives more precise information about cycles in the
heap memory. Further, 3-valued predicates (TVLA)~\cite{srw99,wsr00,srw07} help
in describing properties of the shape graph using three values, viz. false,
true, and don't know. Therefore, both may and must pointer information can be
stored. Shape analysis stores and summarizes heap information precisely, but at
the same time, it is expensive due to the use of predicates for each
node~\cite{cdoy11}.

\item
Another way of summarizing unbounded heap locations is based on the types of
the heap locations. Sundaresan et al. \cite{shrvlgg00} merge unnamed heap nodes if the types
reaching the heap locations are the same. For example, for some variables $\X$
and $\Y$ containing field $\f$, heap locations $\X.\f$ and $\Y.\f$ are merged
and represented as $\C.\f$ if $\X$ and $\Y$ point to objects whose class name is
$\C$.  This method has been used in literature to determine at compile time
which virtual functions may be called at runtime.  This involves determining the
runtime types that reach the receiver object of the virtual function. This
requires data flow analysis to propagate types of the receiver objects from
allocation to the method invocation. These techniques that perform data flow
analysis of types are called {\em type propagation analyses}~\cite{dmm96}. 


\item
Lattner and Adve \cite{la03} point out that if heap objects are distinguished by allocation
sites with a context-insensitive analysis\footnote{A context-sensitive analysis
examines a given procedure separately for different calling contexts.},
precision is lost. This is because it cannot segregate distinct data structure
instances that have been created by the same function i.e. at the same
allocation site via different call paths in the program.  To overcome this
imprecision, Lattner and Adve \cite{la03,lla07} propose to name heap
objects by the entire acyclic call paths through which the heap objects were
created. They compute points-to graphs called {\em Data Structure} graphs, which
use a unification-based approach~\cite{s96}. Here, all heap nodes pointed to by the
same pointer variable via the same field are merged; in other words, every
pointer field points to at most one heap node. The use of acyclic call paths and
unification-based approach help in summarizing the potentially infinite number
of heap nodes that can be created in recursive function calls and loops.


\newcommand{\Greachgraph}
{{
{\psset{unit=.25mm}
\psset{linewidth=.2mm,arrowsize=4pt,arrowinset=0}
		\begin{pspicture}(-3,5)(80,-310)
		\psrelpoint{origin}{v1}{10}{-2}
		\rput(\x{v1},\y{v1}){\rnode{v1}{{$\V$}}}
		\psrelpoint{v1}{v11}{0}{-10}
		\rput(\x{v11},\y{v11}){\rnode{v11}{\pscirclebox[fillstyle=solid,
				fillcolor=white,framesep=2]{}}}
		\psrelpoint{v1}{w1}{25}{0}
		\rput(\x{w1},\y{w1}){\rnode{w1}{{$\W$}}}
		\psrelpoint{w1}{w11}{0}{-10}
		\rput(\x{w11},\y{w11}){\rnode{w11}{\pscirclebox[fillstyle=solid,
				fillcolor=white,framesep=2]{}}}
		\psrelpoint{w1}{x1}{25}{0}
		\rput(\x{x1},\y{x1}){\rnode{x1}{{$\X$}}}
		\psrelpoint{x1}{x11}{0}{-10}
		\rput(\x{x11},\y{x11}){\rnode{x11}{\pscirclebox[fillstyle=solid,
				fillcolor=white,framesep=2]{}}}
		\psrelpoint{x1}{y1}{25}{0}
		\rput(\x{y1},\y{y1}){\rnode{y1}{{$\Y$}}}
		\psrelpoint{y1}{y11}{0}{-10}
		\rput(\x{y11},\y{y11}){\rnode{y11}{\pscirclebox[fillstyle=solid,
				fillcolor=white,framesep=2]{}}}
		\psrelpoint{y1}{z1}{25}{0}
		\rput(\x{z1},\y{z1}){\rnode{z1}{{$\Z$}}}
		\psrelpoint{z1}{z11}{0}{-10}
		\rput(\x{z11},\y{z11}){\rnode{z11}{\pscirclebox[fillstyle=solid,
				fillcolor=white,framesep=2]{}}}

		\psrelpoint{v1}{v2}{0}{-100}
		\rput(\x{v2},\y{v2}){\rnode{v2}{\pscirclebox[fillstyle=solid,
				fillcolor=white,framesep=2]{}}}
		\ncline{->}{v11}{v2}
		\psrelpoint{w1}{w2}{0}{-100}
		\rput(\x{w2},\y{w2}){\rnode{w2}{\pscirclebox[fillstyle=solid,
				fillcolor=white,framesep=2]{}}}
		\ncline{->}{w11}{w2}
		\psrelpoint{x1}{x2}{0}{-100}
		\rput(\x{x2},\y{x2}){\rnode{x2}{\pscirclebox[fillstyle=solid,
				fillcolor=white,framesep=2]{}}}
		\psrelpoint{y1}{y2}{0}{-100}
		\rput(\x{y2},\y{y2}){\rnode{y2}{\pscirclebox[fillstyle=solid,
				fillcolor=white,framesep=2]{}}}
		\ncline{->}{y11}{y2}
		\ncline{->}{y11}{x2}
		\ncput*[nrot=:D,npos=.6]{\footnotesize$\HEAD$}
		\psrelpoint{z1}{z2}{0}{-100}
		\rput(\x{z2},\y{z2}){\rnode{z2}{\pscirclebox[fillstyle=solid,
				fillcolor=white,framesep=2]{}}}
		\ncline{->}{z11}{z2}
		\ncline{->}{z11}{x2}
		\ncput*[nrot=:D,npos=.3]{\footnotesize$\TAIL$}

		\psrelpoint{v2}{v3}{0}{-100}
		\rput(\x{v3},\y{v3}){\rnode{v3}{\pscirclebox[fillstyle=solid,
				fillcolor=white,framesep=2]{}}}
		\psrelpoint{w2}{w3}{0}{-100}
		\rput(\x{w3},\y{w3}){\rnode{w3}{\pscirclebox[fillstyle=solid,
				fillcolor=white,framesep=2]{}}}
		\ncline{->}{w2}{w3}
		\psrelpoint{x2}{x3}{0}{-100}
		\rput(\x{x3},\y{x3}){\rnode{x3}{\pscirclebox[fillstyle=solid,
				fillcolor=white,framesep=2]{}}}
		\ncline{->}{x2}{x3}
		\ncline{->}{x2}{v3}
		\ncput*[nrot=:D]{\footnotesize$\HEAD^{-1}$}
		\psrelpoint{y2}{y3}{0}{-100}
		\rput(\x{y3},\y{y3}){\rnode{y3}{\pscirclebox[fillstyle=solid,
				fillcolor=white,framesep=2]{}}}
		\ncline{->}{y2}{y3}
		\psrelpoint{z2}{z3}{0}{-100}
		\rput(\x{z3},\y{z3}){\rnode{z3}{\pscirclebox[fillstyle=solid,
				fillcolor=white,framesep=2]{}}}
		\ncline{->}{z2}{z3}

		\psrelpoint{v3}{v4}{0}{-100}
		\rput(\x{v4},\y{v4}){\rnode{v4}{\pscirclebox[fillstyle=solid,
				fillcolor=white,framesep=2]{}}}
		\ncline{->}{v3}{v4}
		\psrelpoint{w3}{w4}{0}{-100}
		\rput(\x{w4},\y{w4}){\rnode{w4}{\pscirclebox[fillstyle=solid,
				fillcolor=white,framesep=2]{}}}
		\psrelpoint{x3}{x4}{0}{-100}
		\rput(\x{x4},\y{x4}){\rnode{x4}{\pscirclebox[fillstyle=solid,
				fillcolor=white,framesep=2]{}}}
		\ncline{->}{x3}{x4}
		\ncline{->}{x3}{w4}
		\ncput*[nrot=:D,npos=.6]{\footnotesize$\TAIL^{-1}$}
		\psrelpoint{y3}{y4}{0}{-100}
		\rput(\x{y4},\y{y4}){\rnode{y4}{\pscirclebox[fillstyle=solid,
				fillcolor=white,framesep=2]{}}}
		\ncline{->}{y3}{y4}
		\psrelpoint{z3}{z4}{0}{-100}
		\rput(\x{z4},\y{z4}){\rnode{z4}{\pscirclebox[fillstyle=solid,
				fillcolor=white,framesep=2]{}}}
		\ncline{->}{z3}{z4}

		\end{pspicture}
		}
}}

\newcommand{\Gappend}
{{
{\psset{unit=.25mm}
\psset{linewidth=.2mm,arrowsize=4pt,arrowinset=0}
		\begin{pspicture}(-6,5)(210,-190)
		\psrelpoint{origin}{q}{10}{0}
		\rput(\x{q},\y{q}){\rnode{q}{{$\X$}}}
		\psrelpoint{origin}{j}{-3}{0}
		\rput(\x{j},\y{j}){\rnode{j}{}}
		\psrelpoint{q}{n1}{20}{-35}
		\rput(\x{n1},\y{n1}){\rnode{n1}{
			\begin{tabular}{;{3.5pt/2pt}l;{3.5pt/2pt}l;{3.5pt/2pt}}
			\hdashline[3.5pt/2pt]{\white 1}&{\white 1}\\ \hdashline[3.5pt/2pt]
			\end{tabular}}}
		\ncline{->}{q}{n1}
		\psrelpoint{n1}{n2}{30}{-48}
		\rput(\x{n2},\y{n2}){\rnode{n2}{
			\begin{tabular}{;{3.5pt/2pt}l;{3.pt/2pt}l;{3.pt/2pt}}
			\hdashline[3.5pt/2pt]{\white 1}&{\white 1}\\ \hdashline[3.5pt/2pt]
			\end{tabular}}}
		\psrelpoint{n1}{n3}{15}{0}
		\rput(\x{n3},\y{n3}){\rnode{n3}{}}
		\ncline{->}{n3}{n2}
		\psrelpoint{n2}{n4}{30}{-48}
		\rput(\x{n4},\y{n4}){\rnode{n4}{
			\begin{tabular}{;{3.pt/2pt}l;{3.pt/2pt}l;{3.pt/2pt}}
			\hdashline[3.5pt/2pt]{\white 1}&{\white 1}\\ \hdashline[3.5pt/2pt]
			\end{tabular}}}
		\psrelpoint{n2}{n5}{15}{0}
		\rput(\x{n5},\y{n5}){\rnode{n5}{}}
		\ncline{->}{n5}{n4}

		\psrelpoint{n1}{c1}{-25}{-48}
		\rput(\x{c1},\y{c1}){\rnode{c1}{\pscirclebox[fillstyle=solid,
				fillcolor=white,framesep=9]{}}}
		\psrelpoint{n1}{c2}{-10}{0}
		\rput(\x{c2},\y{c2}){\rnode{c2}{}}
		\ncline{->}{c2}{c1}
		\psrelpoint{n2}{c3}{-25}{-48}
		\rput(\x{c3},\y{c3}){\rnode{c3}{\pscirclebox[fillstyle=solid,
				fillcolor=white,framesep=9]{}}}
		\psrelpoint{n2}{c4}{-10}{0}
		\rput(\x{c4},\y{c4}){\rnode{c4}{}}
		\ncline{->}{c4}{c3}
		\psrelpoint{n4}{c5}{-25}{-48}
		\rput(\x{c5},\y{c5}){\rnode{c5}{\pscirclebox[fillstyle=solid,
				fillcolor=white,framesep=9]{}}}
		\psrelpoint{n4}{c6}{-10}{0}
		\rput(\x{c6},\y{c6}){\rnode{c6}{}}
		\ncline{->}{c6}{c5}


		\psrelpoint{origin}{q1}{90}{0}
		\rput(\x{q1},\y{q1}){\rnode{q1}{{\ttfamily APPEND}}}
		\psrelpoint{origin}{j1}{-3}{0}
		\rput(\x{j1},\y{j1}){\rnode{j1}{}}
		\psrelpoint{q1}{n11}{20}{-35}
		\rput(\x{n11},\y{n11}){\rnode{n11}{
			\begin{tabular}{|l|l|}
			\hline{\white 1}&{\white 1}\\ \hline
			\end{tabular}}}
		\ncline{->}{q1}{n11}
		\psrelpoint{n11}{c21}{-10}{0}
		\rput(\x{c21},\y{c21}){\rnode{c21}{}}
		\nccurve[angleA=135,angleB=45,nodesep=-2,ncurv=0.28]{->}{c21}{c1}
		\psrelpoint{n11}{n21}{30}{-48}
		\rput(\x{n21},\y{n21}){\rnode{n21}{
			\begin{tabular}{|l|l|}
			\hline{\white 1}&{\white 1}\\ \hline
			\end{tabular}}}
		\psrelpoint{n11}{n31}{15}{0}
		\rput(\x{n31},\y{n31}){\rnode{n31}{}}
		\ncline{->}{n31}{n21}
		\psrelpoint{n21}{c41}{-10}{0}
		\rput(\x{c41},\y{c41}){\rnode{c41}{}}
		\nccurve[angleA=135,angleB=45,nodesep=-2,ncurv=0.28]{->}{c41}{c3}
		\psrelpoint{n21}{n41}{30}{-48}
		\rput(\x{n41},\y{n41}){\rnode{n41}{
			\begin{tabular}{|l|l|}
			\hline{\white 1}&{\white 1}\\ \hline
			\end{tabular}}}
		\psrelpoint{n21}{n51}{15}{0}
		\rput(\x{n51},\y{n51}){\rnode{n51}{}}
		\ncline{->}{n51}{n41}
		\psrelpoint{n41}{c61}{-10}{0}
		\rput(\x{c61},\y{c61}){\rnode{c61}{}}
		\nccurve[angleA=135,angleB=45,nodesep=-2,ncurv=0.28]{->}{c61}{c5}
		\psrelpoint{n41}{c7}{25}{-48}
		\rput(\x{c7},\y{c7}){\rnode{c7}{\pscirclebox[fillstyle=solid,
				fillcolor=white,framesep=9]{}}}
		\psrelpoint{n41}{c8}{10}{0}
		\rput(\x{c8},\y{c8}){\rnode{c8}{}}
		\ncline{->}{c8}{c7}
		\psrelpoint{c7}{m}{10}{35}
		\rput(\x{m},\y{m}){\rnode{m}{$\Y$}}
		\ncline{->}{m}{c7}

		\end{pspicture}
		}
}}

\begin{figure}[t]
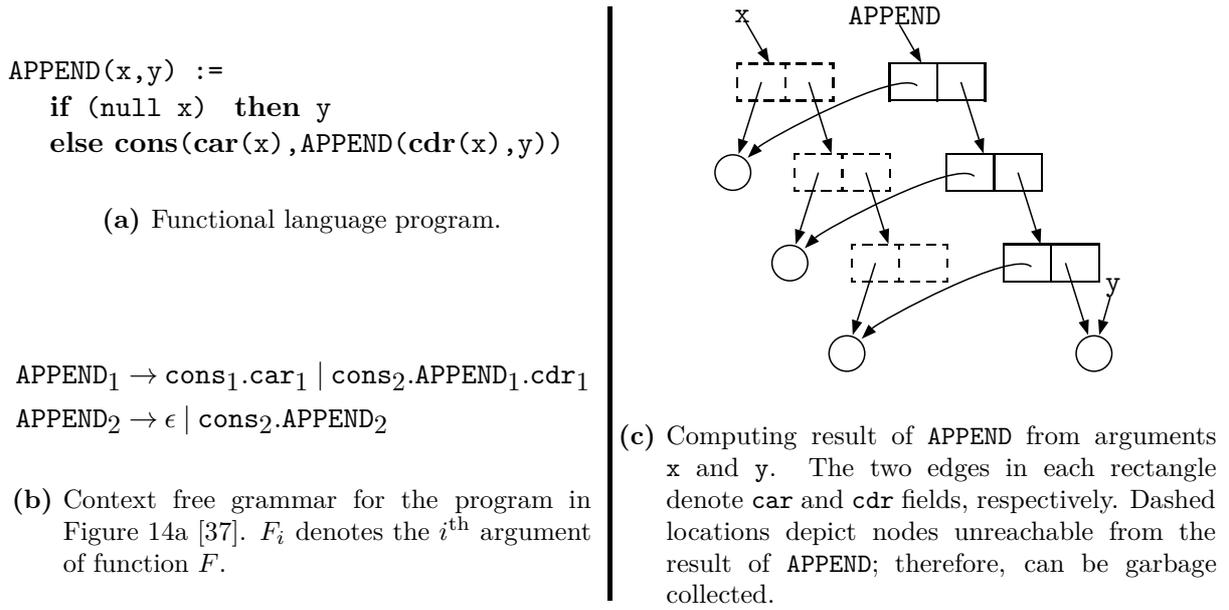

\begin{minipage}[b]{.49\textwidth}
\begin{subfigure}[b]{1\textwidth}
\begin{Verbatim}[commandchars=\\\{\},codes={\catcode`$=3\catcode`_=8}]
APPEND(x,y) := 
  {\bf if} (null x) {\bf then} y
  {\bf else}{\bf cons}({\bf{car}}(x),APPEND({\bf{cdr}}(x),y))
\end{Verbatim}
\caption{Functional language program.}
\label{isy88-1}
\end{subfigure}

\begin{subfigure}[b]{1\textwidth}
\begin{center}
{\white .}\\
{\white .}\\
{\white .}\\
$\begin{aligned}
& \text{\ttfamily APPEND}_1 \rightarrow \cons_1 . \car_1 \mid \cons_2 . \text{\ttfamily APPEND}_1 . \cdr_1 \\
& \text{\ttfamily APPEND}_2 \rightarrow \epsilon \mid \cons_2 . \text{\ttfamily APPEND}_2
\end{aligned}$
\end{center}
\caption{\parbox[t]{.9\textwidth}{Context free grammar for the program in
Figure~\ref{isy88-1} \protect \cite{isy88}.  $F_i$ denotes the $i^{\text{th}}$
argument of function $F$. \\ }}
\label{isy88-2}
\end{subfigure}
\end{minipage}
~
\vrulesep
\begin{subfigure}[b]{.51\textwidth}
\begin{center}
\Gappend
\end{center}
\caption{\parbox[t]{.9\textwidth}{Computing result of {\ttfamily APPEND} from
arguments $\X$ and $\Y$. The two edges in each rectangle denote $\car$ and
$\cdr$ fields, respectively. Dashed locations depict nodes unreachable from the
result of {\ttfamily APPEND}; therefore, can be garbage collected.}}
\label{isy88-3}
\end{subfigure}
\caption{Computing context free grammar for a functional language program in
order to garbage collect unreachable nodes \protect \cite{isy88}.}
\label{isy88}
\figrule
\end{figure}


\item
Another way of summarizing is to build a context free grammar of the
heap~\cite{isy88}. This has been done for functional programs, which consist of
primitive functions like $\cons$, $\car$, and $\cdr$. This grammar has been
used to detect garbage cells in a functional program through compile time
analysis. It is based on the idea that the unshared locations passed as
parameter to a function that are not a part of the final result of the
function, can be garbage collected after the function call. We describe this by
reproducing from the paper the definition of function {\ttfamily APPEND} in
Figure~\ref{isy88-1}. Data structures pointed to by variables $\X$ and $\Y$
(shown in Figure~\ref{isy88-3}) are passed as arguments to {\ttfamily APPEND}
function.  The circular nodes are reachable from the root of the result of the
{\ttfamily APPEND} function; these circular nodes can be identified as
$\X(.\cdr)^*.\car$ and $\Y$. However, the dashed locations, which belong to
$\X$, are not reachable from the root of the result of the {\ttfamily APPEND}
function; these dashed locations can be identified as $\X(.\cdr)^*$.  These
dashed locations can, therefore, be garbage collected.

In order to identify argument locations that are unreachable from the result of
the called function, the paper analyses the usage of each argument of the
called function by constructing a context free language of each argument. The
grammar constructed is shown in Figure~\ref{isy88-2}. Each argument of a
function, say $F(x_1, x_2, \dots, x_i, \dots)$, is represented by a
non-terminal in a context free grammar. Derivation rule for the $i^\text{\em
th}$ argument $x_i$ of function $F$ is $F_i \rightarrow s_1 \mid s_2 \mid \dots
\mid s_k$, where $s_1, s_2, \dots, s_k$ are all the strings obtained from the
function body. The first and the second lines in Figure~\ref{isy88-2} are the
context free grammars of {\ttfamily APPEND}$_1$ and {\ttfamily APPEND}$_2$,
which denote arguments $\X$ and $\Y$ of the program in Figure~\ref{isy88-1}.
The strings on the right hand side of the grammar consist of $\car_1, \cdr_1,
\cons_1,$ and $\cons_2$, and user defined functions.  Each function name is
used with a subscript indicating the position of argument in the function. Let
us study the grammar of function {\ttfamily APPEND} shown in
Figure~\ref{isy88-2}. {\ttfamily APPEND}$_2$ in the second line denotes the
usage of the second argument of {\ttfamily APPEND}, $\Y$ in the function
definition. It would either be used as it is or would be passed as the second
argument to $\cons$ (denoted by $\cons_2$).  {\ttfamily APPEND}$_1$ in the
first line denotes the usage of the first argument of {\ttfamily APPEND}, $\X$
in the function definition.  The strings generated by {\ttfamily APPEND}$_1$
grammar are of the form $\cons_2^k .  \cons_1 .  \car_1 . \cdr_1^k$. By reading
the string in the reverse order, we can see that {\ttfamily APPEND} decomposes
list $\X$, $k$ number of times by the application of $\cdr$, and then a $\car$
selects the element at that position, followed by a $\cons_1$ on the element to
make it the left child of a new location, which itself will be acted on by
$\cons_2$ the same $k$ number of times.  The context free grammar is used to
identify reachable paths from the argument.  For example, using the grammar
{\ttfamily APPEND}$_1$, i.e. argument $\X$, it can be seen that string
$(\cdr_1)^k.\car_1$ (obtained from the reverse of string $\cons_2^k . \cons_1 .
\car_1 .  \cdr_1^k$) denotes the locations $\X(.\cdr)^*.\car$, which are
reachable from the result of {\ttfamily APPEND}.  The rest of the locations in
argument $\X$ are unreachable and can be garbage collected.

Liveness based garbage collection has been performed using grammars also by
Asati et al. \cite{askm14} for creating the notion of a {\em demand\/} that the execution of
an expression makes on the heap memory.  In somewhat similar lines, liveness
based garbage collection for imperative programs has been performed using a
storeless model by Khedker et al. \cite{ksk07} (see Section~\ref{sec:less-patt}).


\begin{figure}[t]
\begin{center}
  {
    \psset{unit=.25mm}
    \psset{linewidth=.2mm}
	\psset{linewidth=.2mm,arrowsize=5pt,arrowinset=0}
    \begin{pspicture}(-10,-195)(95,75)

    \putnode{a}{origin}{40}{60}{1 \psframebox{$\X := \cons(\Y, \Z)$}\white\ 1}
    \putnode{b}{a}{0}{-95}{2 \psframebox{$\V := \car(\X)$}\white\ 2}
    \putnode{c}{b}{0}{-95}{3 \psframebox{$\W := \cdr(\X)$}\white\ 2}
    
    \ncline{->}{a}{b}
    \ncline{->}{b}{c}

    \end{pspicture}
   }
\Greachgraph
\end{center}
\caption{A control flow graph of a program and its equation dependence graph.
Edges in the equation dependence graph have been labelled with $\HEAD$, $\TAIL$,
$\HEAD^{-1}$, and $\TAIL^{-1}$; those shown without labels represent identity
relation (label {\ttfamily id}) \protect \cite{r97}.}
\label{r97}
\figrule
\end{figure}
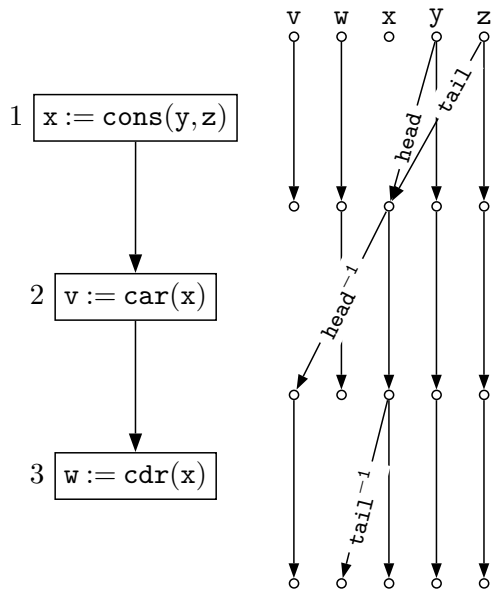


\item 
Another way of building context free grammar of heap access paths is by posing
shape analysis as CFL reachability problem.  This has been done for Lisp like
languages that do not support strong updates~\cite{r97}.  A CFL reachability
problem is different from the graph reachability problem in the sense that a
path between two variables is formed only if the concatenation of the labels on
the edges of the path is a word in the specified context free language.  {\em
Equation dependence graph} is constructed by marking all program variables at
each program point in the program's control flow graph. The edges between these
variables are labelled with $\HEAD$, $\TAIL$, $\HEAD^{-1}$, and $\TAIL^{-1}$. 

We illustrate the use of these labels in the equation dependence graph in
Figure~\ref{r97}. For statement 1, $\X:=\cons(\Y,\Z)$, label $\HEAD$ is marked
on the edge from $\Y$ before statement 1 to $\X$ after statement 1. Similarly,
label $\TAIL$ is marked on the edge from $\Z$ before statement 1 to $\X$ after
statement 1. This denotes that $\X$ derives its $\HEAD$ from $\Y$ and $\TAIL$
from $\Z$.  For program statement 2, $\V:=\car(\X)$, label $\HEAD^{-1}$ is
marked on the edge from $\X$ before statement 2 to $\V$ after statement 2. This
denotes that $\V$ gets its value using the $\HEAD$ of $\Y$. Similarly,
$\TAIL^{-1}$ is labelled for statement 3, $\W:=\cdr(\X)$.

Heap language in terms of access paths is identified by concatenating, in order,
the labels of the edges on the paths of the equation dependence graph.  For
example, the path from $\Z$ before statement 1 to $\W$ after statement 3 shows
that $\W$ gets the value $\Z.\TAIL.{\text{\ttfamily id}}.\TAIL^{-1}$, which is
simply $\Z$. Heap properties can be obtained by solving CFL reachability
problems on the equation dependence graph using the following context free
grammars~\cite{r97}:
\begin{itemize}
\item 
${\text{\em id\_path}} \rightarrow {\text{\em id\_path id\_path}} \mid {\HEAD
\text{ \em id\_path }} \HEAD^{-1} \mid {\TAIL \text{ \em id\_path }} \TAIL^{-1}
\mid {\text{\ttfamily id}} \mid \epsilon$
\\This grammar represents paths in which the number of $\HEAD^{-1}$
($\TAIL^{-1}$) are balanced by a matching number of $\HEAD$ ($\TAIL$), implying
that the heap was used through $\HEAD$$^{-1}$ ($\TAIL$$^{-1}$) as much as it
was constructed using $\HEAD$ ($\TAIL$). 

\item
${\text{\em head\_path}} \rightarrow {\text{\em id\_path }} \HEAD {\text{ \em
id\_path}}$
\\ ${\text{\em tail\_path}} \rightarrow {\text{\em id\_path }} \TAIL {\text{ \em
id\_path}}$ 
\\These grammars represent paths in which the number of $\HEAD$ ($\TAIL$) is
more than the number of $\HEAD^{-1}$ ($\TAIL^{-1}$), implying that the
amount of heap allocated using $\HEAD$ ($\TAIL$) is more than the amount of heap
dereferenced using $\HEAD^{-1}$ ($\TAIL^{-1}$).
\end{itemize}

\end{itemize}

\subsection{Summarization Using Allocation Sites and Other Generic
Instrumentation Predicates}
\label{sec:alloc-other}

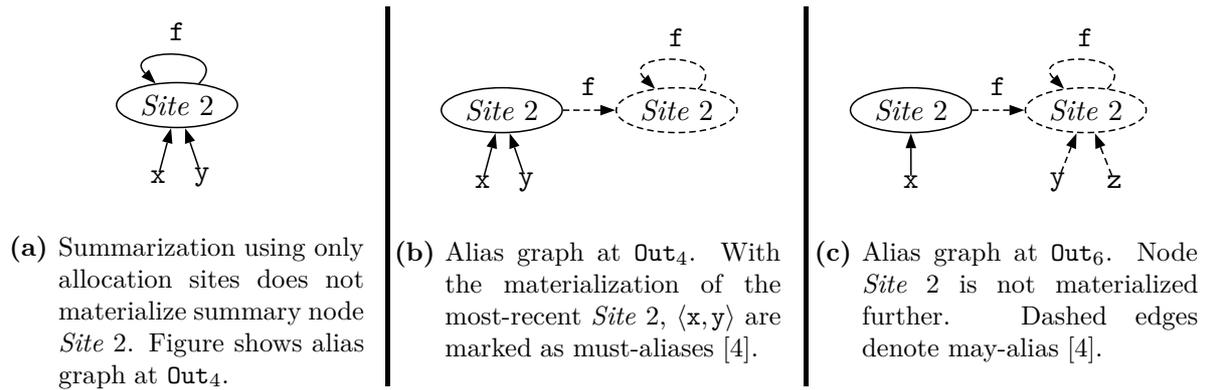
\begin{figure}[t]
\begin{subfigure}[b]{.3\textwidth}
\begin{center}
  {
    \psset{unit=.25mm}
    \begin{pspicture}(10,55)(80,150)

    \putnode{a}{origin}{40}{60}{\Galiastraditional}
    \end{pspicture}
   }
\end{center}
\caption{\parbox[t]{.85\textwidth}{Summarization using only allocation sites
does not materialize summary node {\em Site }2. Figure shows alias graph at
{\Outn}$_4$.}}
\label{br06-0}
\end{subfigure}
~
\vrulesep
\begin{subfigure}[b]{.33\textwidth}
\begin{center}
  {
    \psset{unit=.25mm}
    \begin{pspicture}(10,55)(160,150)

    \putnode{a}{origin}{40}{60}{\Galiastwo}
    \end{pspicture}
   }
\end{center}
\caption{\parbox[t]{.85\textwidth}{Alias graph at {\Outn}$_4$. With the
materialization of the most-recent {\em Site }2, $\langle \X, \Y \rangle$ are
marked as must-aliases~\cite{br06}.\\}}
\label{br06-1}
\end{subfigure}
~
\vrulesep
\begin{subfigure}[b]{.33\textwidth}
\begin{center}
  {
    \psset{unit=.25mm}
    \begin{pspicture}(0,55)(160,150)

    \putnode{a}{origin}{40}{60}{\Galiasthree}
    \end{pspicture}
   }
\end{center}
\caption{\parbox[t]{.85\textwidth}{Alias graph at {\Outn}$_6$. Node {\em Site }2
is not materialized further. Dashed edges denote may-alias~\cite{br06}.\\}}
\label{br06-2}
\end{subfigure}
\caption{Summarization using allocation sites and other generic instrumentation
predicates \protect \cite{br06} for the program in Figure~\ref{eg2-prog} is
shown in Figures~\ref{br06-1} and \ref{br06-2}. For comparison, summarization
using only allocation sites is shown in Figure~\ref{br06-0}.}
\label{br06}
\figrule
\end{figure}

As an attempt to reduce the cost of shape analysis, {\em
recency-abstraction}~\cite{br06} is used as an approximation of heap allocated
storage. This approach does not use the TVLA tool; however, it uses concepts
from 3-valued logic shape analysis~\cite{srw99}. Here, only the most recently
allocated node at an allocation site is kept materialized representing a unique
node. Therefore, its precision level is intermediate between 
\begin{inparaenum}[(a)]
\item one summary node per allocation site and 
\item complex shape abstractions~\cite{srw99}.
\end{inparaenum}
Note that for the program in Figure~\ref{eg2-prog}, Figure~\ref{br06-0} shows
that summarization based only on allocation sites creates a summary node for
objects allocated at site 2.  Here the summary node is not materialized;
therefore, variables $\X$ and $\Y$ point to the summary node itself at
$\Outn_4$. Consequently, allocation site based summarization cannot derive that
$\X$ and $\Y$ are must-aliased. {\em Recency-abstraction} is illustrated in
Figure~\ref{br06-1} for the unbounded graph of Figure~\ref{eg2-unbounded-1}. Due
to materialization of the most recently allocated node, the method is able to
precisely mark $\X$ and $\Y$ as must-aliases at $\Outn_4$. However,
materializing only once is not enough and introduces imprecision at $\Outn_6$.
This is shown in Figure~\ref{br06-2}, where $\Y$ and $\Z$ are marked as
may-aliases (instead of the precise must-alias, as shown by the unbounded
runtime memory graph in Figure~\ref{eg2-unbounded-2}).


\newcommand{\Gsepone}
{{
{\psset{unit=.25mm}
\psset{linewidth=.2mm,arrowsize=4pt,arrowinset=0}
		\begin{pspicture}(-3,0)(75,-50)
		\psrelpoint{origin}{q}{10}{-2}
		\rput(\x{q},\y{q}){\rnode{q}{{$\X$}}}
		\psrelpoint{origin}{j}{-3}{-2}
		\rput(\x{j},\y{j}){\rnode{j}{}}
		\psrelpoint{q}{s}{10}{-38}
		\rput(\x{s},\y{s}){\rnode{s}{\pscirclebox[fillstyle=solid,
				fillcolor=white,framesep=9]{}}}
		\ncline{->}{q}{s}
		\psrelpoint{s}{t}{40}{0}
		\rput(\x{t},\y{t}){\rnode{t}{$\Null$}}
		\ncline{->}{s}{t}
		\naput{$\f$}
		\psrelpoint{s}{u}{10}{38}
		\rput(\x{u},\y{u}){\rnode{u}{{$\Y$}}}
		\ncline{->}{u}{s}

		\end{pspicture}
		}
}}

\newcommand{\Gseptwo}
{{
{\psset{unit=.25mm}
\psset{linewidth=.2mm,arrowsize=4pt,arrowinset=0}
		\begin{pspicture}(0,5)(115,-50)
		\psrelpoint{origin}{q}{10}{-2}
		\rput(\x{q},\y{q}){\rnode{q}{{$\X$}}}
		\psrelpoint{origin}{j}{-3}{-2}
		\rput(\x{j},\y{j}){\rnode{j}{}}
		\psrelpoint{q}{s}{10}{-38}
		\rput(\x{s},\y{s}){\rnode{s}{\pscirclebox[fillstyle=solid,
				fillcolor=white,framesep=9]{}}}
		\ncline{->}{q}{s}
		\psrelpoint{s}{t}{40}{0}
		\rput(\x{t},\y{t}){\rnode{t}{\pscirclebox[fillstyle=solid,
				fillcolor=white,framesep=9]{}}}
		\ncline{->}{s}{t}
		\naput{$\f$}
		\psrelpoint{t}{w}{40}{0}
		\rput(\x{w},\y{w}){\rnode{w}{$\Null$}}
		\ncline{->}{t}{w}
		\naput{$\f$}
		\psrelpoint{t}{u}{0}{38}
		\rput(\x{u},\y{u}){\rnode{u}{{$\XX'$}}}
		\ncline{->}{u}{t}
		\psrelpoint{s}{v}{10}{38}
		\rput(\x{v},\y{v}){\rnode{v}{{$\Y$}}}
		\ncline{->}{v}{s}

		\end{pspicture}
		}
}}

\newcommand{\Gsepthree}
{{
{\psset{unit=.25mm}
\psset{linewidth=.2mm,arrowsize=4pt,arrowinset=0}
		\begin{pspicture}(0,5)(155,-50)
		\psrelpoint{origin}{q}{10}{-2}
		\rput(\x{q},\y{q}){\rnode{q}{{$\X$}}}
		\psrelpoint{origin}{j}{-3}{-2}
		\rput(\x{j},\y{j}){\rnode{j}{}}
		\psrelpoint{q}{s}{10}{-38}
		\rput(\x{s},\y{s}){\rnode{s}{\pscirclebox[fillstyle=solid,
				fillcolor=white,framesep=9]{}}}
		\ncline{->}{q}{s}
		\psrelpoint{s}{t}{40}{0}
		\rput(\x{t},\y{t}){\rnode{t}{\pscirclebox[fillstyle=solid,
				fillcolor=white,framesep=9]{}}}
		\ncline{->}{s}{t}
		\naput{$\f$}
		\psrelpoint{t}{y}{40}{0}
		\rput(\x{y},\y{y}){\rnode{y}{\pscirclebox[fillstyle=solid,
				fillcolor=white,framesep=9]{}}}
		\ncline{->}{t}{y}
		\naput{$\f$}
		\psrelpoint{y}{w}{40}{0}
		\rput(\x{w},\y{w}){\rnode{w}{$\Null$}}
		\ncline{->}{y}{w}
		\naput{$\f$}
		\psrelpoint{t}{u}{0}{38}
		\rput(\x{u},\y{u}){\rnode{u}{{$\XX''$}}}
		\ncline{->}{u}{t}
		\psrelpoint{s}{v}{10}{38}
		\rput(\x{v},\y{v}){\rnode{v}{{$\Y$}}}
		\ncline{->}{v}{s}
		\psrelpoint{y}{z}{0}{38}
		\rput(\x{z},\y{z}){\rnode{z}{{$\XX'$}}}
		\ncline{->}{z}{y}

		\end{pspicture}
		}
}}

\newcommand{\Gsepfour}
{{
{\psset{unit=.25mm}
\psset{linewidth=.2mm,arrowsize=4pt,arrowinset=0}
		\begin{pspicture}(-3,5)(185,-50)
		\psrelpoint{origin}{q}{10}{-2}
		\rput(\x{q},\y{q}){\rnode{q}{{$\X$}}}
		\psrelpoint{origin}{j}{-3}{-2}
		\rput(\x{j},\y{j}){\rnode{j}{}}
		\psrelpoint{q}{s}{0}{-38}
		\rput(\x{s},\y{s}){\rnode{s}{\pscirclebox[fillstyle=solid,
				fillcolor=white,framesep=9]{}}}
		\ncline{->}{q}{s}
		\psrelpoint{s}{t}{40}{0}
		\rput(\x{t},\y{t}){\rnode{t}{\pscirclebox[fillstyle=solid,
				fillcolor=white,framesep=9]{}}}
		\ncline{->}{s}{t}
		\naput{$\f$}
		\psrelpoint{t}{y}{40}{0}
		\rput(\x{y},\y{y}){\rnode{y}{\pscirclebox[fillstyle=solid,
				fillcolor=white,framesep=9]{}}}
		\ncline{->}{t}{y}
		\naput{$\f$}
		\psrelpoint{y}{w}{40}{0}
		\rput(\x{w},\y{w}){\rnode{w}{$\dots$}}
		\ncline{->}{y}{w}
		\naput{$\f$}
		\psrelpoint{w}{p}{40}{0}
		\rput(\x{p},\y{p}){\rnode{p}{$\Null$}}
		\ncline{->}{w}{p}
		\naput{$\f$}
		\psrelpoint{t}{u}{0}{38}
		\rput(\x{u},\y{u}){\rnode{u}{{$\XX'''$}}}
		\ncline{->}{u}{t}
		\psrelpoint{y}{z}{-10}{38}
		\rput(\x{z},\y{z}){\rnode{z}{{$\Y$}}}
		\ncline{->}{z}{y}
		\psrelpoint{y}{z1}{10}{38}
		\rput(\x{z1},\y{z1}){\rnode{z1}{{$\Z$}}}
		\ncline{->}{z1}{y}

		\end{pspicture}
		}
}}

\subsection{Summarization Using Higher-Order Logics}
Heap can be abstracted as logical structures of specialized logic like
separation logic, which are more powerful than simple predicate logic.  Also,
the efficiency of shape analysis can be boosted by representing independent
portions of the heap using formulae in separation logic~\cite{r02}.  To
elaborate, it exploits spatial locality of a code i.e. the fact that each
program statement accesses only a very limited portion of the concrete state.
Using separation logic, the portion of heap that is not accessed by the
statement(s) can be easily separated from the rest and later recombined with the
modified heap after analysing the statement(s). This dramatically reduces the
amount of reasoning that must be performed, specially if the statement is a
procedure call.

Assertions expressed in separation logic may produce infinite sets of concrete
states. A fixpoint computation can be achieved using finitely represented
inductive predicate assertions~\cite{gbc06,cdoy11} like $\ls()$, $\tree()$,
$\dlist()$, representing unbounded number of concrete states, shaped like a
linked list, tree, doubly linked list, respectively. The abstraction comes from
not tracking the precise number of inductive unfoldings from the base case. Note
that unlike logics on storeless model which use access paths and hide locations
in their modeling, separation logic explicates heap locations; therefore,
separation logic is categorized under a store based model.

In separation logic, assertion $A \mapsto B$ denotes memory containing heap
location $A$, which points to heap location $B$.  Assertion $A * B$ denotes
memory represented as a union of two disjoint heaps (i.e. with no common heap
locations)---one satisfying $A$ and the other satisfying $B$. Assertion $A = B$
denotes that $A$ and $B$ have equal values.  Assertion $A \wedge B$ denotes a
heap that satisfies both $A$ and $B$.

\begin{figure}[t]
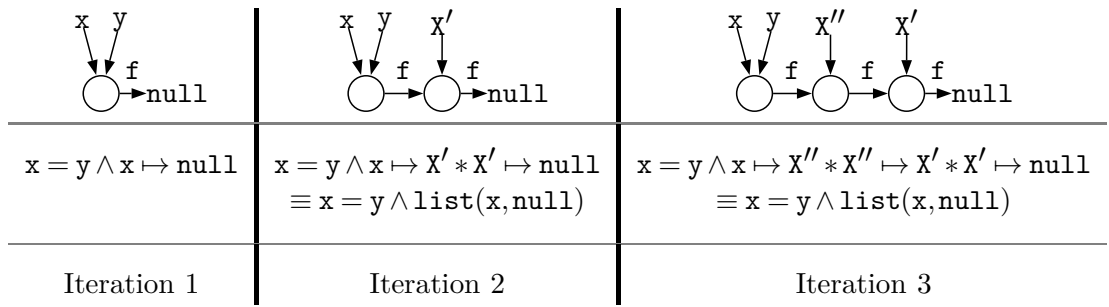
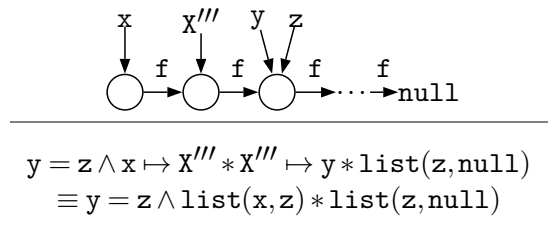

\begin{subfigure}[b]{1\textwidth}
\begin{center}
\begin{tabular}{c!{\vrule width 1.5pt}c!{\vrule width 1.5pt}c}
\Gsepone & 
\Gseptwo &
\Gsepthree 
\\ \arrayrulecolor{gray} \hline 
&& \\[-1ex]

$\X = \Y \wedge \X \mapsto \Null$ & 
$\X = \Y \wedge \X \mapsto \XX' * \XX' \mapsto \Null$ &
$\X = \Y \wedge \X \mapsto \XX'' * \XX'' \mapsto \XX' * \XX' \mapsto \Null$ \\

&
$\equiv \X = \Y \wedge \ls(\X, \Null)$ & 
$\equiv \X = \Y \wedge \ls(\X, \Null)$ \\ 
&& \\[-1ex] 
\arrayrulecolor{gray} \hline 
&& \\[-1ex]

Iteration 1 &
Iteration 2 &
Iteration 3 
\end{tabular}
\end{center}
\caption{\parbox[t]{.9\textwidth}{Heap at $\Outn_4$ obtained after respectively
three iterations of the program. $\XX'$ and $\XX''$ are new variables not used
anywhere in the program code.}}
\label{sep-1}
\end{subfigure}

\begin{subfigure}[b]{1\textwidth}
\begin{center}
\begin{tabular}{c}
\Gsepfour 
\\ \arrayrulecolor{gray} \hline 
\\[-1ex]
$\Y = \Z \wedge \X \mapsto \XX''' * \XX''' \mapsto \Y * \ls(\Z, \Null)$ \\
$\equiv \Y = \Z \wedge \ls(\X, \Z) * \ls(\Z, \Null)$ 
\\[-1ex]
\\ \arrayrulecolor{gray} \hline 
\end{tabular}
\end{center}
\caption{Heap at $\Outn_6$ after fixpoint computation. $\XX'''$ is a new
variable not used anywhere in the program code.}
\label{sep-2}
\end{subfigure}
\caption{Summarization using separation logic \protect \cite{gbc06,cdoy11} for
the program in Figure~\ref{eg2-prog}.}
\label{sep}
\figrule
\end{figure}

We work out the assertions using separation logic for the program in
Figure~\ref{eg2-prog}. In Figure~\ref{sep-1}, we have shown the heap graph and
also the assertions in separation logic at $\Outn_4$ over three iterations of
statements 2, 3, and 4 in a loop.  Assertion in the first iteration says that
$\X$ and $\Y$ hold the same value, which points to a null value. Assertion in
the second iteration says that $\X$ and $\Y$ hold the same value, which points
to a new variable $\XX'$. Separation logic introduces a variable $\XX'$, which
is not used anywhere in the program code. This $\XX'$ points to a null value.
Assertion in the third iteration says that $\X$ and $\Y$ hold the same value,
which points to another new variable $\XX''$, which further points to $\XX'$;
$\XX'$ points to a null value. If we continue in this way, we will get ever
longer formulae. This unboundedness is abstracted using the predicate $\ls()$,
where $\ls(u,v)$ says that there is a linked list segment of unbounded length
from $u$ to $v$. This predicate has the following recursive definition (here
{\ttfamily emp} denotes an empty heap): $$\ls(u,v) \Leftrightarrow
{\text{\ttfamily emp}} \vee \exists w. u \mapsto w * \ls(w,v)$$ With this, we
obtain the abstraction by using the following operation in the second iteration
at $\Outn_4$.
\\ [-1.5em]
$$\text{replace } \quad \X = \Y \wedge \X \mapsto \XX' * \XX' \mapsto \Null \quad
\text{ with } \quad \X = \Y \wedge \ls (\X, \Null)$$
\\ [-1.5em]
Using a similar way of synthesizing, the assertion at $\Outn_6$ (shown in
Figure~\ref{sep-2}) can be obtained to be $\Y = \Z \wedge \ls(\X, \Z) * \ls (\Z,
\Null)$. 

The SpaceInvader tool~\cite{dhy06} also uses separation logic. The tool works on
a subset of separation logic for inferring basic properties of linked list
programs. 


\section{Summarization in Hybrid Heap Model}
\label{sec:hybrid}

For heap applications that need to capture both points-to related properties
(using a store based model) and alias related properties (using a storeless
model), the heap memory is best viewed as a hybrid model combining the storeless
and the store based heap model. This model can also be summarized using various
techniques, like allocation sites, $k$-limiting, variables, and other generic
instrumentation predicates.

\subsection{Summarization Using Allocation Sites and $k$-Limiting}
\label{sec:alloc-k}

Using the hybrid model, alias graphs record may-aliases~\cite{lh88}. Let us
study the abstract memory graph for the program in Figure~\ref{fig:eg1-prog}. We assume
that variable $\X$ is initialised before statement 1
to point to an unbounded memory graph shown in Figure~\ref{heap}. The bounded
representation of this unbounded memory graph is
illustrated in Figure~\ref{lh88} using this technique. This method labels each
node with an access path reaching the node.  If there is more than one access
path reaching a node, then this method arbitrarily chooses any one of the paths
as a label for the node. For example, access paths $\X.\g$ and $\W.\g$ reach the
same node; this node is arbitrarily labelled as $\X.\g$. It can be seen in the
summarized graph in Figure~\ref{lh88} that nodes reachable from $\X$ via fields
$\f$ and $\g$ have been summarized using $k$-limiting; value of $k$ has been set
to 4; therefore, the last node pointed to by variable $\X$ via field $\f$ has the
label $\X.\f.\f.\f(.\f)^+$. This node has a self loop, which denotes that the
node is a summary node of unbounded locations. 

Larus and Hilfinger \cite{lh88} also proposed allocation site based summarization as a way of
naming the nodes. For this, let us study locations pointed to by $\Z$ and $\W$ for
the program in Figure~\ref{fig:eg1-prog}. Memory locations $\Z (.\f)^*$ (or
$\W(.\f)^*$) are allocated at program statement 7.  Figure~\ref{lh88} shows that
these nodes are summarized using allocation sites. A self loop around node,
marked with {\em Site} 7, denotes unbounded dereferences of field $\f$. However,
this summarization spuriously stores the alias relationship $\langle
\X.\f.\f.\g,\W.\f.\f.\g \rangle$.

To handle this imprecision in summarization using allocation sites, Larus and Hilfinger \cite{lh88}
distinguish nodes allocated at the
same site by labeling each newly allocated node with an aggregate of arguments
passed to the allocation function ($\cons$ in Lisp). This hybrid
approach of labeling allocation sites with access paths (arguments of the
allocation function) improves the precision of the graphs.  In order to limit
the abstract graph to a finite size, summary nodes are created using the concept
of $s$-$l$ limiting in which no node has more than $s$ outgoing edges (other
than the nodes representing the bottom element), and no node has a label longer
than $l$.  


\newcommand{\Gallocone}
{{
{\psset{unit=.25mm}
\psset{linewidth=.2mm,arrowsize=4pt,arrowinset=0}
		\begin{pspicture}(-3,-3.5)(38,0)
		\psrelpoint{origin}{q}{10}{-2}
		\rput(\x{q},\y{q}){\rnode{q}{\psovalbox[fillstyle=solid,
				fillcolor=white,framesep=4]{${\X}$}}}
		\psrelpoint{origin}{j}{-23}{-2}
		\rput(\x{j},\y{j}){\rnode{j}{$\X$}}
		\ncline{->}{j}{q}
		\psrelpoint{q}{s}{43}{0}
		\rput(\x{s},\y{s}){\rnode{s}{\psovalbox[fillstyle=solid,
				fillcolor=white,framesep=2.5]{${\X}.\f$}}}
		\ncline{->}{q}{s}
		\naput{$\f$}
		\psrelpoint{s}{r}{59}{0}
		\rput(\x{r},\y{r}){\rnode{r}{\psovalbox[fillstyle=solid,
				fillcolor=white,framesep=3]{${\X}.\f.\f$}}}
		\ncline{->}{s}{r}
		\naput{$\f$}
		\psrelpoint{r}{t}{80}{0}
		\rput(\x{t},\y{t}){\rnode{t}{\psovalbox[fillstyle=solid,
				fillcolor=white,framesep=3]{${\X}.\f.\f.\f$}}}
		\ncline{->}{r}{t}
		\naput{$\f$}
		\psrelpoint{t}{u}{113}{0}
		\rput(\x{u},\y{u}){\rnode{u}{\psovalbox[fillstyle=solid,
				fillcolor=white,framesep=-1]{${\X}.\f.\f.\f(.\f)^+$}}}
		\ncline{->}{t}{u}
		\naput{$\f$}
		\nccurve[angleA=45,angleB=135,nodesep=0,ncurv=2]{->}{u}{u}
		\nbput{$\f$}

		\psrelpoint{q}{v}{0}{-50}
		\rput(\x{v},\y{v}){\rnode{v}{\psovalbox[fillstyle=solid,
				fillcolor=white,framesep=2.5]{${\X}.\g$}}}
		\ncline{->}{q}{v}
		\nbput{$\g$}
		\psrelpoint{r}{w}{0}{-50}
		\rput(\x{w},\y{w}){\rnode{w}{\psovalbox[fillstyle=solid,
				fillcolor=white,framesep=1]{${\X}.\f.\f.\g$}}}
		\ncline{->}{r}{w}
		\nbput{$\g$}
		\psrelpoint{u}{x}{0}{-50}
		\rput(\x{x},\y{x}){\rnode{x}{\psovalbox[fillstyle=solid,
				fillcolor=white,framesep=-1]{${\X}.\f.\f(.\f)^+.\g$}}}
		\ncline{->}{u}{x}
		\nbput{$\g$}

		\psrelpoint{v}{y}{0}{-50}
		\rput(\x{y},\y{y}){\rnode{y}{\psovalbox[fillstyle=solid,
				fillcolor=white,framesep=4]{${\W}$}}}
		\ncline{->}{y}{v}
		\naput{$\g$}
		\psrelpoint{w}{z}{0}{-50}
		\rput(\x{z},\y{z}){\rnode{z}{\psovalbox[fillstyle=solid,
				fillcolor=white,framesep=1.5]{{\em Site }7}}}
		\ncline{->}{z}{w}
		\naput{$\g$}
		\ncline[nodesep=-3]{->}{z}{x}
		\nbput{$\g$}
		\ncline{->}{y}{z}
		\nbput{$\f$}
		\nccurve[angleA=-45,angleB=-135,nodesep=-1,ncurv=3]{->}{z}{z}
		\naput{$\f$}

		\psrelpoint{y}{p}{-35}{0}
		\rput(\x{p},\y{p}){\rnode{p}{{$\W$}}}
		\ncline{->}{p}{y}

		\end{pspicture}
		}
}}

\newcommand{\Galloctwo}
{\scalebox{.8}{
\large
{\psset{unit=.9mm}
\psset{linewidth=.3mm,arrowsize=5pt,arrowinset=0}
		\begin{pspicture}(-3,-3.5)(38,0)
		\psrelpoint{origin}{q}{10}{-2}
		\rput(\x{q},\y{q}){\rnode{q}{\psovalbox[fillstyle=solid,
				fillcolor=white,framesep=2.5]{${\X}$}}}
		\psrelpoint{origin}{j}{-3}{-2}
		\rput(\x{j},\y{j}){\rnode{j}{$\X$}}
		\ncline[linewidth=.4,arrowsize=7pt,arrowinset=0]{->}{j}{q}
		\psrelpoint{q}{s}{19}{0}
		\rput(\x{s},\y{s}){\rnode{s}{\psovalbox[fillstyle=solid,
				fillcolor=white,framesep=1.5]{${\X}.\f$}}}
		\ncline{->}{q}{s}
		\naput{$\f$}
		\psrelpoint{s}{r}{24}{0}
		\rput(\x{r},\y{r}){\rnode{r}{\psovalbox[fillstyle=solid,
				fillcolor=white,framesep=2.5]{${\X}.\f.\f$}}}
		\ncline{->}{s}{r}
		\naput{$\f$}
		\psrelpoint{r}{t}{30}{0}
		\rput(\x{t},\y{t}){\rnode{t}{\psovalbox[fillstyle=solid,
				fillcolor=white,framesep=2.5]{${\X}.\f.\f.\f$}}}
		\ncline{->}{r}{t}
		\naput{$\f$}
		\psrelpoint{t}{u}{37}{0}
		\rput(\x{u},\y{u}){\rnode{u}{\psovalbox[fillstyle=solid,
				fillcolor=white,framesep=2.5]{${\X}.\f.\f.\f(.\f)^+$}}}
		\ncline{->}{t}{u}
		\naput{$\f$}
		\nccurve[angleA=45,angleB=135,nodesep=0,ncurv=2]{->}{u}{u}
		\nbput{$\f$}

		\psrelpoint{q}{v}{0}{-20}
		\rput(\x{v},\y{v}){\rnode{v}{\psovalbox[fillstyle=solid,
				fillcolor=white,framesep=2.5]{${\X}.\g$}}}
		\ncline{->}{q}{v}
		\naput{$\g$}
		\psrelpoint{r}{w}{0}{-20}
		\rput(\x{w},\y{w}){\rnode{w}{\psovalbox[fillstyle=solid,
				fillcolor=white,framesep=2.5]{${\X}.\f.\f.\g$}}}
		\ncline{->}{r}{w}
		\naput{$\g$}
		\psrelpoint{u}{x}{0}{-20}
		\rput(\x{x},\y{x}){\rnode{x}{\psovalbox[fillstyle=solid,
				fillcolor=white,framesep=2]{${\X}.\f.\f.\f(.\f)^+.\g$}}}
		\ncline{->}{u}{x}
		\naput{$\g$}

		\psrelpoint{v}{y}{0}{-20}
		\rput(\x{y},\y{y}){\rnode{y}{\psovalbox[fillstyle=solid,
				fillcolor=white,framesep=2.5]{${\W}$}}}
		\ncline{->}{y}{v}
		\naput{$\g$}
		\psrelpoint{w}{z}{0}{-20}
		\rput(\x{z},\y{z}){\rnode{z}{\psovalbox[fillstyle=solid,
				fillcolor=white,framesep=2]{$\langle {\X}.\f.\f.\g, * \rangle$}}}
		\ncline{->}{z}{w}
		\naput{$\g$}
		\ncline{->}{y}{z}
		\naput{$\f$}
		\psrelpoint{x}{o}{0}{-20}
		\rput(\x{o},\y{o}){\rnode{o}{\psovalbox[fillstyle=solid,
				fillcolor=white,framesep=1.5]{$\langle {\X}.\f.\f(.\f)^+.\g, * \rangle$}}}
		\ncline{->}{o}{x}
		\naput{$\g$}
		\ncline{->}{z}{o}
		\naput{$\f$}
		\nccurve[angleA=-45,angleB=-135,nodesep=0,ncurv=1.5]{->}{o}{o}
		\naput{$\f$}

		\psrelpoint{y}{p}{-15}{0}
		\rput(\x{p},\y{p}){\rnode{p}{{$\W$}}}
		\ncline[linewidth=.4,arrowsize=7pt,arrowinset=0]{->}{p}{y}

		\end{pspicture}
		}
}}

\newcommand{\Gk}
{{
{\psset{unit=.25mm}
\psset{linewidth=.2mm,arrowsize=4pt,arrowinset=0}
		\begin{pspicture}(-3,-3.5)(38,0)
		\psrelpoint{origin}{q}{10}{-2}
		\rput(\x{q},\y{q}){\rnode{q}{\pscirclebox[fillstyle=solid,
				fillcolor=white,framesep=2.5]{$\OO1$}}}
		\psrelpoint{origin}{j}{-23}{-2}
		\rput(\x{j},\y{j}){\rnode{j}{$\X$}}
		\ncline{->}{j}{q}
		\psrelpoint{q}{s}{80}{0}
		\rput(\x{s},\y{s}){\rnode{s}{\pscirclebox[fillstyle=solid,
				fillcolor=white,framesep=2.5]{$\OO2$}}}
		\ncline{->}{q}{s}
		\naput{$\f$}
		\psrelpoint{s}{r}{80}{0}
		\rput(\x{r},\y{r}){\rnode{r}{\pscirclebox[fillstyle=solid,
				fillcolor=white,framesep=2.5]{$\OO3$}}}
		\ncline{->}{s}{r}
		\naput{$\f$}
		\psrelpoint{r}{t}{80}{-30}
		\rput(\x{t},\y{t}){\rnode{t}{\pscirclebox[fillstyle=solid,
				fillcolor=white,framesep=2.5]{$\OO4$}}}
		\ncline{->}{r}{t}
		\nbput{$\f$}
		\psrelpoint{t}{u}{80}{30}
		\rput(\x{u},\y{u}){\rnode{u}{\pscirclebox[fillstyle=solid,
				fillcolor=white,framesep=2.5]{$\OO5$}}}
		\ncline{->}{t}{u}
		\nbput{$\f$}
		\nccurve[angleA=45,angleB=-45,nodesep=-3,ncurv=2]{->}{u}{u}
		\naput{$\f$}

		\psrelpoint{q}{v}{0}{-50}
		\rput(\x{v},\y{v}){\rnode{v}{\pscirclebox[fillstyle=solid,
				fillcolor=white,framesep=2.5]{$\OO9$}}}
		\ncline{->}{q}{v}
		\nbput{$\g$}
		\psrelpoint{r}{w}{0}{-50}
		\rput(\x{w},\y{w}){\rnode{w}{\pscirclebox[fillstyle=solid,
				fillcolor=white,framesep=0]{$\OO10$}}}
		\ncline{->}{r}{w}
		\nbput{$\g$}
		\psrelpoint{u}{x}{0}{-50}
		\rput(\x{x},\y{x}){\rnode{x}{\pscirclebox[fillstyle=solid,
				fillcolor=white,framesep=0]{$\OO11$}}}
		\ncline{->}{u}{x}
		\nbput{$\g$}

		\psrelpoint{v}{y}{0}{-50}
		\rput(\x{y},\y{y}){\rnode{y}{\pscirclebox[fillstyle=solid,
				fillcolor=white,framesep=2.5]{$\OO6$}}}
		\ncline{->}{y}{v}
		\naput{$\g$}
		\psrelpoint{w}{m}{0}{-50}
		\rput(\x{m},\y{m}){\rnode{m}{\pscirclebox[fillstyle=solid,
				fillcolor=white,framesep=2.5]{$\OO7$}}}
		\ncline{->}{y}{m}
		\nbput{$\f$}
		\ncline{->}{m}{w}
		\naput{$\g$}
		\psrelpoint{x}{z}{0}{-50}
		\rput(\x{z},\y{z}){\rnode{z}{\pscirclebox[fillstyle=solid,
				fillcolor=white,framesep=2.5]{$\OO8$}}}
		\ncline{->}{z}{x}
		\naput{$\g$}
		\ncline{->}{m}{z}
		\nbput{$\f$}
		\nccurve[angleA=45,angleB=-45,nodesep=-3,ncurv=2]{->}{z}{z}
		\naput{$\f$}

		\psrelpoint{y}{p}{-35}{0}
		\rput(\x{p},\y{p}){\rnode{p}{{$\W$}}}
		\ncline{->}{p}{y}
		\psrelpoint{r}{n1}{0}{35}
		\rput(\x{n1},\y{n1}){\rnode{n1}{{$\Y$}}}
		\ncline{->}{n1}{r}
		\psrelpoint{u}{n2}{0}{35}
		\rput(\x{n2},\y{n2}){\rnode{n2}{{$\Y$}}}
		\ncline{->}{n2}{u}
		\psrelpoint{m}{o1}{0}{-35}
		\rput(\x{o1},\y{o1}){\rnode{o1}{{$\Z$}}}
		\ncline{->}{o1}{m}
		\psrelpoint{z}{o2}{0}{-35}
		\rput(\x{o2},\y{o2}){\rnode{o2}{{$\Z$}}}
		\ncline{->}{o2}{z}

		\psrelpoint{r}{t2}{80}{30}
		\rput(\x{t2},\y{t2}){\rnode{t2}{\pscirclebox[fillstyle=solid,
				fillcolor=white,framesep=0]{$\OO12$}}}
		\ncline{->}{r}{t2}
		\naput{$\f$}
		\ncline{->}{u}{t2}
		\nbput{$\f$}
		\psrelpoint{t2}{o3}{-15}{35}
		\rput(\x{o3},\y{o3}){\rnode{o3}{{$\U$}}}
		\ncline[nodesepA=3]{->}{o3}{t2}
		\psrelpoint{t2}{o4}{15}{35}
		\rput(\x{o4},\y{o4}){\rnode{o4}{{$\V$}}}
		\ncline[nodesepA=3]{->}{o4}{t2}
		\psrelpoint{t}{o5}{0}{35}
		\rput(\x{o5},\y{o5}){\rnode{o5}{{$\V$}}}
		\ncline[nodesepA=3]{->}{o5}{t}

		\end{pspicture}
		}
}}


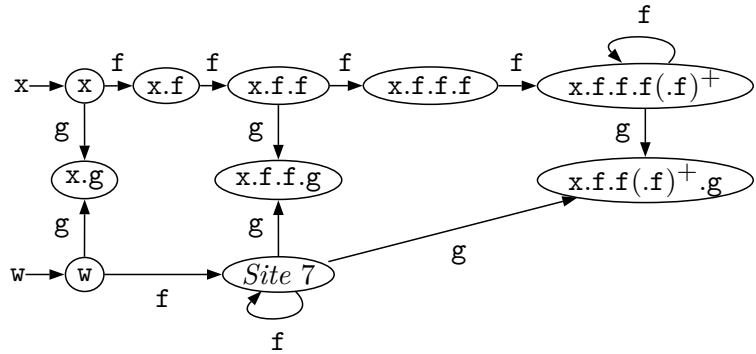
\begin{figure}[t!]
\begin{center}
  {
    \psset{unit=.25mm}
    \begin{pspicture}(-10,-85)(390,115)

    \putnode{a}{origin}{40}{60}{\Gallocone}
    \end{pspicture}
   }
\end{center}
\caption{Summarization using allocation sites and $k$-limiting ($k=4$) on a
hybrid model \protect \cite{lh88} at $\Outn_8$ for the program in
Figure~\ref{fig:eg1-prog}.  Pointer variables $\Y$ and $\Z$ are not shown for
simplicity.}
\label{lh88}
\figrule
\end{figure}


\begin{figure}[t!]
\begin{subfigure}[c]{1\textwidth}
\begin{center}
  {
    \psset{unit=.25mm}
    \begin{pspicture}(-10,-85)(385,145)

    \putnode{a}{origin}{40}{60}{\Gk}
    \end{pspicture}
   }
\end{center}
\caption{\parbox[t]{.9\textwidth}{Illustrating imprecision in store based model.
$k$-limiting ($k=4$) summarized graph at $\Outn_{11}$.  Corresponding to
statements 9, 10, and 11, $\U$ points to $\OO12$, $\Y.\f$ points to both $\OO4$
and $\OO12$; therefore, $\V$ also points to both $\OO4$ and $\OO12$.  Here $\V$ is
imprecisely aliased to $\X.\f.\f.\f$.}}
\label{dd12-0}
\end{subfigure}

\begin{subfigure}[t]{.5\textwidth}
\begin{center}
  {
    \psset{unit=.25mm}

	\shadowbox{
	\begin{tabular}{@{}l|l@{}}
	$\X \rightarrow \{\OO1\}$ 
	& $\W \rightarrow \{\OO6\}$ \\
	$\X \myrightarrow \f \rightarrow \{\OO2\}$ 
	& $\W \myrightarrow \f \rightarrow \{\OO7\}$ \\
	$\X \myrightarrow \f \myrightarrow \f \rightarrow \{\OO3\}$ 
	& $\W \myrightarrow \f \myrightarrow \f \rightarrow \{\OO8\}$ \\ 
	$\X \myrightarrow \f \myrightarrow \f \myrightarrow \f \rightarrow \{\OO4\}$ 
	& $\dots$ \\ 
	$\X \myrightarrow \f \myrightarrow \f \myrightarrow \f \myrightarrow \f \rightarrow \{\OO5\}$ 
	& $\dots$ \\
	& \\ 
	$\X \myrightarrow \g \rightarrow \{\OO9\}$ 
	& $\W \myrightarrow \g \rightarrow \{\OO9\}$ \\
	$\X \myrightarrow \f \myrightarrow \f \myrightarrow \g \rightarrow \{\OO10\}$ 
	& $\W \myrightarrow \f \myrightarrow \g \rightarrow \{\OO10\}$ \\ 
	$\dots$
	& $\dots$ \\ \hline
	$\Y \rightarrow \{\OO3, \OO5\}$ 
	& $\Z \rightarrow \{\OO7, \OO8\}$ \\
	$\dots$
	& $\dots$ \\
	\end{tabular}
	}
   }
\end{center}
\caption{\parbox[t]{.9\textwidth}{$k$-limited ($k=4$) points-to information at
{\Inn}$_9$~\cite{dd12}. $\X \myrightarrow \g$ and $\W \myrightarrow \g$ are
aliased.}}
\label{dd12-1}
\end{subfigure}
~
\begin{subfigure}[t]{.5\textwidth}
\begin{center}
  {
    \psset{unit=.25mm}

	\shadowbox{
	\begin{tabular}{@{}l|l@{}}
	$\X \rightarrow \{\OO1\}$ 
	& $\W \rightarrow \{\OO6\}$ \\
	$\X \myrightarrow \f \rightarrow \{\OO2\}$ 
	& $\W \myrightarrow \f \rightarrow \{\OO7\}$ \\
	$\X \myrightarrow \f \myrightarrow \f \rightarrow \{\OO3\}$ 
	& $\W \myrightarrow \f \myrightarrow \f \rightarrow \{\OO8\}$ \\ 
	$\X \myrightarrow \f \myrightarrow \f \myrightarrow \f \rightarrow \{\OO4, {\bf \OO12}\}$ 
	& $\dots$ \\ 
	$\X \myrightarrow \f \myrightarrow \f \myrightarrow \f \myrightarrow \f \rightarrow \{\OO5\}$ 
	& $\dots$ \\
	& \\ 
	$\X \myrightarrow \g \rightarrow \{\OO9\}$ 
	& $\W \myrightarrow \g \rightarrow \{\OO9\}$ \\
	$\X \myrightarrow \f \myrightarrow \f \myrightarrow \g \rightarrow \{\OO10\}$ 
	& $\W \myrightarrow \f \myrightarrow \g \rightarrow \{\OO10\}$ \\ 
	$\dots$
	& $\dots$ \\ \hline
	$\Y \rightarrow \{\OO3, \OO5\}$ 
	& $\Z \rightarrow \{\OO7, \OO8\}$ \\
	$\Y \myrightarrow \f \rightarrow \{{\bf \OO12}\}$ 
	& $\dots$ \\ 
	$\dots$
	& $\dots$ \\ \hline
	$\U \rightarrow \{{\bf \OO12}\}$ 
	& $\V \rightarrow \{{\bf \OO12}\}$ \\ 
	\end{tabular}
	}
   }
\end{center}
\caption{\parbox[t]{.9\textwidth}{$k$-limited ($k=4$) points-to information at
{\Outn}$_{11}$~\cite{dd12}. Variable $\V$ precisely points to only $\OO12$
(pointed to by $\U$) and is not aliased to $\X.\f.\f.\f$.}}
\label{dd12-2}
\end{subfigure}

\caption{Summarization using $k$-limiting on a hybrid model \protect \cite{dd12}
for the program in Figure~\ref{fig:eg1-prog} is shown in Figures~\ref{dd12-1}
and \ref{dd12-2}. Here $\OO\langle n \rangle$ represents an object name and the
symbol $\rightarrow$ denotes points-to relation. For easy visualization, we have
shown a summarization on a store based model at $\Outn_{11}$ in
Figure~\ref{dd12-0}.}
\label{dd12}
\figrule
\end{figure}


\subsection{$k$-Limiting Summarization}
\label{sec:k}

De and D'Souza \cite{dd12} highlight an imprecision in saving pointer information as graphs.
We illustrate this imprecision using Figure~\ref{dd12-0} for
statements 9, 10, and 11 of our running
program in Figure~\ref{fig:eg1-prog}. 
The problem is caused by the fact
that a summarized object node may represent multiple concrete objects;
therefore, the analysis cannot perform a strong update on such objects. At
$\Inn_9$ of the program, $\Y$ is aliased to the summary node $\X \myrightarrow
\f \myrightarrow \f \myrightarrow \f (\myrightarrow \f)^+$.  Therefore, strong
update cannot be performed in statement 10 i.e.  the pointer of $\Y\myrightarrow
\f$ cannot be removed.  Hence, at $\Outn_{11}$, $\V$ will point to all the
objects previously pointed to by $\Y.\f$ as well as the new location pointed to
by $\U$. Observe that the former is imprecise.

De and D'Souza \cite{dd12} believe that this imprecision is caused by storing points-to
information as graphs. Therefore, instead of using graphs, they use access
paths. Their technique maps $k$-limited access paths (storeless model) to sets
of summarized objects (store based model) (represented as $\OO\langle n \rangle$
in Figure~\ref{dd12-1} and Figure~\ref{dd12-2}). For example, $\X \rightarrow
\{o1\}$ means that the access path $\X$ points to (is mapped to) the object
named $\OO1$. Since the access paths are precise up to $k$ length, like any
$k$-limiting abstraction, it can also perform strong updates up to $k$ length. 

In Figure~\ref{dd12-1} at $\Inn_9$, $\Y$ points to a summarized object $\{\OO3,
\OO5\}$ (pointed to by $\X.\f.\f$ and $\X.\f.\f.\f(.\f)^+$, respectively), as
shown in Figure~\ref{dd12-0}. Program statement 10 updates the pointer
information of $\Y.\f$. Therefore, if $\U$ points to object $\OO12$, then it is
sound to say that $\Y.\f$ will point only to object $\OO12$ at $\Outn_{10}$.
However, it is not sound to say that $\X.\f.\f.\f$ (alias of $\Y.\f$) will point
only to object $\OO12$ since $\Y$ points to multiple access paths, viz.
$\X.\f.\f$ and $\X.\f.\f.\f(.\f)^+$.  Therefore, in Figure~\ref{dd12-2}, at
$\Outn_{10}$, the method strongly updates $\Y.\f$ to $\{\OO12\}$ (pointed to by
$\U$), even though $\Y$ points to multiple objects ($\OO3$ and $\OO5$) at
$\Inn_{10}$. Also, for sound results, $\X.\f.\f.\f$ is not strongly updated, and
$\X.\f.\f.\f$ points to $\OO12$ in addition to the previously pointed object
$\OO4$. Since $\Y.\f$ points only to $\OO12$, at $\Outn_{10}$, access path $\V$
also precisely points only to the new object $\{\OO12\}$ (pointed to by $\U$)
at $\Outn_{11}$.

\subsection{Summarization Using Variables and Other Generic Instrumentation
Predicates}
\label{sec:hybrid-other}

We describe below some application specific predicates that have been used in a
hybrid model.

\begin{itemize}
\item 
In order to remove unreachable parts of the heap across functions in
interprocedural analysis, {\em cutpoints} are marked on the heap~\cite{rbrsr05}.
Cutpoints are objects which separate the local heap of the invoked function
from the rest of the heap. 3-valued logic shape analysis (classified under the
store based model) is used for summarization~\cite{srw99}.  Each cutpoint is
identified by an access path (a feature of a storeless model) which is not
relevant to the function being called. When the function returns, the access
path of the cutpoint object is used to update the caller's local heap with the
effect of the call.  Therefore, irrelevant parts of abstract states that will
not be used during the analysis are removed by modeling the heap using both
storeless and store based representations.

For example, an acyclic list pointed to by $\X$ is passed to the {\em
reverse}$()$ function, which reverses the list performing strong updates. Let us
say, before the function call, $\Y \myrightarrow \g \myrightarrow \g$ and $\X
\myrightarrow \f$ are aliased and $\Y$ is not in scope of function {\em
reverse}$()$. On return of the function, we should be able to derive that $\Y
\myrightarrow \g \myrightarrow \g \myrightarrow \f$ and $\X$ are aliased. To
capture this kind of a relationship, effect of the function on {\em cutpoints}
is tracked. In this example, the second node of list $\X$ is a cutpoint and in
the function {\em reverse}$()$ can be identified with a new alias relationship
between access paths as $\langle\C, \X \myrightarrow \f \rangle$, where $\C$ is
the access path used to label the second node (cutpoint) in the list. On return
of the function {\em reverse}$()$, we will derive $\langle \X, \C \myrightarrow
\f \rangle$ as the alias relationship. Thus, we will be able to restore the
alias relationship between $\X$ and $\Y$ as $\langle \X, \Y \myrightarrow \g
\myrightarrow \g \myrightarrow \f \rangle$ in the calling function. 

\item
Connection analysis (similar to access paths used in a storeless model) along
with store based points-to analysis has been used as an abstraction~\cite{gh98}.
This method first resolves all pointer relationships on the stack using a store
based points-to analysis, which abstracts all heap locations as a single
symbolic location called heap. All pointers reported to be pointing to heap are
then further analysed via a storeless heap analysis, called connection analysis,
and shape analysis. 

\end{itemize}

\section{Design Choices in Heap Abstractions}
\label{sec:perspective}

Given a confounding number of possibilities of combining heap models and
summarization techniques for heap abstractions, it is natural to ask the
question ``which heap abstraction should I use for my analysis?'' This question
is one of the hardest questions to answer because there is no one right answer and the
final choice would depend on a wide range of interdependent, and often
conflicting, requirements of varying importance.

This section attempts to provide some guidelines based on 
\begin{itemize}
\item the properties of heap abstractions,
\item the properties of underlying analyses, and
\item the properties of programs being analysed.
\end{itemize}

The properties of heap abstractions are dominated by the properties of
summarization techniques with the properties of heap models playing a relatively
minor role. Among the properties of summarization, we explore the tradeoffs
between precision and efficiency on the one hand and expressiveness and
automatability on the other. The properties of analyses include flow- and
context-sensitivity, bottom up vs. top down traversals over call graphs, partial
soundness, and demand driven nature.


These guidelines are admittedly incomplete
and somewhat abstract.  
Because of the very nature of heap abstractions and a large
variety of uses they can be put to,  these guidelines may need deeper
examination and may not be applicable directly.

\subsection{Properties of Heap Models}
We believe that in general, 
\begin{itemize}
\item client analyses that explore points-to related properties are easier to
model as store based~\cite{rbrsr05,dhy06}, whereas 
\item analyses that explore alias related properties are easier to model as
storeless~\cite{bil03,rbrsr05,dhy06}.
\end{itemize}
This is because in points-to related properties, heap locations and addresses
contained in locations are important. Store based models are more natural in
such situations because they explicate all locations. On the other hand, alias
related properties can leave the locations implicit which is the case in a
storeless model. The metrics like precision and efficiency are generally not
decided  by the choice of heap model but by the summarization technique used.

\subsection{Properties of Heap Summarization Techniques}
In this section, we compare the summarization techniques with respect to
efficiency, precision, expressiveness, and automatability.

\subsubsection{Precision vs. Efficiency}
In general, if a client analysis requires computing complex heap properties,
like shape of the heap memory, then summarization techniques using variables,
generic instrumentation predicates, and higher-order logics are more precise. On
the other hand for computing simpler heap properties, like finding the pointer
expressions that reach a particular heap location, a client can choose more
efficient summarization techniques like those based on $k$-limiting and
allocation sites.

We describe the other considerations in precision-efficiency tradeoff for
specific summarization techniques.

\begin{itemize}

\item {\em $k$-limiting}.
This technique does not yield very precise results for programs that manipulate
heap locations that are $k$ indirections from some pointer variable of the
program as illustrated in Figures~\ref{fig:k-limiting-storebased} and
\ref{fig:k-limiting-storeless}.  $k$-limiting merges the access paths that are
longer than a fixed constant $k$. Thus the tail of even a non-circular linked
list will be (conservatively) represented as a possibly cyclic data structure.
Due to the summarization of heap locations that are beyond $k$ indirections from
pointer variables, this technique lacks strong update operations on these heap
locations. Consequently, Sagiv et al. \cite{srw96} observe, ``$k$-limiting approach cannot
determine that either list-ness or circular list-ness is preserved by a program
that inserts an element into a list." However, $k$-limiting gives reasonably
precise results if the user program being analysed does not need strong updates.

The efficiency of the analysis is heavily dependent on the value of $k$; larger
values improve the precision but may slow down the analysis
significantly~\cite{aw93}. The analysis may be extremely expensive because as
observed by Sagiv et al. \cite{srw96} ``the number of possible shape graphs is doubly
exponential in $k$." This is because heap locations beyond $k$ indirections
from some pointer variable have to be (conservatively) assumed to be aliased to
every other heap location.  Hence, $k$-limiting is practically feasible only
for small values such as $k \leq 2$~\cite{s06}.  The price to pay is reduced
precision as shown by Chase et al. \cite{cwz90}.  In general it is difficult for a client
analysis to know the best value of $k$ a-priori and it should be guided by
empirical observations on representative programs. 

\item {\em Allocation sites}.
This technique may be imprecise when memory allocation is concentrated within a
small number of user written procedures. In such situations, nodes allocated at
the same allocation site but called from different contexts are merged even
though they may have different properties.  Figure~\ref{lh88} contains an
example of imprecision using allocation sites.  Chase et al. \cite{cwz90} state that
``allocation site based method cannot determine that list-ness is preserved for
either the insert program or the reverse program on a list" because of merging
of nodes.

However, regarding efficiency, Sagiv et al. \cite{srw98} note, ``the techniques based on
allocation sites are more efficient than $k$-limiting summarizations, both from
a theoretical perspective~\cite{cwz90} and from an implementation
perspective~\cite{aw93}." The size of an allocation site based graph is bounded
by the number of allocation sites in the program. Therefore, majority of client
analyses are likely to find this technique space efficient on most practical
programs.

\item {\em Patterns}.
Identifying precise repeating patterns is undecidable in the most general case
because a repeated advance into the heap may arise from an arbitrarily long
cycle of field dereferences~\cite{ma12}.  Therefore, generally the focus remains
on detecting only consecutive repetitions of the same type of field accesses
which may be imprecise.  Also, it seems difficult for an analysis to determine
if an identified repetition will occur an unbounded number of times or only a
bounded number of times.  This approach has been found to be more efficient than
TVLA based shape analysis techniques for discovering liveness of heap
data~\cite{ksk07}.


\item {\em Variables}.  For complex shape graphs, summarization using variables
may be more precise than $k$-limiting. Chase et al. \cite{cwz90} observe that two nodes need
not have similar properties just because they occur $k$ indirections away from
the root variable in an access path. On the other hand, two nodes which are
pointed to by the same set of variables are more likely to have similar properties.
Further, summarization using variables can perform strong nullification in a
larger number of cases; therefore, it may be more precise. However, there are
situations where summarization using variables can also be imprecise: since it
merges nodes not pointed to by any root variable, sometimes nodes are abstracted
imprecisely as  illustrated in Figure~\ref{fig:variables-storebased}.  Contrast
this with the precise summarization of Figure~\ref{fig:alloc-storebased}.

In general this technique has been found to be inefficient. Since each shape
graph node is labelled with a set of root variables in this technique,
Sagiv et al. \cite{srw96} state, ``the number of shape nodes is bounded by
$2^{\mathbb{\mid\text{\em Var}\mid}}$, where {\em Var} is the number of root
pointer variables in the program." They further note, ``unfortunately for some
pathological programs the number of shape nodes can actually grow to be this
large, although it is unlikely to arise in practice."

\item {\em Generic instrumentation predicates}. 
Both the precision and efficiency of a client analysis depends on the chosen
predicate. By identifying one or more suitable predicates, a client analysis can
strike a balance between precision and efficiency. 

The implementation of generic instrumentation predicates using TVLA~\cite{srw99}
has potentially exponential runtime in the number of predicates. Therefore, it
is not suitable for large programs~\cite{cdoy11}.

\item {\em Higher-order logics}.  These techniques have the capability of
computing complex heap properties. With the use of program annotations in the
form of assertions and loop invariants, they can compute surprisingly detailed
heap properties~\cite{jjsk97}.  Unlike TVLA, they can also produce counter
examples for erroneous programs~\cite{as01}.  However, these techniques are
generally used to verify restricted data structures~\cite{bil04}, without
considering the full behaviour of the program and have to be made less detailed
for large programs~\cite{as01} since they are highly inefficient.  An analysis
needs to use simpler and less precise logics in order to improve scalability.
For example, Distefano et al. \cite{dhy06} use a subset of separation logic as the domain of
their analysis; the domain is less powerful because it does not allow nesting of
$*$ and $\wedge$ operators.

These techniques may be highly inefficient as they include higher-order and
undecidable logics.  For example, quantified separation logic is
undecidable~\cite{cyh01}.  For termination, these techniques require program
annotations in the form of assertions and loop
invariants~\cite{jjsk97,as01,bil04}.  Consequently, analyses based on
higher-order logics cannot be made fully automatic. Since the effort of
annotating the program can be significant, these techniques can work
efficiently only on small programs~\cite{jjsk97}. Therefore, these are mostly
used for teaching purposes~\cite{jjsk97} in order to encourage formal reasoning
of small programs. Again, since they are inefficient, these are considered
useful to verify only safety critical applications~\cite{as01} where the effort
of annotating the program is justified by the complex properties that these
techniques can derive.  However, as compared to TVLA, these techniques are
sometimes more scalable due to the use of loop invariants; empirical
measurements show high speedup in these techniques where the use of loop
invariants is more efficient than a fixpoint computation required by
TVLA~\cite{as01}. An advantage of separation logic is its efficiency due to the
following: once the program is analysed for a part of the memory, it can
directly be used to derive properties for the extended memory~\cite{s07}.

\end{itemize}

\subsubsection{Expressiveness vs. Automatability}
Here we discuss degree of expressive power and automation offered by heap
summarization techniques using predicates (for example, $k$-limiting, allocation
sites, variables, pattern, and other user-defined predicates) and those using
higher-order logics.

\begin{itemize}
\item 
{\em Predicates.} Parameterised frameworks like TVLA summarize heap data based
on any desired user-defined predicate. Therefore, they achieve good
expressiveness as per the user's requirements. However, the predefined
predicates (for example, $k$-limiting, allocation sites, variables, pattern)
lack this expressiveness.

Automation of summarization techniques using user-defined predicates in TVLA is
not difficult since TVLA allows only simple predicates.  Also, several automated
tools are already available for predefined predicates. For example,
LFCPA~\cite{kmr12} performs automatic heap analysis using allocation site based
summarization.

\item 
{\em Higher-order logics.} Unlike summarizations based on predicates,
summarizations based on higher-order logics do not need to work with a
predefined user predicate; with the use of heap specialized operators and rules,
the latter can build upon basic predicates to be able to compute complex
properties of the heap. Depending on the underlying logic, a client may find
these summarization techniques to be more powerful and easier to express.

However, summarization techniques using higher-order logics are not fully
automated and need user intervention for inference of non-trivial properties
specially if the technique is based on undecidable logics.

\end{itemize}

\subsection{Properties of Underlying Heap Analysis}

The choice of heap summarization technique is sometimes dependent on the design
dimensions of the underlying analysis that the client uses. We describe some
such dependencies.

\begin{itemize}
\item {\em Flow-sensitive analysis.} 
The precision benefits of a flow-sensitive analysis can be increased by 
\begin{itemize}[$\bullet$]
\item using TVLA whose 3-valued logic enables a more precise meet operation by
distinguishing between the {\em may\/} (i.e. along some paths), {\em must\/}
(i.e. along all paths) and {\em cannot\/} (i.e. along no path) nature of
information discovered.
\item using techniques that aid strong updates: summarization techniques based
on variables~\cite{srw96,srw99} and $k$-limiting~\cite{dd12}, and the
materialization~\cite{srw96,srw99} of summary nodes. 
\end{itemize}

\item {\em Context-sensitive analysis.} A context-sensitive analysis examines a
given procedure separately for different calling contexts. If such a procedure
contains an allocation statement, the allocation site based summarization should
be able to distinguish between the nodes representing different calling
contexts.  This can be achieved by heap cloning~\cite{xr08}.  In the absence of
replication of allocation site based nodes for different calling contexts, the
precision of analysis reduces significantly~\cite{nkh04}. 

\item {\em Bottom-up analysis.} A bottom-up interprocedural analysis traverses
the call graph bottom up by processing callees before callers.  It constructs a
summary of the callee procedures that may access data structures whose
allocation is done in the callers. Thus the allocation site information may not
be available in  a callee's heap summary. Therefore, allocation site based
summarization cannot be used with bottom-up analyses; instead summarization
using patterns has been used for computing procedure
summaries~\cite{mrv11,ma12,gst14}.

\item {\em Partially sound analysis} and {\em demand driven analysis.} Soundness
of an analysis requires covering behaviours of all (possibly an infinite number
of) execution paths. In many situations such as debugging, useful information
may be obtained by covering the behaviour of only some execution paths. Such
partially sound analyses\footnote{Not to be confused with ``soundy'' analyses
which refer to partially unsound analyses that ignore some well identified hard
to analyse constructs~\cite{lsslacgkmv15}.} are often demand driven.  The other
flavour of demand driven analyses (such as assertion verification) may need to
cover all execution paths reaching a particular program point but not all
program points.  In either case, these analyses examine a smaller part of the
input program and hence may be able to afford expensive summarization
techniques.  Here $k$-limiting and higher-order logics based summarization
techniques permit the client to choose a larger value of $k$ and a more complex
logic, respectively thereby improving precision.  Likewise, parametric
frameworks like TVLA can also be used with more complex predicates. Observe
that, allocation site and variable based techniques do not have any inherent
parameter for which the analysis may be improved.

\end{itemize}

\subsection{Properties of Programs} 
The suitability of a technique depends on various properties of the input
program. These are discussed below. 

\begin{itemize}
\item {\em $k$-limiting.} If the input program contains a small number of
indirections from pointer variables, $k$-limiting summarization based on a
suitable choice of empirically observed $k$ would give reasonable results.

\item {\em Allocation sites.} For input programs where allocations are made from
sites that are distributed over the program, rather than being made from a small
set of procedures, summarization using allocation sites will be able to preserve
heap properties efficiently.  

\item {\em Patterns.} For input programs containing simple repeating patterns,
summarization techniques based on patterns can produce useful summaries.  

\item {\em Variables.} In our opinion, summarizations based on variables are
precise in generally all types of programs using the heap; however they are
usually not as efficient as techniques using $k$-limiting, allocation sites, and
patterns.  

\item {\em Higher-order logics.} Techniques based on logics are inefficient and
need manual intervention. Therefore, their usefulness may be limited
on small input programs.
\end{itemize}

\section{Heap Analyses and Their Applications}
\label{app:heap-analysis-applications}

In this section, we categorize applications of heap analyses and
list common heap analyses in terms of the properties that they discover. 

\subsection{Applications of Heap Analyses}
\label{sec:app2}

We present the applications of heap analyses under the following three broad
categories:
\begin{itemize}[--]
\item {\em Program understanding.} 
Software engineering techniques based on heap analysis are used to maintain or
reverse engineer programs for understanding and debugging them.  Heap related
information like shape, size, reachability, cyclicity, and others are collected
for this purpose.  Program slicing of heap manipulating programs~\cite{k13} can
help in program understanding by extracting the relevant part of a program.

\item {\em Verification and validation.} Heap analysis is used for detecting
memory errors at compile time (for example, dereferencing null pointers,
dangling pointers, memory leaks, freeing a block of memory more than once, and
premature deallocation)~\cite{gh98,syks03,hsp05,mk11}.  Sorting programs that
use linked lists have been verified using heap analyses~\cite{lrsw00}.

\item {\em Optimization.} Modern compilers use heap analysis results to produce
code that maximizes performance.  An optimization of heap manipulating programs
is the garbage collection of accessible yet unused objects~\cite{ksk07,askm14}
which are otherwise beyond the scope of garbage collection that depends purely
on runtime information.  Transformation of sequential heap manipulating
programs for better parallel execution involves heap analysis~\cite{bdk10}.
Heap analysis also helps in performing data prefetching based on future uses
and updates on heap data structures in the program~\cite{gh98}. Data locality
of dynamically allocated data has been identified and exploited using heap
analysis by Castillo et al. \cite{ctcnaz06}.
\end{itemize}

\subsection{Heap Analyses}
\label{sec:analyses}
A compile time program analysis that needs to discover and verify properties of
heap data could perform one or more of the following analyses.

\begin{itemize}
\item {\em Shape analysis}~\cite{gh96,srw99,wsr00} also called {\em storage
analysis} discovers invariants that describe the data structures in a program
and identifies alias relationships between paths in the heap. Its applications
include program understanding and debugging~\cite{drs98}, compile time detection
of memory and logical errors, establishing shape properties, code optimizations,
and others.

\item {\em Liveness analysis} of  heap data statically identifies last uses of
objects in a program to discover reachable but unused heap locations to aid
garbage collection performed at runtime~\cite{isy88,syks03,ksk07,askm14}.

\item {\em Escape analysis} is a method for determining whether an object is
visible outside a given procedure. It is used for 
\begin{inparaenum}[(a)]
\item scalar replacement of fields,
\item removal of synchronization, and 
\item stack allocation of heap objects~\cite{km05}.
\end{inparaenum}

\item {\em Side-effect analysis} finds the heap locations that are used (read
from or written to) by a program statement. This analysis can optimize code by
eliminating redundant loads and stores~\cite{mrr02}.

\item {\em Def-use analysis} finds point pairs of statements that initialize a
heap location and then read from that location. This analysis is used to check
for the uses of undefined variables and unused variables~\cite{mrr02}.

\item {\em Heap reachability analysis} finds whether a heap object can be
reached from a pointer variable via field dereferences for detecting memory
leaks at compile time~\cite{bcs13}. 

\item {\em Call structure analysis} disambiguates virtual calls in
object-oriented languages and function pointers. Presence of heap makes this
disambiguation non-trivial. Instead of relying on a call graph constructed with
a relatively less precise points-to analysis, the program call graph can be
constructed on-the-fly with pointer analysis~\cite{wl04,pk13,scdfy13}.
Receiver objects of a method call can also be disambiguated in order to
distinguish between calling contexts using
object-sensitivity~\cite{mrr02,sbl11} and type propagation
analysis~\cite{shrvlgg00}.
\end{itemize}

\section{Engineering Approximations for Efficiency}
\label{app:engg.approx}

Given the vital importance of pointer analysis and the inherent difficulty of
performing precise pointer analysis for practical
programs~\cite{lr92,r94,h97,c03}, a large number of investigations involve a
significant amount of engineering approximations~\cite{khe13}.  A detailed
description of these is beyond the scope of this paper because its focus is on
building the basic concepts of various modeling and summarization techniques
for heap.  Here we merely list some notable efforts in engineering
approximations used in heap analysis. 

Since heap data is huge at compile time Calcagno et al. \cite{cdoy11} perform
compositional/modularized analysis, i.e. using function summaries.  Heap data
can also be restricted by propagating the part of the heap that is sufficient
for a procedure~\cite{rbrsr05,dhy06,gbc06,cdoy11}. Amount of heap data
collection can be controlled by a demand-driven analysis using client
intervention~\cite{gl03,scdfy13}. Rountev et al. \cite{rrl99} restrict the scope of program where
high precision is required.  For example, they determine program fragments where
accuracy is vital (like regions of code, pointer variables) and find ways to
make the results precise for only for those critical regions.  They have also
performed safe analysis for incomplete programs.  Limiting the analysis to live
and defined variables of the program has also helped in achieving scalability
without any loss of precision~\cite{amss06,dd12,kmr12}. An inexpensive
flow-insensitive heap analysis over an SSA form \cite{fks00} of a program seeks
a middle ground between a flow-sensitive and a flow-insensitive heap analysis.
Incremental computations~\cite{vr01} and efficient encoding of information by
using BDDs \cite{wl04} are amongst other engineering techniques employed for
efficient heap analysis.

Given a large body of work on building efficient
approximations, Michael Hind observes that
although the problem of pointer analysis is undecidable, ``fortunately many
approximations exists'' and goes on to note that ``unfortunately too many
approximations exist''~\cite{h01}.  We view this trend as unwelcome because a
large fraction of pointer analysis community seems to believe that compromising
on precision is necessary for scalability and efficiency. Amer Diwan adds, ``It
is easy to make pointer analysis that is very fast and scales to large programs.
But are the results worth anything?"~\cite{h01}.

In our opinion, a more desirable approach is to begin with a careful and
precise modeling of the desired heap properties even if it is not computable.
Then the analysis can be gradually refined into a computable version which can
further be refined to make it scalable and efficient to make it practically
viable. Tom Reps notes that ``There are some interesting precision/efficiency
trade-offs: for instance, it can be the case that a more precise pointer
analysis runs more quickly than a less precise one''~\cite{h01}. Various
implementations~\cite{lc11,sbl11,kmr12} show that this top-down approach does
not hinder efficiency. In fact increased precision in pointer information not
only causes a subsequent (dependent) analysis to produce more precise results,
it also causes the subsequent analysis to run faster~\cite{sh97}.


\section{Related Surveys}
\label{sec:related}

We list below some investigations that survey heap abstractions, either as the
main goal or as one of the important subgoals of the paper.

Hind \cite{h01}, Ryder \cite{r03}, and Smaragdakis and Balatsouras \cite{sb15} present a theoretical discussion on
some selective pointer analysis metrics like efficiency, precision, client
requirements, demand driven approaches, handling of incomplete programs, and
others.  They also discuss some chosen dimensions that influence the precision
of heap analyses like flow-sensitivity, context-sensitivity, field-sensitivity,
heap modeling, and others.
Smaragdakis and Balatsouras \cite{sb15} present some of these aspects in the form of a tutorial.
Hind \cite{h01} provide an excellent compilation of literature on pointer analysis
which are presented without describing their algorithms.

Sridharan et al. \cite{scdfy13} present a high-level survey of alias analyses that they have
found useful from their industrial experiences.
Hind and Pioli \cite{hp00} give an empirical comparison of precision and efficiency of five
pointer analysis algorithms.
Ghiya \cite{g98} provides a collection of literature on stack and heap pointer
analyses and highlights their key features.
Sagiv et al. \cite{srw07} and Nielson et al. \cite{nnh05} have a detailed chapter on shape analysis
and abstract interpretation.

There are short sections on literature surveys~\cite{r08,s12}, which categorize
a variety of heap analyses into storeless and store based models. Chakraborty \cite{s12}
points out that heap models cannot always be partitioned into storeless and
store based only; some literature use hybrid model. 

We have not come across a comprehensive survey 
which seeks a unifying theme among a plethora of heap abstractions.

\section{Conclusions}
\label{sec:conclusions}

A simplistic compile time view of heap memory consists of an unbounded number of
unnamed locations relating to each other in a seemingly arbitrary
manner. On the theoretical side, this offers deep intellectual
challenges for building suitable abstractions of heap for more sophisticated
compile time views of the heap memory. On the practical
side, the quality of the result of a heap analysis is largely decided
by the heap abstraction used. It is not surprising, therefore, that
heap abstraction is a fundamental and vastly studied component of heap
analysis. What is surprising, however, is that a quest of a unifying
theme in heap abstractions has not received adequate attention which, in
our opinion, it deserves.

This paper is an attempt to fill this void by separating the heap model
as a representation of heap memory, from a summarization technique used
for bounding it. This separation has allowed us to explore and compare
a comprehensive list of algorithms used in the literature making it
accessible to a large community of researchers. We observe that the heap
models can be classified as storeless, store based, and hybrid. The
summarization techniques use $k$-limiting, allocation sites, patterns,
variables, other generic instrumentation predicates, and higher-order
logics.

We have also studied the design choices in heap abstractions by
comparing and contrasting various techniques used in literature
with respect to client requirements like efficiency, precision,
expressiveness, automatability, dimensions of the underlying analysis,
and user program properties. We hope that these comparisons can be
helpful for a client to decide which abstraction to use for designing
a heap analysis. It is also expected to pave way for creating new
abstractions by mix-and-match of models and summarization techniques.

We observe in passing that, as program analysts, we still face the
challenge of creating summarizations that are efficient, scale to large
programs, and yield results that are precise enough to be practically
useful.


\section*{Acknowledgements}
An invigorating discussion in the Dagstuhl Seminar on Pointer
Analysis~\cite{dagstuhl-seminar} sowed the seeds of this survey paper.  We
would like to thank Amitabha Sanyal, Supratik Chakraborty, and Alan Mycroft for
their comments on this paper as also for enlightening discussions related to
heap analysis from time to time. Anders M{\o}ller helped us in improving the
description of Pointer Assertion Logic Engine. Rohan Padhye, Alefiya Lightwala,
and Prakash Agrawal gave valuable feedback on the paper, helped in rewording
some text, and pointed out some errors in the examples. We would also like to
thank the anonymous reviewers for their rigorous and extensive reviews and
thought-provoking questions and suggestions.  

Vini Kanvar is partially supported by TCS Fellowship.

\bibliographystyle{plain}
\bibliography{heap-abstractions}

\appendix

\section{Heap and Stack Memory in C/C++ and Java}
\label{app:cjava}

In this section, we briefly compare the programming constructs related to
pointer variables in C/C++ and Java programs. 

{\em Referencing variables on stack and heap.}
In C/C++, both stack and heap allow pointer variables. Java does not allow stack
directed pointers. C/C++ allows pointers to variables on the stack through the
use of addressof operator $\&$; Java does not have this operator.  Both C/C++
and Java allow pointers/references to objects on the heap using $\malloc$
function (in C/C++) and $\new$ operator (in C++ and Java).

{\em Dereferencing pointers.}
Every variable on the stack, whether it contains a reference or a value, always
has a name because all the objects allocated on the stack have compile time
names associated with them. Heap allocated data items do not possess names and
are all anonymous. The only way to access heap items is using pointer
dereferences. C/C++ has explicit pointers. Pointer variables in C/C++ are
dereferenced using star operator ($*$), for example, $\Y := *\X$. Fields of a
pointer to an aggregate data type ($\union$, $\struct$, or $\class$) can be
accessed using star operator ($*$) and dot operator ($.$), for example, $(*\X) .
\f$, or using arrow operator ($\verb+->+$), for example, $\X \verb+->+ \f$;
both are equivalent pointer dereferences of the member field $\f$ of pointer
variable $\X$. In Java, fields are dereferenced using the dot operator ($.$),
for example, $\X . \f$.

{\em Analysis of scalar and aggregate pointers.}
In Java, a pointer variable cannot point to an object of scalar data type such
as integer or floating point number; pointer variables point to an object of
only aggregate data types in Java such as structures, classes etc.  However,
C/C++ allows pointers to both scalars and aggregate structures. In C++, pointer
analysis of scalar variables is comparatively straightforward (due to type
restrictions) as compared to the pointer analysis of aggregate variables.  For
example, a program statement $\X := *\X$ is syntactically invalid---the scalar
pointer $\X$ cannot advance to a location of a different data type. On the other
hand an aggregate pointer can be advanced subject to its type compatibility
making it difficult to find properties of such pointers. For example, program
statement $\X := \X \verb+->+ \f$ in a loop allows the aggregate pointer $\X$ to
point to any location after $\X$ through field $\f$.  Further, cycles in
recursive data structures, cause infinite number of paths that refer to the same
memory location. This makes the analysis of an aggregate pointer challenging
over a scalar pointer.

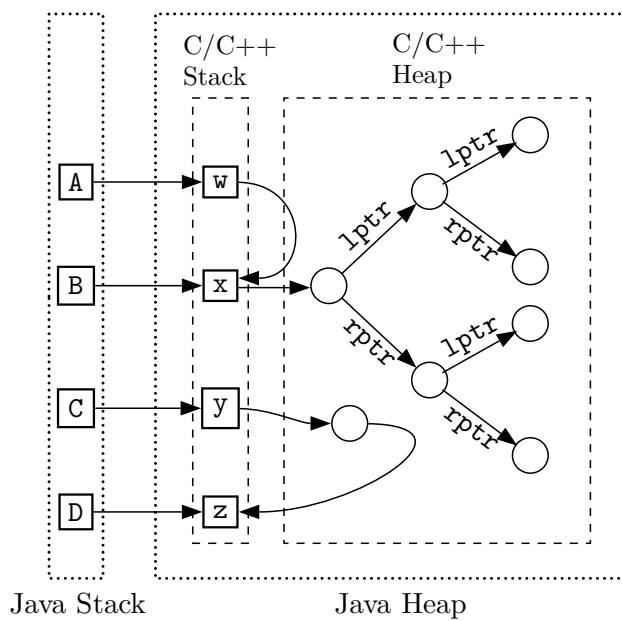
\begin{figure}[t]
\begin{center}
  {
    \psset{unit=.25mm}
    \psset{linewidth=.2mm,arrowsize=5.5pt,arrowinset=0}

\begin{pspicture}(-70,-236)(271,93)
{
\renewcommand{\lptr}{{\ttfamily lptr}}
\renewcommand{\rptr}{{\ttfamily rptr}}
\psframe[linewidth=.4mm,fillstyle=solid,linestyle=dotted,dotsep=1.5pt](10,-211)(260,90)
\psframe[linewidth=.2mm,fillstyle=solid,linestyle=dashed,dash=3pt 3pt](78,-192)(240,45)
\psframe[linewidth=.2mm,fillstyle=solid,linestyle=dashed,dash=3pt 3pt](30,-192)(60,45)
\psframe[linewidth=.4mm,fillstyle=solid,linestyle=dotted,dotsep=1.5pt](-16,-211)(-46,90)
\rput(160,64){\small \begin{tabular}{l}C/C++ \\ Heap\end{tabular}} 
\rput(50,63){\small \begin{tabular}{l}C/C++ \\ Stack\end{tabular}} 
\rput(-30,-225){\begin{tabular}{l}Java Stack\end{tabular}} 
\rput(140,-225){\begin{tabular}{l}Java Heap\end{tabular}} 
\putnode{x}{origin}{45}{-55}{\psframebox[fillstyle=solid,
			fillcolor=white,linewidth=.3mm]{\ttfamily x}}
\putnode{w}{x}{0}{55}{\psframebox[framesep=4.3,fillstyle=solid,
			fillcolor=white,linewidth=.3mm]{\ttfamily w}}
\putnode{y}{x}{0}{-65}{\psframebox[framesep=4.95,fillstyle=solid,
			fillcolor=white,linewidth=.3mm]{\ttfamily y}}
\putnode{z}{y}{0}{-55}{\psframebox[framesep=4.3,fillstyle=solid,linewidth=.3mm]{\ttfamily z}}
\putnode{W}{w}{-76}{0}{\psframebox[framesep=4,fillstyle=solid,linewidth=.3mm]{\ttfamily A}}
\putnode{X}{x}{-76}{0}{\psframebox[framesep=4.8,fillstyle=solid,
			fillcolor=white,linewidth=.3mm]{\ttfamily B}}
\putnode{Y}{y}{-76}{0}{\psframebox[framesep=4.8,fillstyle=solid,
			fillcolor=white,linewidth=.3mm]{\ttfamily C}}
\putnode{Z}{z}{-76}{0}{\psframebox[framesep=4,fillstyle=solid,linewidth=.3mm]{\ttfamily D}}

\putnode{a}{x}{57}{0}{\pscirclebox[fillstyle=solid,fillcolor=white,
		framearc=.3,framesep=9]{}}
\putnode{b}{a}{53}{50}{\pscirclebox[fillstyle=solid,fillcolor=white,
		framearc=.3,framesep=9]{}}
\putnode{i}{a}{53}{-50}{\pscirclebox[fillstyle=solid,fillcolor=white,
		framearc=.3,framesep=9]{}}
\putnode{c}{b}{53}{30}{\pscirclebox[fillstyle=solid,fillcolor=white,
		framearc=.3,framesep=9]{}}
\putnode{f}{b}{53}{-40}{\pscirclebox[fillstyle=solid,fillcolor=white,
		framearc=.3,framesep=9]{}}
\putnode{j}{i}{53}{30}{\pscirclebox[fillstyle=solid,fillcolor=white,
		framearc=.3,framesep=9]{}}
\putnode{m}{i}{53}{-40}{\pscirclebox[fillstyle=solid,fillcolor=white,
		framearc=.3,framesep=9]{}}
\ncline[nodesep=-3]{->}{a}{i}
\bput[1pt]{:U}{\rptr}
\ncline[nodesep=-3]{->}{a}{b}
\aput[1pt]{:U}{\lptr}
\ncline[nodesep=-3]{->}{b}{c}
\aput[1pt]{:U}{\lptr}
\ncline[nodesep=-3]{->}{b}{f}
\bput[1pt]{:U}{\rptr}
\ncline[nodesep=-3]{->}{i}{m}
\bput[1pt]{:U}{\rptr}
\ncline[nodesep=-3]{->}{i}{j}
\aput[1pt]{:U}{\lptr}

\putnode{i}{a}{11}{-73}{\pscirclebox[fillstyle=solid,fillcolor=white,
		framearc=.3,framesep=9]{}}
\ncline[offset=-1]{->}{x}{a}
\nccurve[angleA=-1,angleB=180,linecolor=black]{->}{y}{i}
\nccurve[angleA=-1,angleB=0,linecolor=black,ncurv=1.5]{->}{i}{z}
\nccurve[angleA=-1,angleB=0,linecolor=black,ncurv=1.5,offsetB=-4]{->}{w}{x}
\ncline{->}{W}{w}
\ncline{->}{X}{x}
\ncline{->}{Y}{y}
\ncline{->}{Z}{z}
}
\end{pspicture}
}
\end{center}
\caption{C/C++ memory framework modeled as a Java memory framework.}
\label{fig:memmodel}
\figrule
\end{figure}

{\em Mapping C/C++ memory to the Java memory.} 
As explained before, C/C++ heap and stack pointers can point to locations on
both stack and heap. On the other hand, Java stack pointers can point only to
Java heap locations. In spite of this difference in memory modeling, stack and
heap memory in C/C++ can be modeled like a Java memory. To achieve this, C/C++
memory is viewed as consisting of two partitions of the memory---addresses of
variables and the rest of the memory (stack and heap together)~\cite{ksk07}.
Here, the first partition of the C/C++ memory (i.e.  the addresses of variables)
works like the Java stack. The second partition of the C/C++ memory consisting
of the rest of the memory (stack and heap together) works like the Java heap.

Figure~\ref{fig:memmodel} illustrates a C/C++ memory snapshot, which has been
modeled as Java memory (in dotted lines). Pointer variables $\W$, $\X$, $\Y$,
and $\Z$ are on the C/C++ stack and pointer variables $\A$, $\B$, $\C$, and
$\D$ are on the Java stack. C/C++ pointers point to stack variables $\X$ and
$\Z$ in the figure.  The stack and heap of C/C++ are represented as the Java
heap. Java stack is the set of addresses of C/C++ locations (viz. $\W$, $\X$,
$\Y$, and $\Z$) stored in $\A$, $\B$, $\C$, and $\D$, respectively. To overcome
the difference of pointer dereferences ($*$) and addressof ($\&$) operator in
C/C++ which are absent in Java, Khedker et al. \cite{ksk07} model these two C/C++
constructs as follows:
\begin{itemize}
\item Pointer dereference ($*$) is considered as a field dereference $\deref$,
which has not been used elsewhere in the program. For example~\cite{ksk07},
$(*\X).\f$ in C/C++ is viewed as $\X.\deref.\f$ in Java.  \item The addresses
of C/C++ variables are represented by the Java stack (as shown in
figure~\ref{fig:memmodel}, where $\A$ denotes $\&\W$, $\B$ denotes $\&\X$, $\C$
denotes $\&\Y$, and $\D$ denotes $\&\Z$). For example~\cite{ksk07}, $\Y.\f$
in Java is modeled as $\&\Y.\deref.\f$ in C/C++.
\end{itemize}


\end{document}